\documentclass[10pt]{article}

\usepackage[dvips]{epsfig} 
\usepackage{amssymb,amsmath,amsthm,mathrsfs} 
\usepackage{graphicx}
\usepackage{lineno}

\usepackage[authoryear,sort&compress]{natbib}
\usepackage{url}
\usepackage{enumerate}
\usepackage{colordvi,color,pspicture}
\usepackage{psfrag}

\usepackage{setspace}

\newcommand{\U}{\mathcal{U}}
\newcommand{\I}{\mathbf{I}}

\newcommand{\E}{\mathbf{E}}
\newcommand{\Var}{\mathbf{Var}}
\newcommand{\Cov}{\mathbf{Cov}}

\newcommand{\ah}{\widehat{\alpha}}
\newcommand{\bh}{\widehat{\beta}}

\theoremstyle{plain}
\newtheorem{theorem}{Theorem}[section]

\theoremstyle{definition}

\theoremstyle{remark}

\newtheorem{remark}[theorem]{Remark}

\begin{document}


\title{Technical Report \# KU-EC-08-7:\\
Class-Specific Tests of Spatial Segregation Based on Nearest Neighbor Contingency Tables}
\author{
Elvan Ceyhan
\thanks{Department of Mathematics, Ko\c{c} University, Sar{\i}yer, 34450, Istanbul, Turkey. {\it E-mail:} elceyhan@ku.edu.tr}
}
\date{\today}
\maketitle


\begin{abstract}
\noindent
The spatial interaction between two or more classes (or
species) has important consequences in many fields and might cause
multivariate clustering patterns such as segregation or association.
The spatial pattern of \emph{segregation} occurs when members of a
class tend to be found near members of the same class (i.e.,
conspecifics), while \emph{association} occurs when members of a
class tend to be found near members of the other class or classes.
These patterns can be tested using a nearest neighbor contingency
table (NNCT). The null hypothesis is randomness in the nearest
neighbor (NN) structure, which may result from --- among other
patterns --- random labeling (RL) or complete spatial randomness
(CSR) of points from two or more classes (which is called
\emph{CSR independence}, henceforth). In this article, we consider
Dixon's class-specific tests of segregation and introduce a new
class-specific test, which is a new decomposition of Dixon's overall
chi-square segregation statistic. We demonstrate that the tests we
consider provide information on different aspects of the spatial
interaction between the classes and are conditional under the
CSR independence pattern, but not under the RL pattern. We analyze
the distributional properties and prove the consistency of these
tests; compare the empirical significant levels (Type I error rates)
and empirical power estimates of the tests using extensive Monte Carlo
simulations.
We demonstrate that the new class-specific tests also have
comparable performance with the currently available tests based on NNCTs
in terms of Type I error and power.
For illustrative purposes, we use three example data sets.
We also provide guidelines for using these tests.
\end{abstract}

\noindent
{\small {\it Keywords:} Association; clustering; completely mapped data;
complete spatial randomness; random labeling; spatial point pattern

\vspace{.25 in}



\newpage


\section{Introduction}
\label{sec:intro}
Spatial patterns have important implications in various fields,
such as epidemiology, population biology, and ecology.
A spatial point pattern includes the locations of some measurements,
such as the coordinates of trees in a region of interest.
These locations are referred to as \emph{events} by some authors,
in order to distinguish them from arbitrary points in the region of interest (\cite{diggle:2003}).
However in this article such a distinction is not necessary, 
as we only consider the locations of events.
Hence \emph{points} will refer to the locations of events, henceforth.
It is of practical interest to investigate the patterns
of one type of points with respect to other types.
See, for example, \cite{pielou:1961}, \cite{whipple:1980}, \cite{dixon:1994, dixon:NNCTEco2002}.
In fact, most point patterns marked point patterns generated by marked point processes,
which define the distributions of the ``marks" or ``class labels" to the locations
of the points and perhaps are the most common spatial point patterns
(\cite{diggle:2003}, \cite{gavrikov:1995}, \cite{penttinen:1992}, and \cite{schlather:2004}).
For convenience and generality,
we call the different types of points as ``classes",
but the class can stand for any characteristic of an observation at
a particular location.
For example, the spatial segregation pattern
has been investigated for \emph{plant species} (\cite{diggle:2003}),
\emph{age classes} of plants (\cite{hamill:1986}),
\emph{fish species} (\cite{herler:2005}),
and \emph{sexes} of dioecious plants (\cite{nanami:1999}).
Many of the epidemiologic applications are for a two-class system of
case and control labels (\cite{waller:2004}).

In the analysis of multivariate point patterns,
the null pattern is usually one of the two (random) pattern
types: \emph{random labeling} (RL) or \emph{complete spatial randomness} (CSR)
of two or more classes (i.e., \emph{CSR independence}).
We consider two major types of spatial patterns as alternatives:
\emph{association} and \emph{segregation}.
{\em Association} occurs if the nearest neighbor (NN) of an individual is more
likely to be from another class than to be from the same class.
{\em Segregation} occurs if the
NN of an individual is more likely to be of the same
class as the individual than to be of another class (see, e.g., \cite{pielou:1961}).

Many tests of spatial segregation have been proposed in the literature (\cite{kulldorff:2006}).
These include comparison of Ripley's $K$ or $L$-functions (\cite{ripley:2004}),
comparison of NN distances (\cite{diggle:2003}, \cite{cuzick:1990}),
and analysis of nearest neighbor contingency tables (NNCTs)
which are constructed using the NN frequencies of classes
(\cite{pielou:1961}, \cite{meagher:1980}).
\cite{pielou:1961} proposed tests (of segregation, symmetry, niche specificity,
etc.) and \cite{dixon:1994} introduced an
overall test of segregation, cell- and class-specific tests based on NNCTs
for the two-class case and \cite{dixon:NNCTEco2002} extended his tests to multi-class case.
\cite{ceyhan:2008cell} proposed new cell-specific and overall segregation tests
which are more robust to the differences in the relative abundance of classes
and have better performance in terms of size and power.
The tests (i.e., inference) based on Ripley's $K$ or $L$-functions are only appropriate
when the null pattern can be assumed to be the CSR independence pattern,
if the null pattern is the RL of points from an inhomogeneous Poisson pattern,
they are not appropriate (\cite{kulldorff:2006}).
But, there are also variants of $K(t)$ that explicitly correct for inhomogeneity
(see \cite{baddeley:2000b}).
Cuzick and Edward's $k$-NN tests are designed for testing bivariate spatial interaction
and mostly used for spatial clustering of cases or controls in epidemiology.
Diggle's $D$-function is a modified version of Ripley's $K$-function (\cite{diggle:2003})
and is appropriate for the case in which the null pattern is the RL
of points from any point pattern.
However, Ripley's and Diggle's functions are designed to analyze
univariate or bivariate spatial interaction at various scales
(i.e., inter-point distances).
On the other hand,
Dixon's overall segregation test is a compound summary statistic and applicable
to test (small-scale) multivariate spatial interaction between classes in a given study area;
(base-)class-specific test provides
information on the NN distribution of a base class (if $(X,Y)$ is a pair of points in
which $Y$ is the closest point to $X$,
then $X$ is the base point and $Y$ is the NN point),
while NN-class-specific test (which is introduced in this article)
provides the (base) distribution of classes to which members of a class serve as NNs.
Hence, the NNCT-tests answer different questions;
i.e., they provide information about different aspects of the spatial
interaction (at small scales) compared to each other and the other tests in literature.

\cite{pielou:1961} used the usual Pearson's $\chi^2$ test of
independence for testing spatial segregation.
Due to the ease in computation and interpretation,
Pielou's test of segregation has been used frequently (\cite{meagher:1980}) for
both completely mapped or sparsely sampled data,
although it is not appropriate for such data (\cite{meagher:1980}, \cite{dixon:1994}).
For example, Pielou's test is used for testing the segregation
between males and females in dioecious species (see, e.g., \cite{herrera:1988}
and \cite{armstrong:1989}), and between different species (\cite{good:1982}).

In this article, we discuss Dixon's overall and class-specific test of segregation
and introduce a new class-specific test.
We only consider \emph{completely mapped data};
i.e., for our data sets, the locations of all events in a defined space are observed.
We provide the null and alternative patterns in Section \ref{sec:null-alt};
describe the NNCTs in Section \ref{sec:NNCT};
provide the base- and NN-class-specific tests in Section \ref{sec:class-spec};
empirical significance levels for the two-class case in Section \ref{sec:monte-carlo-2Cl},
for the three-class case in Section \ref{sec:monte-carlo-3Cl};
empirical power comparisons for the segregation and association alternatives
for the two-class case in Section \ref{sec:emp-power-2Cl},
for the three-class case in Section \ref{sec:emp-power-3Cl};
examples in Section \ref{sec:examples};
and our conclusions and guidelines for using the tests in Section \ref{sec:disc-conc}.

\section{Null and Alternative Patterns}
\label{sec:null-alt}
In this section, for simplicity,
we describe the spatial point patterns for two classes only;
the extension to multi-class case is straightforward.

In the univariate (i.e., one-class) spatial point pattern analysis,
the null hypothesis is usually \emph{complete spatial randomness} (\emph{CSR}) (\cite{diggle:2003}).
Given a spatial point pattern $\mathcal P=\{X_i\cdot \I(X_i \in D), i=1,\ldots,n: D \subseteq \mathbb{R}^2 \}$
where $X_i$ stands for the location of event $i$ (i.e., point $i$)
and $\I(X_i \in D)$ is the indicator function
which denotes the Bernoulli random variable denoting the event that
point $i$ is in region $D$.
The pattern $\mathcal P$ exhibits CSR if
given $n$ events in domain $D$, the events are an independent
random sample from the uniform distribution on $D$.
This implies that there is no spatial interaction;
i.e., the locations of these points have no influence on one another.
Furthermore, when the reference region $D$ is large,
the number of points in any planar region with area
$A(D)$ follows (approximately) a Poisson distribution with
intensity $\lambda$ and mean $\lambda \cdot A(D)$.

To investigate the spatial interaction between two or more classes
in a bivariate process, usually there are two benchmark hypotheses:
(i) \emph{independence}, which implies two classes of points are generated by a pair of independent
univariate processes and
(ii) \emph{random labeling} (\emph{RL}), which implies that the class labels
are randomly assigned to a given set of locations in the region of interest (\cite{diggle:2003}).
In complete spatial randomness (CSR) independence, points from each of the two classes satisfy the CSR pattern
in the region of interest.
On the other hand, RL is the pattern in which,
given a fixed set of points in a region,
class labels are assigned to these fixed
points randomly so that the labels are independent of the locations.
So, RL is less restrictive than CSR independence.
CSR independence is a process defining the spatial distribution
of event locations, while RL is a process, conditioned on locations,
defining the distribution of labels on these locations.

We consider either pattern as our null hypotheses in this article.
That is, when the points are assumed to be uniformly
distributed over the region of interest, then the null hypothesis we
consider is
$$H_o: CSR\;\,independence$$
and when only the labeling (marking) of a set of fixed points,
where the allocation of the points might be regular, aggregated, clustered,
or of lattice type, is considered, our null hypothesis is
$$H_o: RL.$$
We discuss the differences in practice and theory for both cases.
The distinction between CSR independence and RL is very important
when defining the appropriate null model for each empirical case;
i.e., the null model depends on the particular ecological context.
\cite{goreaud:2003} assert that CSR (independence) implies that the two classes are \emph{a priori}
the result of different processes (e.g., individuals of different species or age cohorts),
whereas RL implies that some processes affect \emph{a posteriori}
the individuals of a single population
(e.g., diseased vs. non-diseased individuals of a single species).
Although CSR independence and RL are not same,
they lead to the same null model (i.e., randomness in NN structure) in tests using NNCT,
which does not require spatially-explicit information.
We provide the differences in the proposed tests
under either null hypotheses.

As clustering alternatives, we consider two major types of spatial patterns:
\emph{association} and \emph{segregation}.
{\em Association} occurs if the NN of an individual is more
likely to be from another class than to be of the same class as the individual.
For example, in plant biology, the two classes of points
might represent the coordinates of mutualistic plant species,
so the species depend on each other to survive.
As another example, one class of points
might be the geometric coordinates of
parasitic plants exploiting  the other plant
whose coordinates are of the other class.
In epidemiology, one class of points might
be the geographical coordinates of contaminant sources,
such as a nuclear reactor, or a factory emitting toxic waste,
and the other class of points might be the coordinates of the residences of cases
(incidences) of certain diseases, e.g., some type of cancer caused by the contaminant.

{\em Segregation} occurs if the
NN of an individual is more likely to be of the same
class as the individual than to be from a different class;
i.e., the members of the same class tend to be clumped or clustered
(see, e.g., \cite{pielou:1961}).
For instance, one type of plant might not grow well
around another type of plant, and vice versa.
In plant biology, points from one class might represent
the coordinates of trees from a species with large canopy,
so that other plants (whose coordinates are the points from the other class)
that need light cannot grow (well or at all) around these trees.
See, for instance, \cite{dixon:1994} and \cite{coomes:1999}.

The two patterns of segregation or association are not symmetric in the sense that,
when two classes are segregated (associated), they do not necessarily
exhibit the same degree of segregation (association).
Many different forms of segregation (and association) are possible.
Although it is not possible to list all
segregation types, its existence can be tested by an analysis of the
NN relationships between the classes (\cite{pielou:1961}).

\section{Nearest Neighbor Contingency Tables}
\label{sec:NNCT}
NNCTs are constructed using the NN frequencies of classes.
We describe the construction of NNCTs for two classes from a binomial spatial process;
extension to multi-class case is straightforward.
Consider two classes with labels $\{1,2\}$.
Let $N_i$ be the number of points from class $i$ for $i \in \{1,2\}$ and
$n$ be the total sample size, so $n=N_1+N_2$.
If we record the class
of each point and the class of its NN, the NN
relationships fall into four distinct categories:
$(1,1),\,(1,2);\,(2,1),\,(2,2)$ where in cell $(i,j)$,
class $i$ is the \emph{base class},
while class $j$ is the class of \emph{NN} of class $i$.
That is, the $n$ points constitute $n$ (base,NN) pairs.
Then each pair can be categorized with respect to
the base label (row categories) and NN label (column categories).
Denoting $N_{ij}$ as the frequency of cell $(i,j)$ for $i,j \in
\{1,2\}$, we obtain the NNCT in Table \ref{tab:NNCT-2x2} where $C_j$
is the sum of column $j$; i.e., number of times class $j$ points
serve as NNs for $j \in \{1,2\}$.
Furthermore, $N_{ij}$ is the cell count for
cell $(i,j)$ that is the sum of all (base,NN) pairs each of which
has label $(i,j)$.
Note also that
$n=\sum_{i,j}N_{ij}$; $n_i=\sum_{j=1}^2\, N_{ij}$; and
$C_j=\sum_{i=1}^2\, N_{ij}$.
By construction, if $N_{ij}$ is larger (smaller) than expected,
then class $j$ serves as NN more (less) to class $i$ than expected,
which implies (lack of) segregation if $i=j$ and (lack of) association of class $j$
with class $i$ if $i\not=j$.
Furthermore, we adopt the convention that variables denoted by upper (lower) case letters
are random (fixed) quantities.
Hence, column sums and cell counts are random, while row sums and the overall sum are fixed
quantities in a NNCT.
However if we want to have the overall sum to be random also,
we might consider a Poisson spatial process.

\begin{table}[ht]
\centering
\begin{tabular}{cc|cc|c}
\multicolumn{2}{c}{}& \multicolumn{2}{c}{NN class}& \\
\multicolumn{2}{c|}{}& class 1 &  class 2 & sum  \\
\hline
&class 1 &    $N_{11}$  &   $N_{12}$  &   $n_1$  \\
\raisebox{1.5ex}[0pt]{base class}
&class 2 &    $N_{21}$ &  $N_{22}$    &   $n_2$  \\
\hline
& sum    &    $C_1$   & $C_2$         &   $n$  \\
\end{tabular}
\caption{
\label{tab:NNCT-2x2}
NNCT for two classes.}
\end{table}

Note that, under segregation, the diagonal entries,
$N_{ii}$ for $i=1,2$, tend to be larger than expected;
while under association, the off-diagonals tend to be larger than expected.
The general two-sided alternative is that some cell counts are different
than expected under the null hypothesis (i.e., CSR independence or RL).

\cite{pielou:1961} used $\chi^2$ test of independence to test for segregation.
The assumption for the use
of chi-square test for NNCTs is the independence between cell-counts
(and rows and columns also), which is violated under CSR independence or RL.
This problem was first
noted by \cite{meagher:1980} who identify the main source
of it to be reflexivity of (base, NN) pairs.
A (base,NN) pair $(X,Y)$ is \emph{reflexive} if $(Y,X)$ is also a (base,NN) pair.
As an alternative,
they suggest using Monte Carlo simulations for Pielou's test.
\cite{dixon:1994} derived the appropriate asymptotic sampling
distribution of cell counts using Moran join count statistics (\cite{moran:1948}) and
hence the appropriate test which also has a $\chi^2$-distribution asymptotically.

\section{Class-Specific Tests of Spatial Segregation}
\label{sec:class-spec}
In this section, we discuss the row-wise partitioning
of the overall segregation test due to \cite{dixon:EncycEnv2002}
and introduce the column-wise partitioning.
These class-specific tests can also be viewed as
resulting from combining the cells at each row or column, respectively.

\subsection{Base-Class-Specific Test of Spatial Segregation}
\label{sec:dix-class-spec}

Dixon's overall test of segregation tests the hypothesis that all
cell counts in the NNCT are equal to their expected values (\cite{dixon:1994}).
Under RL, these expected cell counts are as
\begin{equation}
\label{eqn:Exp[Nij]}
\E[N_{ij}]=
\begin{cases}
n_i(n_i-1)/(n-1) & \text{if $i=j$,}\\
n_i\,n_j/(n-1)     & \text{if $i \not= j$,}
\end{cases}
\end{equation}
where $n_i$ is a realization of $N_i$, i.e., is the fixed sample size for
class $i$ for $i=1,2,\ldots,q$.
Observe that the expected cell counts depend only on the size of
each class (i.e., row sums), but not on column sums.
In the multi-class case with $q$ classes, \cite{dixon:NNCTEco2002} suggests the quadratic form
\begin{equation}
\label{eqn:dix-chisq-qxq}
C=(\mathbf{N}-\E[\mathbf{N}])'\Sigma^-(\mathbf{N}-\E[\mathbf{N}])
\end{equation}
where $\mathbf{N}$ is the $q^2\times1$ vector of $q$ rows
of cell counts concatenated row-wise,
$\E[\mathbf{N}]$ is the vector of $\E[N_{ii}]$ which are as in Equation (\ref{eqn:Exp[Nij]}),
$\Sigma$ is the $q^2 \times q^2$ variance-covariance matrix for
the cell count vector $\mathbf{N}$ with diagonal entries being equal to
$\Var[N_{ii}]$ and off-diagonal entries being $\Cov[N_{ij},\,N_{kl}]$
for $(i,j) \neq (k,l)$.
The explicit forms of the variance and covariance terms are provided in (\cite{dixon:NNCTEco2002}).
Furthermore, $\Sigma^-$ is a generalized inverse of $\Sigma$ (\cite{searle:2006})
and $'$ stands for the transpose of a vector or matrix.
The overall test statistic $C$ can be partitioned into $q$
class-specific (or species-specific) test statistics.
Consider the NNs of type $i$ base points,
i.e., \emph{row $i$} in the NNCT
denoted by $\mathbf N_i=(N_{i1},N_{i2},\ldots,N_{iq})'$.
Then
\begin{equation}
\label{eqn:base-class}
C_B(i)=(\mathbf N_i-\E[\mathbf N_i])'\Sigma_i^-(\mathbf N_i-\E[\mathbf N_i])
\end{equation}
where $\E[\mathbf N_i]$ is the vector of expected cell counts
for row $i$ and $\Sigma_i^-$ is the generalized inverse
of variance-covariance matrix of $\mathbf{N_i}$.
Asymptotically $C_B(i)$ has $\chi^2_{(q-1)}$ distribution
since $\Sigma_i$ has rank $(q-1)$.
Furthermore, $C_B(i)$ are not independent,
hence do not sum to $C$.
Each $C_B(i)$ tests whether each row $i$ is similar to the one under RL;
that is, the frequencies of the NNs for \emph{base-class} $i$ is as expected under RL:
\begin{equation}
\label{eqn:null-base-class}
H_o:\E[\mathbf N_i]=\bigl (\E[N_{i1}],\E[N_{i2}],\ldots,\E[N_{iq}] \bigr)'
\end{equation}
where $\E[N_{ij}]$ are as in Equation \eqref{eqn:Exp[Nij]}.
Therefore, we will call this type of class-specific test as \emph{base-class-specific
test of segregation} for class $i$, and hence the notation $C_B(i)$.

The tests in Equations \eqref{eqn:dix-chisq-qxq} and \eqref{eqn:base-class} depend on the quantities $Q$ and $R$
where $Q$ is the number of points with shared  NNs,
which occurs when two or more points share a NN and
$R$ is twice the number of reflexive pairs.
Then $Q=2\,(Q_2+3\,Q_3+6\,Q_4+10\,Q_5+15\,Q_6)$
where $Q_i$ is the number of points that serve
as a NN to other points $i$ times.
Notice that under RL, $Q$ and $R$ are fixed quantities,
as they depend only on the location of the points, not the types of NNs.

\subsection{NN-Class-Specific Tests of Spatial Segregation}
\label{sec:NN-class-spec-segreg}
Another way to partition $C$ in Equation \eqref{eqn:dix-chisq-qxq}
is by columns instead of rows.
Consider the frequency of type $j$ points serving as NN to all other types
(including class $j$), i.e.,
\emph{column} $j$, namely, $\mathbf C_j=(N_{1j},N_{2j},\ldots,N_{qj})'$.

Under RL, for each column $j$, we have
\begin{equation}
\label{eqn:NN-class}
C_{NN}(j)=(\mathbf C_j-\E[\mathbf C_j])'\Sigma_j^{-1}(\mathbf C_j-\E[\mathbf C_j]),
\end{equation}
where $\E[\mathbf C_j]$ is the vector of expected cell counts for column $j$
and $\Sigma_j^{-1}$ is the inverse of variance-covariance matrix of $\mathbf C_j$.
The test statistics
$C_{NN}(j)$ are asymptotically distributed as $\chi^2_q$,
since $\Sigma_j$ has rank $q$.
The test statistics $C_{NN}(j)$ are dependent, hence do not sum to $C$.

If significant, these class-specific tests imply that
class $j$ serves more (or less) frequently as \emph{NN} to other classes
(including class $j$ itself) than expected under RL:
\begin{equation}
\label{eqn:null-NN-class}
H_o:\E[\mathbf C_j]= \bigl (\E[N_{1j}],\E[N_{2j}],\ldots,\E[N_{qj}] \bigr )'
\end{equation}
where $\E[N_{ij}]$ are as in Equation \eqref{eqn:Exp[Nij]}.
Hence the notation $C_{NN}(j)$.
Moreover, we call them \emph{NN-class-specific tests of segregation}
which are also suggestive of deviation from RL,
which might indicate segregation or association.

\begin{remark}
\label{rem:QandR}
\textbf{The Status of $Q$ and $R$ under CSR Independence and RL:}
Under CSR independence, the expected cell counts in Equation \eqref{eqn:Exp[Nij]},
the overall test statistic in Equation \eqref{eqn:dix-chisq-qxq}, and
the base-class-specific test in Equation \eqref{eqn:base-class}
and NN-class-specific test in Equation \eqref{eqn:NN-class}
and the relevant discussions are similar to the RL case.
The only difference is that under RL the quantities $Q$ and $R$
are fixed, while under CSR independence they are random.
That is, under CSR independence, $C_B(i)$ asymptotically has $\chi^2_{(q-1)}$ distribution 
and $C_{NN}(j)$ asymptotically has $\chi^2_q$ distribution conditional on $Q$ and $R$,
because the variances and covariances used in Equations \eqref{eqn:base-class}
and \eqref{eqn:NN-class}, and all the quantities depending on these
expectations also depend on $Q$ and $R$.
Hence under the CSR independence model they are conditional variances and
covariances obtained by conditioning on $Q$ and $R$.
The unconditional variances and covariances can be obtained
by using the  conditional expectation formula incorporating $Q$ and $R$.
Letting $W=(Q,R)'$, we have
\begin{eqnarray}
\Var[N_{ij}] &=& \E_W[\Var[N_{ij}|W]]+\Var_W[\E[N_{ij}|W]]\nonumber \\
             &=& \E_W[\Var[N_{ij}|W]]\nonumber
\end{eqnarray}
since $\E[N_{ij}|W]$ is independent of $Q$ and $R$.
Similarly,
\begin{eqnarray}
\Cov[N_{ij},N_{kl}] &=& \E_W[\Cov[N_{ij},N_{kl}|W]]+\Cov_W[\E[(N_{ij},N_{kl})|W]]\nonumber \\
             &=& \E_W[\Cov[N_{ij},N_{kl}|W]]\nonumber
\end{eqnarray}
since $\E[(N_{ij},N_{kl})|W]$ is independent of $Q$ and $R$.
Furthermore, $\Var[N_{ij}|W]$ and $\Cov[N_{ij},N_{kl}|W]$ are linear
in $Q$ and $R$ (\cite{dixon:NNCTEco2002}).
Hence the unconditional variances and covariances can be found by replacing $Q$ and $R$
with their expectations.

Unfortunately, given the difficulty of calculating the
expectations of $Q$ and $R$ under CSR independence,
it is reasonable and convenient to use test statistics employing the
conditional variances and covariances even when assessing their
behavior under the CSR independence model.
\cite{cox:1981} calculated analytically that $\E[R|N]= 0.6215 N$ for a planar
Poisson process.
Alternatively, we can estimate the expected values of $Q$ and $R$ empirically,
substitute these estimates in the expressions.
For example, for homogeneous planar Poisson pattern,
we have $\E[Q|N] \approx 0.63 N$ and $\E[R|N] \approx 0.62 N$
(estimated empirically by 1000000 Monte Carlo simulations for various values of $N$ on unit square).
Notice that $\E[R|N]$ agrees with the analytical result of \cite{cox:1981}.
$\square$
\end{remark}

\begin{remark}
\label{rem:class-spec-2Cl}
\textbf{Comparison of Base- and NN-Class Specific Tests:}
Notice that the base-class-specific test for class $i$ tests whether the
NN frequencies of base class $i$ follows the expected
under CSR independence (conditional on $Q$ and $R$) or RL.
That is, base-class-specific test for class (or species) $i$ is concerned
with how other classes interact as NNs with base class $i$ in the NN structure.
For example, if $(q-1)$ plant species are introduced to
a region with species $i$ being already there,
the base-class-specific test for species $i$ is more appropriate
and informative as it provides information on
how the newly introduced species interact with the already existent base species $i$.
On the other hand, the NN-class-specific test for class $j$ tests whether the
base frequencies for which class $j$ serves as NN follows the expected
under CSR independence (conditional on $Q$ and $R$) or RL.
So, NN-class-specific test for class $j$ is concerned with
how class $j$ points interact as NNs with other base classes in the NN structure.
For example, if species $j$ was introduced to a region
with $(q-1)$ species already present, NN-class-specific
test for species $j$ is more appropriate and informative,
as it provides information on how the newly introduced
species interact with the already existent $(q-1)$ base species.  $\square$
\end{remark}

\subsection{The Two-Class Case}
\label{sec:two-class}
In the two-class case, \cite{dixon:1994} calculates
$Z_{ii}=(N_{ii}-\E[N_{ii}])\big /\sqrt{\Var[N_{ii}]}$ for both $i \in
\{1,2\}$ and then combines these test statistics into a statistic
that is asymptotically distributed as $\chi^2_2$.
The suggested test statistic
is given by
\begin{equation}
\label{eqn:dix-chisq-2x2}
C=\mathbf{Y}'\Sigma^{-1}\mathbf{Y}=
\left[
\begin{array}{c}
N_{11}-\E[N_{11}] \\
N_{22}-\E[N_{22}]
\end{array}
\right]' \left[
\begin{array}{cc}
\Var[N_{11}] & \Cov[N_{11},N_{22}] \\
\Cov[N_{11},N_{22}] & \Var[N_{22}] \\
\end{array}
\right]^{-1} \left[
\begin{array}{c}
N_{11}-\E[N_{11}] \\
N_{22}-\E[N_{22}]
\end{array}
\right].
\end{equation}
Notice that this is equivalent to $\displaystyle C=\frac{Z_{AA}^2+Z_{BB}^2-2\,r\,Z_{AA}Z_{BB}}{1-r^2}$
where $\displaystyle Z_{AA}=\frac{N_{11}-\E[N_{11}]}{\sqrt{\Var[N_{11}]}}$,
$\displaystyle Z_{BB}=\frac{N_{22}-\E[N_{22}]}{\sqrt{\Var[N_{22}]}}$,
and $\displaystyle r=\Cov[N_{11},N_{22}]\big /\sqrt{\Var[N_{11}]\Var[N_{22}]}$.
The overall test statistic $C$ can be
partitioned into two base-class-specific test statistics $C_B(i)$
which asymptotically have $\chi^2_1$,
the chi-square distribution with 1 degree of freedom.

When $C$ is partitioned by columns $\mathbf C_j$,
we get the NN-class-specific test
$C_{NN}(j)$ which  are
asymptotically distributed as $\chi^2_2$.
However, in the two-class case,
$C_{NN}(1)=C_{NN}(2)=C$, since each column $\mathbf C_j$
contains all the information about the NNCT.

\begin{remark}
\label{rem:q>2-class-case}
\textbf{The Multi-Class Case with $q>2$ Classes:}
In the multi-class case with more than 2 classes,
the overall test statistic $C$ can be
partitioned into $q$ base-class-specific test statistics $C_B(i)$
which asymptotically have $\chi^2_{q-1}$.
When $C$ is partitioned by columns $\mathbf C_j$,
we get $q$ NN-class-specific tests $C_{NN}(j)$ which  are
asymptotically distributed as $\chi^2_q$.
In the case of $q>2$,
$C_{NN}(j)$ and $C_{NN}(l)$ are very likely to be different for $j \neq l$
and $C_{NN}(j)$ and $C$ are also very likely to be different for each $j$.
$\square$
\end{remark}

\subsection{Asymptotic Structures for the NNCT-Tests}
\label{sec:asy-structure}
There are two major types of asymptotic structures for spatial data (\cite{lahiri:1996}).
In the first, any two observations are required to be at least a fixed distance apart,
hence as the number of observations increase, the region on which the process
is observed eventually becomes unbounded.
This type of sampling structure is called \emph{increasing domain asymptotics}.
In the second type, the region of interest is a fixed
bounded region and more and more points are observed in this region.
Hence the minimum distance between data points tends to zero
as the sample size tends to infinity.
This type of structure is called \emph{infill asymptotics}, due to \cite{cressie:1993}.

Under RL, the sampling structure in our asymptotic sampling distribution
could be either one of increasing domain or infill asymptotics,
as we only consider the class sizes and hence the total sample size
tending to infinity regardless of the size of the study region.
That is, under RL, the locations on which labels are assigned randomly could be realizations from a process
whose asymptotic structure follows either infill or increasing domain asymptotics.
On the other hand,
under CSR independence, we can not have the increasing domain asymptotics in the usual sense,
since for any bounded subset of the region,
the points will not necessarily follow the uniform distribution as there is a restriction
on the interpoint distances.
But if we let the area of the region go to infinity
while uniformness is preserved for any bounded subset of the region,
this type of ``modified" increasing domain asymptotics is applicable for the CSR independence pattern.
Furthermore, under CSR independence,
the number of points from uniform distribution on a bounded region
going to infinity will give rise to the infill asymptotics.
However, the values and distributions of $Q$ and $R$
are affected by infill and increasing domain asymptotics.
In both asymptotic structures, under CSR independence, the tests will still be conditional on $Q$ and $R$,
whose asymptotic distribution will also depend on the type of the asymptotics.
For the actual conditional result, the only possible asymptotics for which the
current arguments might be appropriate is what might be called \emph{pattern stamp asymptotics}.
In this form of asymptotics, the study region (assumed to be rectangular or at least something that
will tile the plane) and its locations are repeated (stamped).
This multiple stamping of the same pattern of locations maintains $Q$ and $R$ at fixed constants.
However, this is not either of the usual asymptotics for the spatial data nor a realistic case in practice.


\begin{theorem}
(Consistency)
Under RL, the overall test in Equation \eqref{eqn:dix-chisq-qxq}
which rejects for $C>\chi^2_{q(q-1)}(1-\alpha)$ where
$\chi^2_{q(q-1)}(1-\alpha)$ is the $100(1-\alpha)^{th}$ percentile
of $\chi^2_{q(q-1)}$ distribution,
the test against $H_o$ in Equation \eqref{eqn:null-base-class}
which rejects for $C_B(i)>\chi^2_{(q-1)}(1-\alpha)$,
and the test against $H_o$ in Equation \eqref{eqn:null-NN-class} which rejects for
$C_{NN}(j)>\chi^2_q(1-\alpha)$ are consistent.
Under CSR independence, consistency follows conditional on $Q$ and $R$.
\end{theorem}

{\bf Proof:}
The consistency of the overall test is proved in (\cite{ceyhan:overall}).
Under RL, the null hypothesis in Equation \eqref{eqn:null-base-class} is equivalent
to $H_o:\lambda=0$ where $\lambda$ is the non-centrality parameter
for the $\chi^2_{(q-1)}$ distribution.
Any deviation from the null case corresponds to $H_a:\lambda>0$,
then under any specific alternative the power tends to unity
by the consistency of usual $\chi^2$ tests.
The consistency of $C_{NN}(j)$ under RL and of both tests under CSR independence follow similarly.
$\blacksquare$

\section{Empirical Significance Levels for the Two-Class Case}
\label{sec:monte-carlo-2Cl}
In this section, we provide the empirical significance levels
for base- and NN-class-specific tests and Dixon's overall test of segregation
for the two-class case under CSR independence and RL.

\subsection{Empirical Significance Levels for the Two-Class Case under CSR Independence}
\label{sec:CSR-emp-sign-2Cl}
First, we consider the two-class case with classes $X$ and $Y$.
We generate $n_1$ points from class $X$ and $n_2$ points from class $Y$
both of which are independently uniformly distributed on the unit square, $(0,1) \times (0,1)$.
Hence, all $X$ points are independent of each other and so are $Y$ points;
and $X$ and $Y$ are independent data sets.
This will imply randomness in the NN structure,
which is the null hypothesis for our NNCT-tests.
Thus, we simulate the CSR independence pattern for the performance of the tests
under the null case.
We generate $X$ and $Y$ points for some combinations of $n_1,n_2 \in \{10,30,50,100\}$.
We repeat the sample generation $N_{mc}=10000$ times for each sample size combination
for a reasonable precision in the results within a reasonable time period.
For each Monte Carlo replication, we construct the NNCT for classes $X$ and $Y$,
then compute the base- and NN-class-specific tests for each class
and Dixon's overall segregation test.
Out of these 10000 samples the number of significant results by
each test is recorded.
The nominal significance level used in all these tests is $\alpha=0.05$.
Then empirical sizes are calculated as,
e.g., $\widehat \alpha_i^B=m_i/N_{mc}$ where $m_i$ is the number of significant base-class-specific
test statistics for class $i$.
The empirical sizes for NN-class-specific tests and
Dixon's overall segregation test are calculated similarly.

We present the empirical significance
levels for the NNCT-tests in Table \ref{tab:class-spec-null},
where $\ah_i^B$ is the empirical significance level for
(Dixon's) base-class-specific test and
$\ah_i^{NN}$ is for the NN-class-specific
test for $i \in \{1,2\}$ and $\ah_D$ is for Dixon's overall test of segregation.
The empirical sizes significantly smaller (larger) than 0.05 are marked with $^c$ ($^{\ell}$),
which indicate the corresponding test is conservative (liberal).
The asymptotic normal approximation to proportions are used in determining the significance of
the deviations of the empirical sizes from the nominal level of 0.05.
For the tests regarding the proportions,
we also use $\alpha=0.05$ to test against empirical size being equal to 0.05.
Notice that in the two-class case
$\ah_1^B\not=\ah_2^B$ but $\ah_1^{NN}=\ah_2^{NN}=\ah_D$,
so only $\ah_1^{NN}$ is presented.

Observe that (Dixon's) base-class-specific test for class $Y$
(i.e., the larger class in unequal sample size combinations)
yields about the desired significance levels in rejecting $H_o: CSR\;\,independence$
for all sample size combinations.
However, when one of the samples is small the tests are conservative,
but base-class-specific test for class $X$ (i.e., the class with the smaller sample size)
is more conservative than others.
For large classes with unequal sizes (i.e., when the relative abundance of classes is different),
the base-class-specific test tends to be slightly liberal for the smaller class.

\begin{table}[ht]
\centering
\begin{tabular}{|c||c|c||c|}
\hline
\multicolumn{4}{|c|}{Empirical significance levels} \\
\multicolumn{4}{|c|}{under CSR independence} \\
\hline
sizes & \multicolumn{2}{|c||}{base} & \multicolumn{1}{|c|}{NN*} \\
\hline
$(n_1,n_2)$  & $\ah_1^B$ & $\ah_2^B$ & $\ah_1^{NN}$ \\
\hline
(10,10) & .0454$^c$ & .0465 & .0432$^c$ \\
\hline
(10,30) & .0306$^c$ & .0485 & .0440$^c$ \\
\hline
(10,50) & .0270$^c$ & .0464 & .0482 \\
\hline
(30,30) & .0507 & .0505 & .0464 \\
\hline
(30,50) & .0590$^{\ell}$ & .0522 & .0443$^c$ \\
\hline
(50,50) & .0465 & .0469 & .0508 \\
\hline
(50,100) & .0601$^{\ell}$ & .0533 & .0560$^{\ell}$ \\
\hline
(100,100) & .0493 & .0463$^c$ & .0504 \\
\hline
\end{tabular}
\caption{
\label{tab:class-spec-null}
The empirical significance levels for the NNCT-tests
in the two-class case under $H_o: CSR\;\,independence$ with
$N_{mc}=10000$, $n_1,n_2$ in $\{10,30,50,100\}$ at the nominal level of $\alpha=0.05$.
($^c$: empirical size significantly less than 0.05; i.e., the test is conservative.
$^{\ell}$: empirical size significantly larger than 0.05; i.e., the test is liberal.
*only $\ah_1^{NN}$ is presented since $\ah_1^{NN}=\ah_2^{NN}=\ah_{D}$.
base = base-class-specific test, NN = NN-class-specific test.)}
\end{table}


\subsection{Empirical Significance Levels for the Two-Class Case under RL}
\label{sec:RL-emp-sign-2Cl}
Recall that the segregation tests we consider are conditional under the CSR independence pattern.
To evaluate their empirical size performance better,
we also perform Monte Carlo simulations under the RL pattern, for which
the tests are not conditional.
For the RL pattern we consider the following three cases,
in each of which, we first determine the locations of points for which labels
are to be assigned randomly.
Then we apply the RL procedure to these points for the appropriate sample size combinations.

\noindent
\textbf{RL Case (1):}
First, we generate $n=(n_1+n_2)$ points iid $\U((0,1) \times (0,1))$,
the uniform distribution on the unit square,
for some combinations of $n_1,n_2 \in \{10,30,50,100\}$.
The locations of these points are taken to be the fixed locations
for which we assign the labels randomly.
Thus, we simulate the RL pattern for the performance of the tests
under the null case.
For each sample size combination $(n_1,n_2)$,
we randomly choose $n_1$ points (without replacement) and label them
as $X$ and the remaining $n_2$ points as $Y$ points.
We repeat the RL procedure $N_{mc}=10000$ times for each sample size combination.
For each RL iteration, we construct the NNCT for classes $X$ and $Y$,
and then compute the base- and NN-class-specific tests for each class
and Dixon's overall segregation test.
Out of these 10000 samples the number of significant tests by each test is recorded.
The nominal significance level used in all these tests is $\alpha=0.05$.
Based on these significant results, empirical sizes are calculated as
the ratio of number of significant test statistics to the number of Monte
Carlo replications, $N_{mc}$.

\noindent
\textbf{RL Case (2):}
We generate $n_1$ points iid $\U((0,2/3) \times (0,2/3))$ and
$n_2$ points iid $\U((1/3,1) \times (1/3,1))$
for some combinations of $n_1,n_2 \in \{10,30,50,100\}$.
The locations of these points are taken to be the fixed locations
for which we assign labels randomly.
The RL process is applied to these fixed points $N_{mc}=10000$
times for each sample size combination.
The empirical sizes for the tests are calculated similarly as in RL case (1).

\noindent
\textbf{RL Case (3):}
We generate $n_1$ points iid $\U((0,1) \times (0,1))$
$n_2$ points iid $\U((2,3) \times (0,1))$
for some combinations of $n_1,n_2 \in \{10,30,50,100\}$.
RL procedure and the empirical sizes for the tests
are calculated similarly as in the previous RL cases.

\begin{figure}
\rotatebox{-90}{ \resizebox{2. in}{!}{\includegraphics{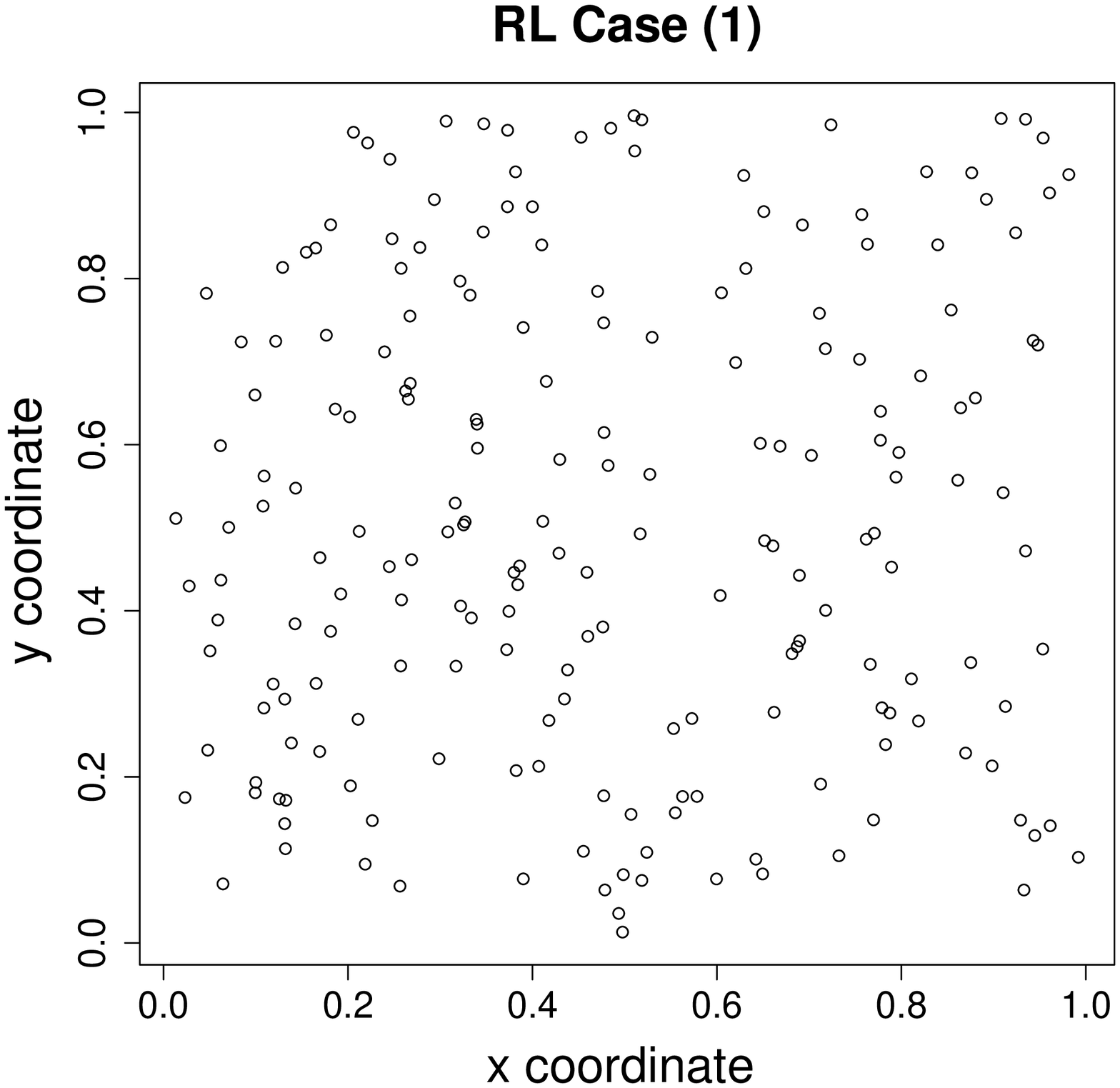} }}
\rotatebox{-90}{ \resizebox{2. in}{!}{\includegraphics{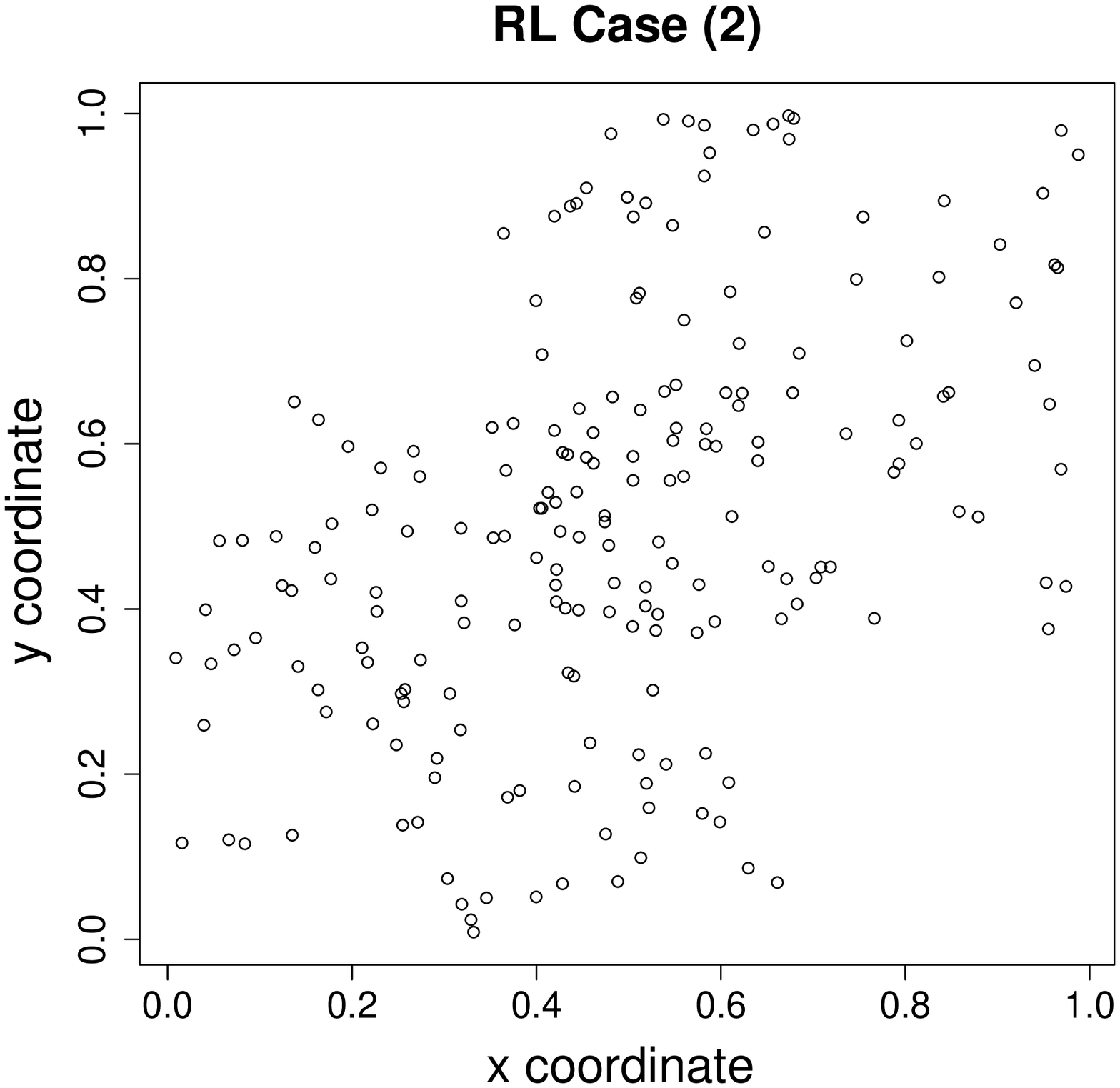} }}
\rotatebox{-90}{ \resizebox{2. in}{!}{\includegraphics{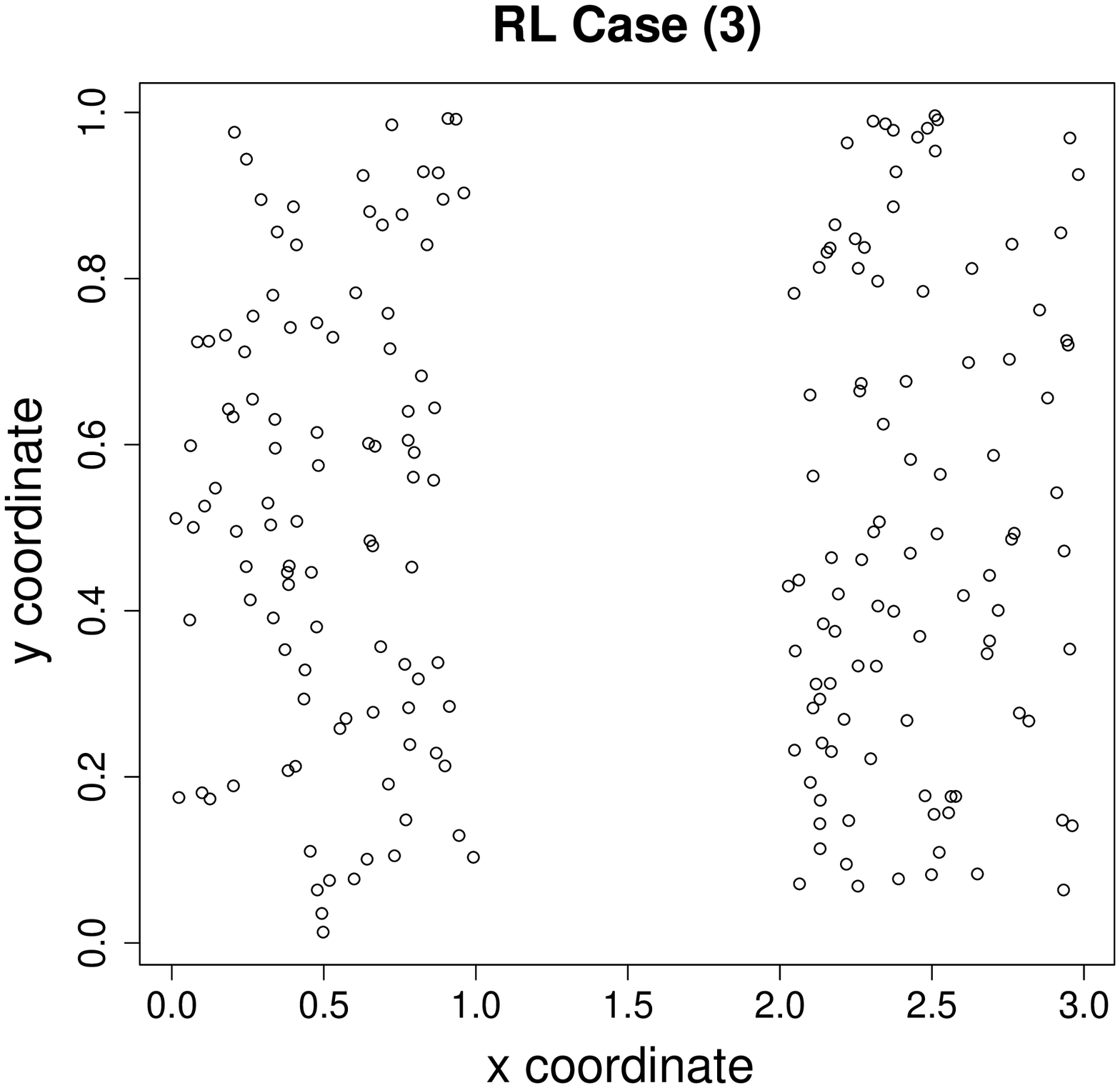} }}
 \caption{
\label{fig:RLcases}
The fixed locations for which RL procedure is applied for RL Cases (1)-(3) with $n_1=n_2=100$
in the two-class case. Notice that $x$-axis for RL Case (3) is differently scaled.
}
\end{figure}

The locations for which the RL procedure is applied in RL Cases (1)-(3) are plotted
in Figure \ref{fig:RLcases} for $n_1=n_2=100$.
We present the empirical significance
levels for the NNCT-tests in Table \ref{tab:class-spec-null-RL},
where the empirical significance level labeling is as in Table \ref{tab:class-spec-null}.
The empirical sizes are marked with $^c$ and $^{\ell}$
for conservativeness and liberalness as in Section \ref{sec:CSR-emp-sign-2Cl}.
Observe that when both samples are small, base-class-specific tests are liberal under RL Case (2),
conservative under other RL cases,
and all other tests (NN-class-specific tests and overall segregation test) are conservative under all RL cases.
When one sample is small, the tests tend to be conservative,
especially for the base-class-specific test for the smaller class.
When both samples are large, the base-class-specific test
is liberal for the smaller class, and other tests are about the desired level.

Comparing Tables \ref{tab:class-spec-null} and \ref{tab:class-spec-null-RL}
we observe that the empirical sizes are about the same under the CSR independence and RL patterns.
However, the tests are conservative when at least one sample is small,
regardless of whether the null case is CSR independence or RL.

\begin{table}[ht]
\centering
\begin{tabular}{|c||c|c|c||c|c|c||c|c|c|}
\hline
\multicolumn{10}{|c|}{Empirical significance levels under RL} \\
\hline
 & \multicolumn{3}{|c||}{RL Case (1)} & \multicolumn{3}{|c||}{RL Case (2)} &
 \multicolumn{3}{|c|}{RL Case (3)} \\
\hline
sizes & \multicolumn{2}{|c|}{base} & \multicolumn{1}{|c||}{NN*} &
\multicolumn{2}{|c|}{base} & \multicolumn{1}{|c||}{NN*} &
\multicolumn{2}{|c|}{base} & \multicolumn{1}{|c|}{NN*} \\
\hline
$(n_1,n_2)$  & $\ah_1^B$ & $\ah_2^B$ & $\ah_1^{NN}$
  & $\ah_1^B$ & $\ah_2^B$ & $\ah_1^{NN}$
  & $\ah_1^B$ & $\ah_2^B$ & $\ah_1^{NN}$ \\
\hline
(10,10) & .0356$^{c}$ & .0341$^{c}$ & .0491 & .0624$^{\ell}$ & .0657$^{\ell}$ & .0446$^c$
& .0444$^{c}$ & .0481$^{\ell}$ & .0404$^c$\\
\hline
(10,30) & .0311$^c$ & .0699$^{\ell}$ & .0466 & .0297$^c$ & .0341$^c$ & .0327$^c$
& .0281$^c$ & .0447$^c$ & .0324$^c$\\
\hline
(10,50) & .0264$^c$ & .0472 & .0507 & .0251$^c$ & .0384$^c$ & .0508 & .0260$^c$
& .0404$^c$ & .0500\\
\hline
(30,30) & .0579 & .0547 & .0497 & .0513 & .0523 & .0469 & .0549$^{\ell}$ & .0553$^{\ell}$ & .0484\\
\hline
(30,50) & .0621$^{\ell}$ & .0608$^{\ell}$ & .0444$^c$ & .0626$^{\ell}$ & .0594$^{\ell}$ & .0411$^c$
& .0677$^{\ell}$ & .0685$^{\ell}$ & .0445$^c$\\
\hline
(50,50) & .0512 & .0524 & .0497 & .0509 & .0511 & .0501 & .0504 & .0506 & .0488\\
\hline
(50,100) & .0625$^{\ell}$ & .0512 & .0482  & .0566$^{\ell}$ & .0421$^c$ & .0460
& .0590$^{\ell}$ & .0484 & .0479\\
\hline
(100,100) & .0538$^{\ell}$ & .0534 & .0525 & .0439$^c$ & .0453$^c$ & .0505 & .0495
& .0476 & .0534\\
\hline
\end{tabular}
\caption{
\label{tab:class-spec-null-RL}
The empirical significance levels
for the two-class case under $H_o:RL$ for RL Cases (1)-(3) with
$N_{mc}=10000$, $n_1,n_2$ in $\{10,30,50,100\}$ at the nominal level of $\alpha=0.05$.
($^c$: empirical size significantly less than 0.05; i.e., the test is conservative.
$^{\ell}$: empirical size significantly larger than 0.05; i.e., the test is liberal.
*only $\ah_1^{NN}$ is presented since $\ah_1^{NN}=\ah_2^{NN}=\ah_{D}$.
base = base-class-specific test, NN = NN-class-specific test.)}
\end{table}

%

\section{Empirical Significance Levels for the Three-Class Case}
\label{sec:monte-carlo-3Cl}
In this section, we provide the empirical significance levels
for the base- and NN-class-specific tests and Dixon's overall test of segregation
for the three-class case under CSR independence and RL.

\subsection{Empirical Significance Levels for the Three-Class Case under CSR Independence}
\label{sec:CSR-emp-sign-3Cl}
In the two-class case, NN-class-specific tests are equal to the overall test of segregation.
But for the $q$-class case with $q > 2$,
neither $C_{NN}(j)$ are likely to be equal to each other nor any of them are likely to be equal to $C$ for $j=1,2,\ldots,q$.
Therefore, in order to better compare the performance of NN-class-specific tests
with base-class-specific tests,
we consider the three-class case with classes $X$, $Y$, and $Z$ under CSR independence.
We generate $n_1,\,n_2,\,n_3$ points distributed independently uniformly
on the unit square $(0,1) \times (0,1)$
from classes $X,\,Y$, and $Z$, respectively.
That is, each data set from classes $X,\,Y$, and $Z$ enjoy within sample and
between sample independence.
We generate data points for some combinations of $n_1,n_2,n_3 \in \{10,30,50,100\}$;
and for each sample size combination, we generate data sets $X,\,Y$, and $Z$ for
$N_{mc}=10000$ times.
The empirical sizes and the significance of their deviation from 0.05 are calculated
as in Section \ref{sec:CSR-emp-sign-2Cl}.

The empirical significance levels for the three-class case
are presented in Table \ref{tab:class-spec-null-CSR-3Cl}, 
where empirical sizes significantly smaller (larger) than 0.05 are marked with $^c$ ($^{\ell}$),
which indicate the corresponding test is conservative (liberal).
Notice that when at least one class is small (i.e., $n_i \leq 10$)
tests are usually conservative, especially the base-class-specific tests for the smaller classes.
When all classes are large, the tests are about the desired level,
and conservative for a few sample size combinations.
For the three-class case,
we conclude that the NN-class-specific tests exhibit better performance
than the base-class-specific tests in terms of empirical size.

\begin{table}[ht]
\centering
\begin{tabular}{|c||c|c|c||c|c|c||c|}
\hline
\multicolumn{8}{|c|}{Empirical significance levels under CSR independence } \\
\hline
sizes & \multicolumn{3}{|c||}{base} & \multicolumn{3}{|c||}{NN} & \multicolumn{1}{|c|}{overall} \\
\hline
$(n_1,n_2,n_3)$  & $\ah_1^B$ & $\ah_2^B$ & $\ah_3^B$ &
$\ah_1^{NN}$ & $\ah_2^{NN}$ & $\ah_3^{NN}$ & $\ah_D$ \\
\hline
(10,10,10) & .0401$^c$ & .0432$^c$ & .0436$^c$ & .0422$^c$ & .0422$^c$ & .0400$^c$ & .0421$^c$\\
\hline
(10,10,30) & .0416$^c$ & .0422$^c$ & .0493 & .0448$^c$ & .0436$^c$ & .0450$^c$ & .0445$^c$ \\
\hline
(10,10,50) & .0356$^c$ & .0350$^c$ & .0449$^c$ & .0489 & .0495 & .0463$^c$ & .0510 \\
\hline
(10,30,30) & .0369$^c$ & .0522 & .0514 & .0487 & .0433$^c$ & .0503 & .0439$^c$\\
\hline
(10,30,50) & .0525 & .0432$^c$ & .0446$^c$ & .0473 & .0434$^c$ & .0444$^c$ & .0450$^c$ \\
\hline
(10,50,50) & .0582$^{\ell}$   & .0492   & .0487   & .0422$^c$   & .0519   & .0528   & .0566$^{\ell}$\\
\hline
(30,30,30) & .0487 & .0426$^c$ & .0466 & .0511 & .0492 & .0498 & .0497  \\
\hline
(30,30,50) & .0488 & .0476 & .0485 & .0457$^c$ & .0498 & .0479 & .0463$^c$  \\
\hline
(30,50,50) & .0467 & .0508 & .0504 & .0461$^c$ & .0492 & .0520 & .0486 \\
\hline
(50,50,50) & .0504 & .0509 & .0493 & .0525 & .0496 & .0483 & .0497 \\
\hline
(50,50,100) & .0513 & .0481 & .0499 & .0479 & .0503 & .0485 & .0488 \\
\hline
(50,100,100) & .0455$^c$ & .0496 & .0491 & .0489 & .0463 & .0487 & .0495 \\
\hline
(100,100,100) & .0502 & .0515 & .0496 & .0484 & .0482 & .0482 & .0456$^c$ \\
\hline
\end{tabular}
\caption{\label{tab:class-spec-null-CSR-3Cl}
The empirical significance levels 
for the three-class case under $H_o: CSR\;\,independence$ with
$N_{mc}=10000$, $n_1,n_2,n_3$ in $\{10,30,50,100\}$ at the nominal level of $\alpha=0.05$.
($^c$: the empirical size is significantly smaller than 0.05; i.e., the test is conservative.
$^{\ell}$: the empirical size is significantly larger than 0.05; i.e., the test is liberal.
base = base-class-specific test, NN = NN-class-specific test, overall = overall test.)
}
\end{table}


\subsection{Empirical Significance Levels for the Three-Class Case under RL}
\label{sec:RL-emp-sign-3Cl}
To remove the confounding effect of conditional nature of the tests under CSR independence,
we also perform Monte Carlo simulations under the RL pattern.
For RL with $q > 2$ classes, we consider two cases.
In each case, first we determine the locations of points for which labels
are to be assigned randomly.
Then we apply the RL process to these points for various appropriate
sample size combinations.

\noindent
\textbf{RL Case (1):}
First, we generate $n=(n_1+n_2+n_3)$ points iid $\U((0,1) \times (0,1))$
for some combinations of $n_1,n_2,n_3 \in \{10,30,50,100\}$.
The locations of these points are taken to be the fixed locations
for which we assign the labels randomly.
Thus, we simulate the RL pattern for the performance of the tests
under the null case.
For each sample size combination $(n_1,n_2,n_3)$
we pick $n_1$ points (without replacement) and label them as $X$,
pick $n_2$ points from the remaining points (without replacement) and label them as $Y$ points,
and label the remaining $n_3$ points as $Z$ points.
We repeat the RL procedure $N_{mc}=10000$ times for each sample size combination.
For each RL iteration, we construct the NNCT for classes $X$, $Y$, and $Z$
and then compute the test statistics.
Out of these 10000 samples the number of significant results by the tests is recorded.
The nominal significance level used in all these tests is $\alpha=0.05$.
Based on these significant results, empirical sizes are calculated as before.

\noindent
\textbf{RL Case (2):}
We generate $n_1$ points iid $\U((0,1) \times (0,1))$,
$n_2$ points iid $\U((2,3) \times (0,1))$,
and $n_3$ points iid $\U((1,2) \times (2,3))$
for some combinations of $n_1,n_2,n_3 \in \{10,30,50,100\}$.
RL procedure is performed
and the empirical sizes for the tests
are calculated similarly as in RL Case (1).

The locations for which the RL procedure is applied in RL Cases (1)-(2) are plotted
in Figure \ref{fig:RL3Clcases} for $n_1=n_2=n_3=100$.
We present the empirical significance
levels for the NNCT-tests in Table \ref{tab:class-spec-null-RL-3Cl}, 
where the empirical significance level labeling is as in Table \ref{tab:class-spec-null-CSR-3Cl}.
The empirical sizes are marked with $^c$ and $^{\ell}$
for conservativeness and liberalness as in Section \ref{sec:CSR-emp-sign-2Cl}.
Observe that when all sample sizes are equal each test statistic is
about the desired significance level in rejecting $H_o:RL$ for both RL Cases.
When at least one class has a small sample size ($n_i \leq 10$),
the tests are liberal, conservative, or at the desired level.
In particular, the base-class-specific test tends to be conservative for class $X$,
and NN-class-specific test tends to be conservative for class $Y$.

\begin{figure}
\centering
\rotatebox{-90}{ \resizebox{2.5 in}{!}{\includegraphics{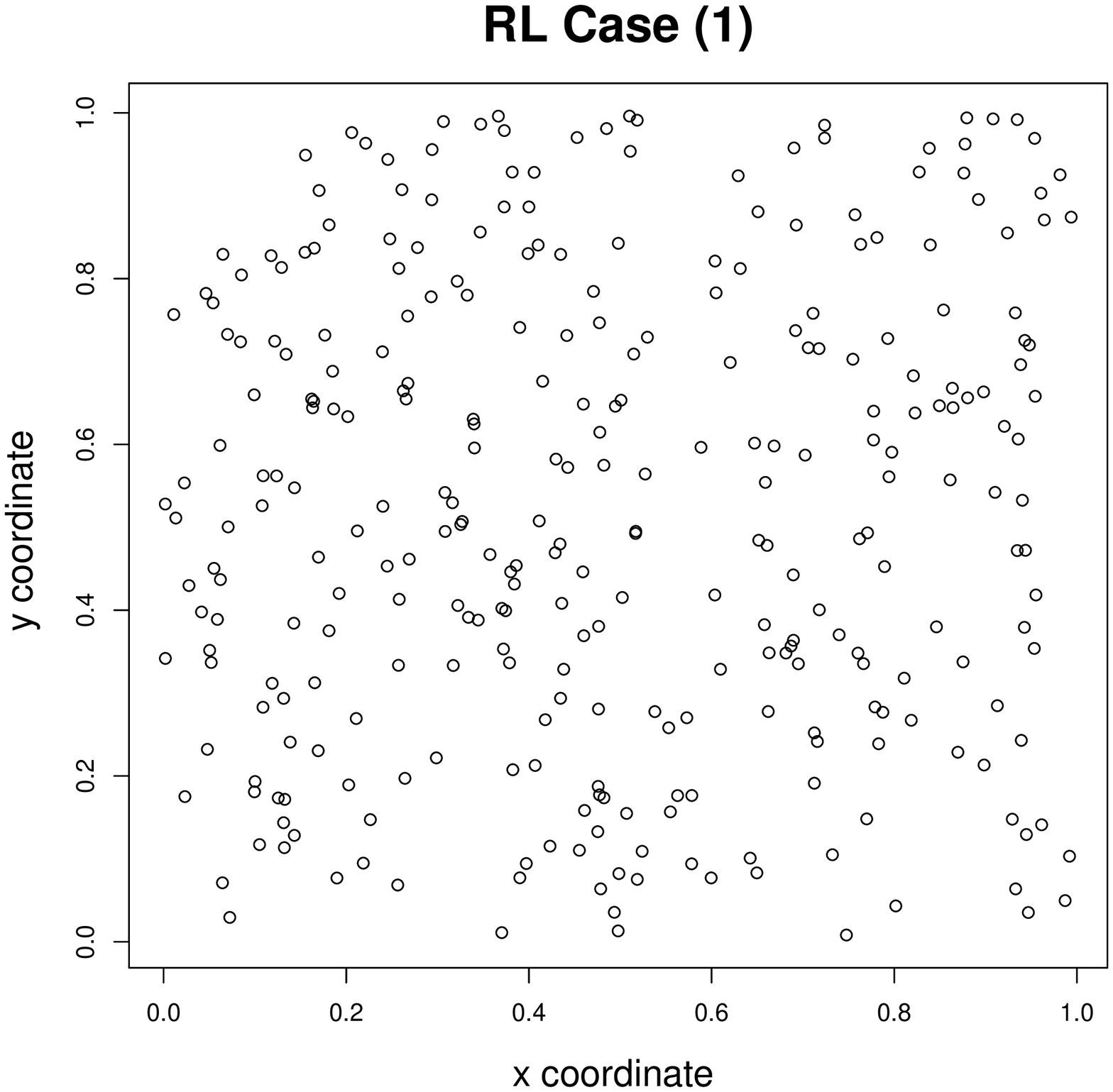} }}
\rotatebox{-90}{ \resizebox{2.5 in}{!}{\includegraphics{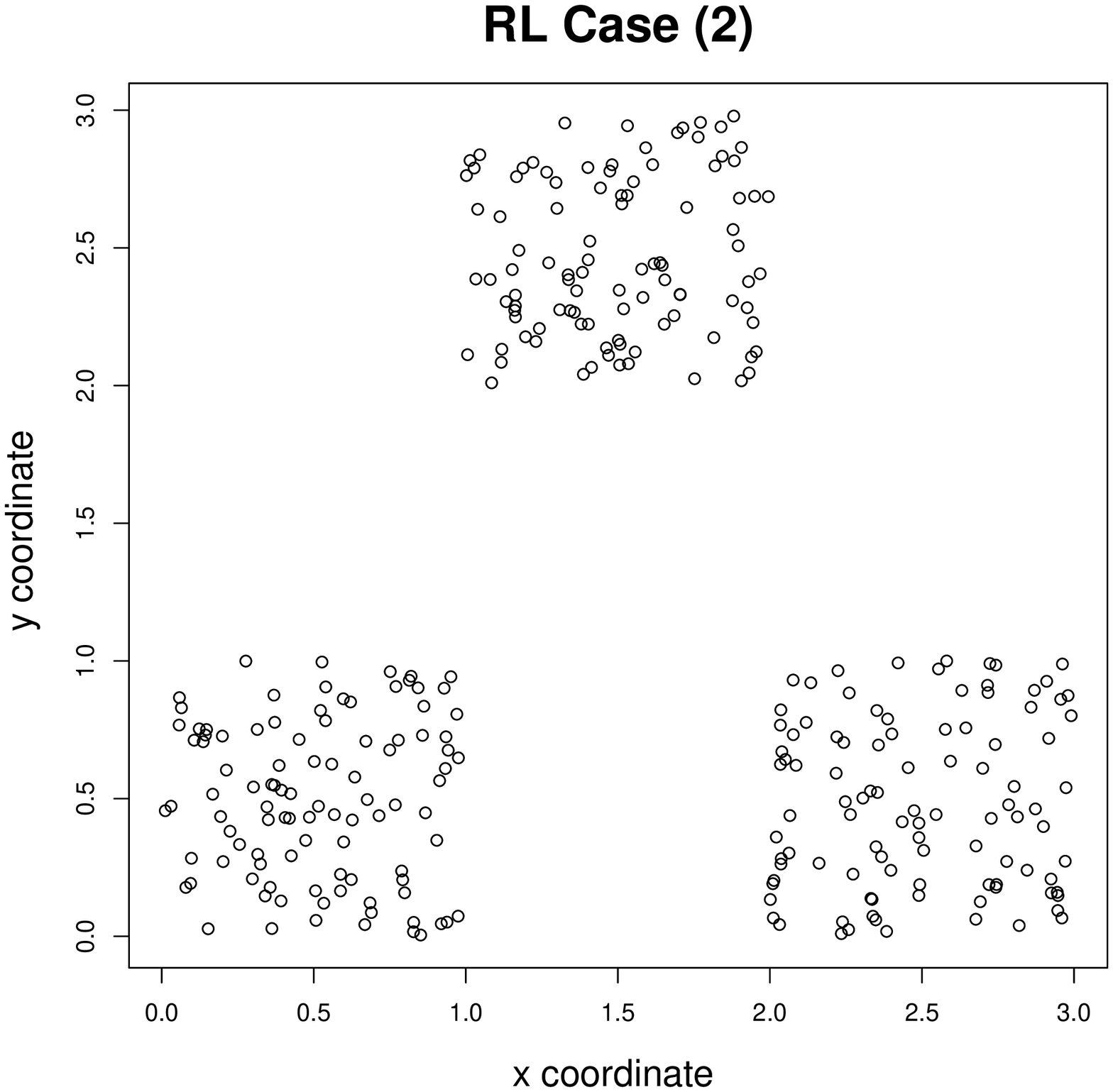} }}
 \caption{
\label{fig:RL3Clcases}
The fixed locations for which RL procedure is applied for RL Cases (1)-(2) with $n_1=n_2=n_3=100$
in the two-class case. Notice that $x$-axis for RL Case (2) is differently scaled.
}
\end{figure}

Comparing Tables \ref{tab:class-spec-null-CSR-3Cl} and \ref{tab:class-spec-null-RL-3Cl}
we observe that the empirical sizes are about the same under the CSR independence and RL patterns.
Hence we conclude that the conservativeness
or liberalness of the tests for small sample sizes
does not result from conditioning on $Q$ and $R$ under the CSR independence pattern,
as we get similar results under the RL pattern also.

\begin{remark}
\label{rem:MC-emp-size}
\textbf{Main Result of Monte Carlo Simulations for Empirical Sizes:}
Based on the simulation results presented in Sections \ref{sec:monte-carlo-2Cl}
and \ref{sec:monte-carlo-3Cl}, we conclude that none of the tests we consider
have the desired level when at least one sample size is small so that
the cell count(s) in the corresponding NNCT have a high probability of being $\leq 5$.
This corresponds to the case that at least one sample size is $\leq 10$ in our simulation study.
However, when relative abundances (i.e., sample sizes) are very different,
cell counts are also more likely to be $\leq 5$,
compared to cell counts for similar sample sizes (roughly,
we assume sample sizes are similar when $\max_{i} (n_i)/ \min_{i} (n_i) \leq 2$.)
The tests tend to be conservative when the row or column they pertain to
contain cell(s) whose count(s) are $\leq 5$.
For larger samples which imply the cell counts are larger than 5,
the tests yield empirical sizes that are about the desired nominal level.
\cite{dixon:1994} recommends Monte Carlo randomization
test when some cell count(s) are $\leq 5$ in a NNCT
for the overall segregation test.
We extend this recommendation for the base- and NN-class-specific tests, in the sense that,
if the corresponding row or column contains cells whose counts are $\leq 5$,
we recommend Monte Carlo randomization,
otherwise we recommend the asymptotic approximation for the class-specific tests.
Thus, when sample sizes are small
(hence the corresponding cell counts are $\leq 5$),
the asymptotic approximation of the tests is not
appropriate, in particular for the base-class-specific test
for the smaller class.
So when at least one sample size is small,
the power comparisons should be carried out using the Monte Carlo
critical values.
On the other hand, for large samples,
the power comparisons can be made using the asymptotic
or Monte Carlo critical values.
$\square$
\end{remark}

\begin{table}[ht]
\centering
\begin{tabular}{|c||c|c|c||c|c|c||c|}
\hline
\multicolumn{8}{|c|}{Empirical significance levels under RL Case (1)} \\
\hline
sizes & \multicolumn{3}{|c||}{base} & \multicolumn{3}{|c||}{NN} & \multicolumn{1}{|c|}{overall} \\
\hline
$(n_1,n_2,n_3)$  & $\ah_1^B$ & $\ah_2^B$ & $\ah_3^B$ &
$\ah_1^{NN}$ & $\ah_2^{NN}$ & $\ah_3^{NN}$ & $\ah_D$ \\
\hline
(10,10,10) & .0453$^c$ & .0411$^c$ & .0374$^c$ & .0360$^c$ & .0401$^c$ & .0393$^c$ & .0377$^c$ \\
\hline
(10,10,30) & .0298$^c$ & .0319$^c$ & .0483 & .0474 & .0477 & .0402$^c$ & .0455$^c$\\
\hline
(10,10,50) & .0346$^c$ & .0304$^c$ & .0478 & .0497 & .0459$^c$ & .0436$^c$ & .0477\\
\hline
(10,30,30) & .0466 & .0528 & .0522 & .0497 & .0444$^c$ & .0476 & .0477\\
\hline
(10,30,50) & .0563$^{\ell}$ & .0491 & .0467 & .0551$^{\ell}$ & .0470 & .0477 & .0518\\
\hline
(10,50,50) & .0403$^c$ & .0471 & .0481 & .0554$^{\ell}$ & .0460$^c$ & .0445$^c$ & .0500\\
\hline
(30,30,30) & .0447$^c$ & .0465 & .0431$^c$ & .0506 & .0473 & .0479 & .0471\\
\hline
(30,30,50) & .0479 & .0491 & .0529 & .0485 & .0508 & .0519 & .0489\\
\hline
(30,50,50) & .0503 & .0507 & .0526 & .0509 & .0472 & .0501 & .0473\\
\hline
(50,50,50) & .0482 & .0533 & .0528 & .0482 & .0506 & .0481 & .0469\\
\hline
(50,50,100) & .0488 & .0481 & .0494 & .0481 & .0468 & .0462$^c$ & .0450$^c$\\
\hline
(50,100,100) & .0473 & .0486 & .0465 & .0472$^c$ & .0496 & .0459$^c$ & .0476\\
\hline
(100,100,100) & .0516 & .0525 & .0518 & .0485 & .0508 & .0510 & .0510\\
\hline
\multicolumn{8}{|c|}{Empirical significance levels under RL Case (2)} \\
\hline
(10,10,10) & .0371$^c$ & .0384$^c$ & .0397$^c$ & .0358$^c$ & .0339$^c$ & .0355$^c$ & .0372$^c$\\
\hline
(10,10,30) & .0553$^{\ell}$ & .0550$^{\ell}$ & .0457$^c$ & .0457$^c$ & .0447$^c$ & .0426$^c$ & .0470\\
\hline
(10,10,50) & .0368$^c$ & .0365$^c$ & .0458$^c$ & .0497 & .0523 & .0477 & .0501\\
\hline
(10,30,30) & .0433$^c$ & .0498 & .0485 & .0485 & .0463$^c$ & .0435$^c$ & .0420$^c$\\
\hline
(10,30,50) & .0503 & .0431$^c$ & .0473 & .0513 & .0461$^c$ & .0431$^c$ & .0469\\
\hline
(10,50,50) & .0401$^c$ & .0493 & .0476 & .0510 & .0483 & .0489 & .0507\\
\hline
(30,30,30) & .0484 & .0452$^c$ & .0493 & .0489 & .0464 & .0504 & .0472\\
\hline
(30,30,50) & .0429$^c$ & .0462$^c$ & .0454$^c$ & .0462$^c$ & .0433$^c$ & .0485 & .0436$^c$\\
\hline
(30,50,50) & .0488 & .0502 & .0510 & .0482 & .0479 & .0476 & .0475\\
\hline
(50,50,50) & .0539$^{\ell}$ & .0511 & .0525 & .0468 & .0501 & .0531 & .0513\\
\hline
(50,50,100) & .0518 & .0506 & .0507 & .0490 & .0498 & .0506 & .0497\\
\hline
(50,100,100) & .0504 & .0500 & .0483 & .0483 & .0511 & .0488 & .0516\\
\hline
(100,100,100) & .0542$^{\ell}$ & .0506 & .0477 & .0463 & .0506 & .0479 & .0457$^c$\\
\hline

\end{tabular}
\caption{\label{tab:class-spec-null-RL-3Cl}
The empirical significance levels for the three-class case
under $H_o:RL$ for RL Cases (1)-(2) with
$N_{mc}=10000$, $n_1,n_2,n_3$ in $\{10,30,50,100\}$ at the nominal level of $\alpha=0.05$.
($^c$: the empirical size is significantly smaller than 0.05; i.e., the test is conservative.
$^{\ell}$: the empirical size is significantly larger than 0.05; i.e., the test is liberal.
base = base-class-specific test, NN = NN-class-specific test, overall = overall test.)
}
\end{table}

%

\begin{remark}
\label{rem:MC-crit-value}
\textbf{Monte Carlo Critical Values:}
When sample sizes are small so that some cell count(s)
are expected to be $\leq 5$ with a high probability,
then it will not be appropriate to use
the asymptotic approximation (hence the asymptotic critical values)
for the overall and class-specific tests of segregation (see Remark \ref{rem:MC-emp-size}).
That is, when some cell counts are small,
the tests tend to be either liberal or conservative.
In order to better evaluate the empirical power performance of the tests,
for each sample size combination, we record the test statistics at each Monte Carlo
simulation under the CSR independence cases of Sections \ref{sec:CSR-emp-sign-2Cl} and \ref{sec:CSR-emp-sign-3Cl}.
We find the 95$^{th}$ percentiles of the recorded test statistics at each sample size combination
(not presented) and use them as ``Monte Carlo critical values" for
the power estimation in the following sections.
For example, for base-class-specific test for class $X$ in the two-class case,
the $\chi^2$ test statistics are recorded for $(n_1,n_2)=(30,50)$
under the CSR independence pattern as in Section \ref{sec:CSR-emp-sign-2Cl},
then the 95$^{th}$ percentile of these statistics is used
as the Monte Carlo critical value for $(n_1,n_2)=(30,50)$.
That is, under a segregation or association alternative with $(n_1,n_2)=(30,50)$,
a calculated test statistic is deemed significant if greater than or equal
this Monte Carlo critical value.
$\square$
\end{remark}

\section{Empirical Power Analysis for the Two-Class Case}
\label{sec:emp-power-2Cl}
We consider three cases for each of segregation and association alternatives
in the two-class case.

\subsection{Empirical Power Analysis under Segregation Alternatives in the Two-Class Case}
\label{sec:power-comp-seg-2Cl}
For the segregation alternatives, we generate
$X_i \stackrel{iid}{\sim} \U((0,1-s)\times(0,1-s))$ and
$Y_j \stackrel{iid}{\sim} \U((s,1)\times(s,1))$
for $i=1,\ldots,n_1$ and $j=1,\ldots,n_2$.
Notice that the level of segregation is determined by the magnitude of $s \in (0,1)$.
We consider the following three segregation alternatives:
\begin{equation}
\label{eqn:seg-alt-2Cl}
H_S^I: s=1/6, \;\;\; H_S^{II}: s=1/4, \text{ and } H_S^{III}: s=1/3.
\end{equation}

Observe that, from $H_S^I$ to $H_S^{III}$ (i.e., as $s$ increases),
the segregation between $X$ and $Y$ gets stronger
in the sense that $X$ and $Y$ points tend to form one-class clumps or clusters.
We calculate the power estimates using the asymptotic critical values
based on the corresponding $\chi^2$-distributions and using the Monte Carlo critical values.
The power estimates based on the asymptotic and Monte Carlo critical values
are presented in Table \ref{tab:emp-power-seg}, 
where $\bh^B_i$ are for (Dixon's) base-class-specific tests for classes $i=1,2$,
and $\bh^{NN}_1$ is for NN-class-specific test for class 1,
$\bh^D$ is for Dixon's overall segregation test.
We only present $\bh^{NN}_1$ since $\bh^{NN}_1=\bh^{NN}_2=\bh^D$.
Observe that, for both class-specific tests,
as $n=(n_1+n_2)$ gets larger, the power estimates get larger;
for similar $n=(n_1+n_2)$ values, the power estimate is larger
for classes with similar relative abundance (i.e., for similar sample sizes $n_1 \approx n_2$);
and as the segregation gets stronger, the power estimates get larger
for each sample size combination.
Furthermore, power estimates based on the Monte Carlo critical values
tend to be larger compared to the ones using the asymptotic critical values.
When sample sizes are similar, NN-class-specific tests tend to
have higher power.
On the other hand, when relative abundances (i.e., sample sizes) are very different,
the base-class-specific test for classes $X$ and $Y$ have the highest and lowest
power estimates, respectively.

\begin{table}[ht]
\centering
\begin{tabular}{|c|c||c|c||c||c||c|c|}
\hline
\multicolumn{8}{|c|}{Empirical power estimates under the segregation alternatives} \\
\hline
& sizes & \multicolumn{3}{|c||}{Asymptotic} & \multicolumn{3}{|c|}{Monte Carlo} \\
\hline & $(n_1,n_2)$ & $\bh_1^B$ & $\bh_2^B$ & $\bh_1^{NN}$* &
$\bh_1^B$ & $\bh_2^B$ & $\bh_1^{NN}$* \\
\hline \hline
 & $(10,10)$ & .0734 & .0698 & .0775 & .0904 & .0861 & .0836 \\
\cline{2-8}
 & $(10,30)$ & .1436 & .1540 & .1414 & .1915 & .1556 & .1530\\
\cline{2-8}
 & $(10,50)$ & .1639 & .1615 & .2193 & .2311 & .1699 & .2217\\
\cline{2-8}
 & $(30,30)$ & .2883 & .2783 & .2904 & .2883 & .2783 & .3003\\
\cline{2-8}\raisebox{0.ex}[0pt]{$H_S^I$}
 & $(30,50)$ & .4491 & .4045 & .3911 & .4136 & .3958 & .4096\\
\cline{2-8}
 & $(50,50)$ & .5091 & .5016 & .5546 & .5347 & .5237 & .5541\\
\cline{2-8}
 & $(50,100)$ & .7686 & .6689 & .7425 & .7540 & .6629 & .7307\\
\cline{2-8}
 & $(100,100)$ & .8761 & .8730 & .9121 & .8778 & .8891 & .9119\\

\hline \hline

 & $(10,10)$ & .2057 & .2044 & .2305 & .2439 & .2419 & .2364\\
\cline{2-8}
 & $(10,30)$ & .4601 & .4133 & .4555 & .5365 & .4181 & .4745\\
\cline{2-8}
 & $(10,50)$ & .5420 & .4477 & .6174 & .6368 & .4627 & .6187\\
\cline{2-8}
 & $(30,30)$ & .7783 & .7769 & .8141 & .7783 & .7769 & .8207\\
\cline{2-8}\raisebox{0.ex}[0pt]{$H_S^{II}$}
 & $(30,50)$ & .9262 & .8775 & .9126 & .9095 & .8712 & .9199\\
\cline{2-8}
 & $(50,50)$ & .9543 & .9551 & .9777 & .9594 & .9593 & .9777\\
\cline{2-8}
 & $(50,100)$ & .9977 & .9866 & .9975 & .9974 & .9857 & .9972\\
\cline{2-8}
 & $(100,100)$ & .9998 & .9999 & 1.000 & .9998 & .9999 & 1.000\\

\hline \hline

 & $(10,10)$ & .5144 & .5121 & .5817 & .5641 & .5610 & .5852\\
\cline{2-8}
 & $(10,30)$ & .8873 & .7833 & .8787 & .9172 & .7871 & .8892\\
\cline{2-8}
 & $(10,50)$ & .9353 & .8002 & .9528 & .9570 & .8116 & .9530\\
\cline{2-8}
 & $(30,30)$ & .9929 & .9915 & .9969 & .9929 & .9915 & .9971\\
\cline{2-8} \raisebox{0.ex}[0pt]{$H_S^{III}$}
 & $(30,50)$ & .9999 & .9979 & .9997 & .9999 & .9976 & .9998\\
\cline{2-8}
 & $(50,50)$ & .9999 & 1.000 & 1.000 & 1.000 & 1.000 & 1.000\\
\cline{2-8}
 & $(50,100)$ & 1.000 & 1.000 & 1.000 & 1.000 & 1.000 & 1.000\\
\cline{2-8}
 & $(100,100)$ & 1.000 & 1.000 & 1.000 & 1.000 & 1.000 & 1.000\\
\cline{2-8} \hline

\end{tabular}
\caption{
\label{tab:emp-power-seg}
The empirical power estimates
for the tests under the segregation alternatives, $H_S^I$,
$H_S^{II}$, and $H_S^{III}$ for the two-class case with
$N_{mc}=10000$, for some combinations of $n_1,n_2 \in
\{10,30,50,100\}$ at the nominal level of $\alpha=0.05$.
(*only $\bh_1^{NN}$ is presented since $\bh_1^{NN}=\bh_2^{NN}=\bh_{D}$.
Asymptotic = with asymptotic critical values, Monte Carlo = with Monte Carlo critical values.)}
\end{table}


\subsection{Empirical Power Analysis under Association Alternatives in the Two-Class Case}
\label{sec:power-comp-assoc-2Cl}
For the association alternatives, we consider three cases.
First, we generate $X_i \stackrel{iid}{\sim} \U((0,1)\times(0,1))$ for $i=1,2,\ldots,n_1$.
Then we generate $Y_j$ for $j=1,2,\ldots,n_2$ as follows.
For each $j$, we pick an $i$ randomly, then generate $Y_j$ as
$X_i+R_j\,(\cos T_j, \sin T_j)'$ where
$R_j \stackrel{iid}{\sim} \U(0,r)$ with $r \in (0,1)$ and
$T_j \stackrel{iid}{\sim} \U(0,2\,\pi)$.
In the pattern generated, appropriate choices of
$r$ will imply association between classes $X$ and $Y$.
That is, it will be  more likely to have $(X,Y)$ or $(Y,X)$
NN pairs than the same-class NN pairs (i.e., $(X,X)$ or $(Y,Y)$).
The three values of $r$ we consider constitute
the following three association alternatives;
\begin{equation}
\label{eqn:assoc-alt-2Cl}
H_A^{I}: r=1/4,\;\;\; H_A^{II}: r=1/7, \text{ and } H_A^{III}: r=1/10.
\end{equation}
Observe that, from $H_A^I$ to $H_A^{III}$ (i.e., as $r$ decreases),
the association between $X$ and $Y$ gets stronger
in the sense that $X$ and $Y$ points tend to occur
together more and more frequently.
By construction, for similar sample sizes the association between $X$ and $Y$ are
at about the same degree as association between $Y$ and $X$.
For very different samples, smaller sample is associated with the larger
but the abundance of the larger sample confounds its association with the smaller.

The empirical power estimates, based on the asymptotic critical values
and Monte Carlo critical values, are presented in Table \ref{tab:emp-power-assoc}.
Observe that when $n_1 \approx n_2$,
for both class-specific tests and Dixon's overall segregation test,
as $n_i$ gets larger, the power estimates get larger;
and as the association gets stronger, the power estimates get larger
for each sample size combination.
Furthermore, empirical power estimates based on Monte Carlo critical values tend to be
larger for NN-class-specific tests,
smaller for base-class-specific tests.

When at least one class has a small size (i.e., $n_i \leq 10$),
the base-class-specific test for the smaller class, say class $i$, fails to detect any
deviation from CSR independence for row $i$.
But the base-class-specific test for the larger class
is robust to the differences in relative abundance (i.e., sample sizes) and behaves as expected.
Likewise, the NN-class-specific test is not affected by the differences in relative abundance
as long as the smaller class is associated with the larger
(the corresponding simulation results are not presented).
In our set-up this corresponds to the case that class $Y$ is much larger than class $X$.
If on the contrary, the smaller class is associated with the larger
(i.e., class $X$ is much larger than class $Y$),
then NN-class-specific test has very poor power performance.

When sample sizes are similar, the base-class-specific tests
for class $Y$ (the class less associated with the other) have higher power under weak association,
while NN-class-specific tests have higher power under strong association.
When sample sizes are very different,
the base-class-specific tests for classes $Y$  and $X$ have highest
and lowest power estimates, respectively.
Base-class-specific test for class $i$ measures the association of
other classes with class $i$,
while NN-class-specific test for class $j$ measures the association of
class $j$ with other classes.

\begin{table}[ht]
\centering
\begin{tabular}{|c|c||c|c||c||c||c|c|}
\hline
\multicolumn{8}{|c|}{Empirical power estimates under the association alternatives} \\
\hline
& sizes & \multicolumn{3}{|c||}{Asymptotic} & \multicolumn{3}{|c|}{Monte Carlo} \\
\hline & $(n_1,n_2)$ & $\bh_1^B$ & $\bh_2^B$ & $\bh_1^{NN}$* &
$\bh_1^B$ & $\bh_2^B$ & $\bh_1^{NN}$* \\
\hline
\hline

 & $(10,10)$ & .1349 & .1776 & .1105 & .1352 & .1785 & .1282\\
\cline{2-8}
 & $(10,30)$ & .0002 & .4366 & .3007 & .0008 & .4451 & .3334\\
\cline{2-8}
 & $(10,50)$ & .0002 & .4947 & .3318 & .0010 & .5070 & .3435\\
\cline{2-8}
 & $(30,30)$ & .1413 & .2434 & .1697 & .1388 & .2397 & .1795\\
\cline{2-8}\raisebox{0.ex}[0pt]{$H_A^I$}
 & $(30,50)$ & .1833 & .3984 & .2903 & .1736 & .3903 & .3075\\
\cline{2-8}
 & $(50,50)$ & .1149 & .2421 & .1738 & .1153 & .2421 & .1729\\
\cline{2-8}
 & $(50,100)$ & .1813 & .4411 & .3410 & .1398 & .4236 & .3230\\
\cline{2-8}
 & $(100,100)$ & .0853 & .2309 & .1677 & .0855 & .2310 & .1663\\
\hline
\hline

 & $(10,10)$ & .2499 & .2569 & .1834 & .2500 & .2571 & .2010\\
\cline{2-8}
 & $(10,30)$ & .0000 & .6463 & .4956 & .0001 & .6550 & .5461\\
\cline{2-8}
 & $(10,50)$ & .0000 & .7062 & .5500 & .0004 & .7138 & .5591\\
\cline{2-8}
 & $(30,30)$ & .4053 & .4457 & .4141 & .3987 & .4383 & .4298\\
\cline{2-8}\raisebox{0.ex}[0pt]{$H_A^{II}$}
 & $(30,50)$ & .4896 & .6957 & .6332 & .4661 & .6873 & .6559\\
\cline{2-8}
 & $(50,50)$ & .4034 & .4961 & .4616 & .4034 & .4961 & .4609\\
 \cline{2-8}
 & $(50,100)$ & .5527 & .7991 & .7559 & .4783 & .7858 & .7428\\
\cline{2-8}
 & $(100,100)$ & .3868 & .5575 & .5013 & .3868 & .5575 & .4987\\
\hline
\hline

 & $(10,10)$ & .3038 & .2918 & .2222 & .3038 & .2920 & .2401\\
\cline{2-8}
 & $(10,30)$ & .0000 & .7364 & .6003 & .0000 & .7428 & .6480\\
\cline{2-8}
 & $(10,50)$ & .0000 & .7907 & .6512 & .0001 & .7983 & .6582\\
\cline{2-8}
  & $(30,30)$ & .6092 & .6011 & .6157 & .5991 & .5915 & .6289\\
\cline{2-8}\raisebox{0.ex}[0pt]{$H_A^{III}$}
  & $(30,50)$ & .7211 & .8491 & .8386 & .6880 & .8441 & .8535\\
\cline{2-8}
 & $(50,50)$ & .6842 & .6891 & .7285 & .6842 & .6891 & .7280\\
\cline{2-8}
 & $(50,100)$ & .8024 & .9442 & .9433 & .7490 & .9376 & .9383\\
\cline{2-8}
 & $(100,100)$ & .7207 & .7973 & .8030 & .7207 & .7973 & .8011\\
\cline{2-8}

\hline


\end{tabular}
\caption{
\label{tab:emp-power-assoc}
The empirical power estimates for the tests under the association alternatives
$H_A^I$, $H_A^{II}$, and $H_A^{III}$ for the two-class case
with $N_{mc}=10000$, for some combinations of $n_1,n_2 \in
\{10,30,50,100\}$ at the nominal level of $\alpha=0.05$.
(*only $\bh_1^{NN}$ is presented since $\bh_1^{NN}=\bh_2^{NN}=\bh_{D}$.
Asymptotic = with asymptotic critical values, Monte Carlo = with Monte Carlo critical values.)}
\end{table}


\section{Empirical Power Analysis in the Three-Class Case}
\label{sec:emp-power-3Cl}
We consider three cases for each of segregation and association alternatives in the
three-class case.

\subsection{Empirical Power Analysis under Segregation Alternatives in the Three-Class Case}
\label{sec:power-comp-seg-3Cl}
For the segregation alternatives, we generate
$X_i \stackrel{iid}{\sim} \U((0,1-2s)\times(0,1-2s))$,
$Y_j \stackrel{iid}{\sim} \U((2s,1)\times(2s,1))$, and
$Z_{\ell} \stackrel{iid}{\sim} \U((s,1-s)\times(s,1-s))$
for $i=1,\ldots,n_1$, $j=1,\ldots,n_2$, and $\ell=1,\ldots,n_3$.
Notice the level of segregation is determined by the magnitude of $s \in (0,1/2)$.
We consider the following three segregation alternatives:
\begin{equation}
\label{eqn:seg-alt-3Cl}
H^1_S: s=1/12, \;\;\; H^2_S: s=1/8, \text{ and } H^3_S: s=1/6.
\end{equation}


Observe that, from $H^1_S$ to $H^3_S$ (i.e., as $s$ increases), the segregation gets stronger
in the sense that $X$, $Y$, and $Z$ points tend to form one-class clumps or clusters.
Furthermore, under each segregation alternative, $X$ and $Y$ are more segregated
compared to $X$ and $Z$ or $Y$ and $Z$.

The empirical power estimates using the asymptotic and Monte Carlo critical values are provided
in Tables \ref{tab:emp-power-seg-class-asy-3Cl} and \ref{tab:emp-power-seg-class-emp-3Cl}, respectively.
Observe that, in both tables, at each sample size combination,
as the segregation gets stronger, the power estimates get larger.
Furthermore, the power estimates for the similar sample sizes
tend to be larger compared to the different sample sizes when
the total sizes $n=(n_1+n_2+n_3)$ are similar.
The power estimates also confirm that classes $X$ and $Y$ are more
segregated compared to classes $X$ and $Z$ or $Y$ and $Z$.
Furthermore, power estimates using the Monte Carlo critical values
tend to be higher compared to the ones using the asymptotic critical values.
In fact, based on Remark \ref{rem:MC-emp-size},
when at least one of the samples is small (i.e., $n_i \leq 10$)
the estimates based on the Monte Carlo critical values would be more appropriate.
For larger samples, asymptotic and Monte Carlo based empirical power estimates
are both reliable.

For large samples, Dixon's overall test tends to have the highest power estimates,
but class-specific tests provide more information about the pattern.
For all (small or large) similar relative abundance values,
NN-class-specific tests have slightly better power performance,
while for very different relative abundances,
base-class-specific tests have slightly better power performance.
However, overall, we can conclude that in the three-class case,
both base- and NN-class-specific tests and Dixon's overall test have about the same
performance in terms of power.

{\small
\begin{table}[ht]
\centering
\begin{tabular}{|c|c||c|c|c||c|c|c||c|}
\hline
\multicolumn{9}{|c|}{Empirical power estimates under the segregation alternatives} \\
\hline
& sample sizes & \multicolumn{7}{|c|}{with asymptotic critical values} \\
\hline
& $(n_1,n_2,n_3)$
& $\bh_1^B$ & $\bh_2^B$ & $\bh_3^B$ & $\bh_1^{NN}$ & $\bh_2^{NN}$ & $\bh_3^{NN}$ & $\bh_D$\\
\hline
 & (10,10,10) & .0592 & .0604 & .0404 & .0564 & .0552 & .0537 & .0606\\
\cline{2-9}
 & (10,10,30) & .0457 & .0510 & .0482 & .0567 & .0614 & .0583 & .0619\\
\cline{2-9}
 & (10,10,50) & .0433 & .0411 & .0511 & .0597 & .0590 & .0597 & .0611\\
\cline{2-9}
 & (10,30,30) & .1130 & .0872 & .0624 & .0961 & .0905 & .0782 & .0960 \\
\cline{2-9}
 & (10,30,50) & .1180 & .0793 & .0659 & .0832 & .0895 & .0806 & .0994\\
\cline{2-9}
 & (10,50,50) & .1341 & .1207 & .0819 & .1216 & .1314 & .1091 & .1416\\
\cline{2-9} \raisebox{0.ex}[0pt]{$H^1_S$}
 & (30,30,30) & .1864 & .1932 & .0593 & .1785 & .1811 & .0688 & .1567\\
\cline{2-9}
 & (30,30,50) & .1764 & .1748 & .0669 & .1563 & .1489 & .0807 & .1617\\
\cline{2-9}
 & (30,50,50) & .2646 & .2396 & .0695 & .2295 & .2497 & .0868 & .2343\\
\cline{2-9}
 & (50,50,50) & .3439 & .3434 & .0669 & .3393 & .3348 & .0775 & .3160 \\
\cline{2-9}
 & (50,50,100) & .3088 & .3098 & .1028 & .2980 & .2929 & .1028 & .3202\\
\cline{2-9}
 & (50,100,100) & .4882 & .4829 & .1311 & .4467 & .5206 & .1474 & .5109\\
\cline{2-9}
 & (100,100,100) & .7056 & .7023 & .1221 & .7164 & .7226 & .1181 & .7288\\
\hline
\hline

& (10,10,10)  & .1287 & .1250 & .0445 & .1234 & .1259 & .0688 & .1146 \\
\cline{2-9}
& (10,10,30)  & .0885 & .0933 & .0709 & .1058 & .1082 & .0858 & .1293\\
\cline{2-9}
& (10,10,50)  & .0923 & .0978 & .0916 & .1217 & .1250 & .1093 & .1440 \\
\cline{2-9}
& (10,30,30)  & .2853 & .2622 & .1119 & .2246 & .2796 & .1571 & .2856\\
\cline{2-9}
& (10,30,50)  & .2706 & .2681 & .1493 & .2000 & .2895 & .1993 & .3159\\
\cline{2-9}
& (10,50,50)  & .3298 & .4056 & .2308 & .2810 & .4638 & .3071 & .4623\\
\cline{2-9}\raisebox{0.ex}[0pt]{$H^2_S$}
 & (30,30,30) & .5506 & .5535 & .1008 & .5526 & .5609 & .0976 & .5337 \\
\cline{2-9}
 & (30,30,50) & .5149 & .5127 & .1507 & .4949 & .4822 & .1423 & .5476\\
\cline{2-9}
 & (30,50,50) & .6891 & .6719 & .1616 & .6480 & .7186 & .1749 & .7241\\
\cline{2-9}
 & (50,50,50) & .8345 & .8351 & .1627 & .8456 & .8453 & .1538 & .8643 \\
\cline{2-9}
 & (50,50,100) & .7998 & .8036 & .3103 & .8022 & .8024 & .3003 & .8832\\
\cline{2-9}
 & (50,100,100) & .9358 & .9501 & .3942 & .9195 & .9706 & .4054 & .9752\\
\cline{2-9}
 & (100,100,100) & .9960 & .9955 & .3501 & .9973 & .9970 & .3412 & 1.000\\
\hline
\hline

& (10,10,10)  & .2862 & .2868 & .0587 & .2991 & .3003 & .0847 & .2829\\
\cline{2-9}
& (10,10,30)  & .2280 & .2270 & .1475 & .2647 & .2652 & .1704 & .3623\\
\cline{2-9}
& (10,10,50)  & .2630 & .2705 & .2264 & .3175 & .3217 & .2692 & .4273\\
\cline{2-9}
& (10,30,30)  & .6026 & .6428 & .2816 & .4934 & .6860 & .3310 & .6950\\
\cline{2-9}
& (10,30,50)  & .5599 & .6880 & .3861 & .4510 & .7225 & .4713 & .7573\\
\cline{2-9}
& (10,50,50)  & .6525 & .8467 & .5624 & .5629 & .8960 & .6604 & .8934\\
\cline{2-9}\raisebox{0.ex}[0pt]{$H^3_S$}
 & (30,30,30) & .9275 & .9326 & .2345 & .9342 & .9388 & .2208 & .9433\\
\cline{2-9}
 & (30,30,50) & .9138 & .9130 & .3731 & .9049 & .9080 & .3595 & .9568\\
\cline{2-9}
 & (30,50,50) & .9764 & .9780 & .4227 & .9706 & .9872 & .4356 & .9917\\
\cline{2-9}
 & (50,50,50) & .9979 & .9969 & .4178 & .9985 & .9983 & .4075 & .9994\\
\cline{2-9}
 & (50,50,100) & .9954 & .9961 & .7195 & .9958 & .9966 & .7530 & .9998\\
\cline{2-9}
 & (50,100,100) & .9997 & 1.000 & .8195 & .9996 & 1.000 & .8456 & 1.000\\
\cline{2-9}
 & (100,100,100) & 1.000 & 1.000 & .7733 & 1.000 & 1.000 & .7960 & 1.000\\
\cline{2-9}

\hline
\end{tabular}
\caption{
\label{tab:emp-power-seg-class-asy-3Cl}
The empirical power estimates for the tests under the segregation alternatives,
$H^1_S$, $H^2_S$, and $H^3_S$ for the three-class case
with $N_{mc}=10000$, for some combinations of $n_1,n_2 \in
\{10,30,50,100\}$ at the nominal level of $\alpha=0.05$.}
\end{table}
}


{\small
\begin{table}[ht]
\centering
\begin{tabular}{|c|c||c|c|c||c|c|c||c|}
\hline
\multicolumn{9}{|c|}{Empirical power estimates under the segregation alternatives} \\
\hline
& sample sizes & \multicolumn{7}{|c|}{with Monte Carlo critical values} \\
\hline
& $(n_1,n_2,n_3)$
& $\bh_1^B$ & $\bh_2^B$ & $\bh_3^B$ & $\bh_1^{NN}$ & $\bh_2^{NN}$ & $\bh_3^{NN}$ & $\bh_D$\\
\hline
 & (10,10,10) & .0664 & .0631 & .0637 & .0721 & .0721 & .0468 & .0700\\
\cline{2-9}
 & (10,10,30) & .0621 & .0691 & .0657 & .0519 & .0584 & .0500 & .0699\\
\cline{2-9}
 & (10,10,50) & .0619 & .0597 & .0653 & .0469 & .0439 & .0572 & .0595\\
\cline{2-9}
 & (10,30,30) & .0990 & .1010 & .0774 & .1283 & .0822 & .0600 & .1048 \\
\cline{2-9}
 & (10,30,50) & .0889 & .0994 & .0904 & .1179 & .0885 & .0728 & .1082 \\
\cline{2-9}
 & (10,50,50) & .1086 & .1306 & .1079 & .1596 & .1182 & .0805 & .1387\\
\cline{2-9} \raisebox{0.ex}[0pt]{$H^1_S$}
 & (30,30,30) & .1753 & .1823 & .0690 & .1881 & .2100 & .0623 & .1576 \\
\cline{2-9}
 & (30,30,50) & .1694 & .1494 & .0846 & .1776 & .1764 & .0686 & .1696 \\
\cline{2-9}
 & (30,50,50) & .2394 & .2511 & .0823 & .2733 & .2381 & .0692 & .2381 \\
\cline{2-9}
 & (50,50,50) & .3361 & .3379 & .0791 & .3427 & .3403 & .0680 & .3170 \\
\cline{2-9}
 & (50,50,100) & .3076 & .2924 & .1084 & .3051 & .3182 & .1030 & .3244 \\
\cline{2-9}
 & (50,100,100) & .4516 & .5303 & .1495 & .5088 & .4856 & .1331 & .5114 \\
\cline{2-9}
 & (100,100,100) & .7207 & .7295 & .1213 & .7049 & .6985 & .1229 & .7398 \\
\hline
\hline

& (10,10,10)  & .1375 & .1385 & .0803 & .1472 & .1393 & .0518 & .1314 \\
\cline{2-9}
& (10,10,30)  & .1158 & .1255 & .0976 & .0969 & .1010 & .0717 & .1406 \\
\cline{2-9}
& (10,10,50)  & .1260 & .1267 & .1190 & .0930 & .0987 & .1015 & .1415 \\
\cline{2-9}
& (10,30,30)  & .2299 & .3013 & .1565 & .3058 & .2522 & .1092 & .3036 \\
\cline{2-9}
& (10,30,50)  & .2086 & .3132 & .2149 & .2705 & .2884 & .1609 & .3294 \\
\cline{2-9}
& (10,50,50)  & .2609 & .4620 & .3039 & .3836 & .4011 & .2272 & .4579\\
\cline{2-9}\raisebox{0.ex}[0pt]{$H^2_S$}
 & (30,30,30) & .5495 & .5628 & .0980 & .5527 & .5768 & .1042 & .5353 \\
\cline{2-9}
 & (30,30,50) & .5163 & .4836 & .1481 & .5153 & .5136 & .1534 & .5620 \\
\cline{2-9}
 & (30,50,50) & .6598 & .7194 & .1686 & .6960 & .6708 & .1609 & .7283 \\
\cline{2-9}
 & (50,50,50) & .8431 & .8478 & .1564 & .8332 & .8333 & .1656 & .8645 \\
\cline{2-9}
 & (50,50,100) & .8110 & .8023 & .3100 & .7967 & .8115 & .3109 & .8856 \\
\cline{2-9}
 & (50,100,100) & .9217 & .9728 & .4085 & .9426 & .9506 & .3977 & .9752 \\
\cline{2-9}
 & (100,100,100) & .9975 & .9971 & .3510 & .9960 & .9955 & .3509 & .9985 \\
\hline
\hline

& (10,10,10)  & .3188 & .3198 & .0973 & .3240 & .3076 & .0653 & .3165\\
\cline{2-9}
& (10,10,30)  & .2823 & .2912 & .1822 & .2413 & .2383 & .1501 & .3782 \\
\cline{2-9}
& (10,10,50)  & .3245 & .3239 & .2794 & .2633 & .2705 & .2455 & .4229\\
\cline{2-9}
& (10,30,30)  & .4995 & .7096 & .3301 & .6164 & .6312 & .2766 & .7150 \\
\cline{2-9}
& (10,30,50)  & .4610 & .7492 & .4887 & .5599 & .7119 & .4052 & .7677 \\
\cline{2-9}
& (10,50,50)  & .5434 & .8955 & .6565 & .7050 & .8451 & .5568 & .8893\\
\cline{2-9}\raisebox{0.ex}[0pt]{$H^3_S$}
 & (30,30,30) & .9332 & .9396 & .2211 & .9282 & .9408 & .2387 & .9435\\
\cline{2-9}
 & (30,30,50) & .9140 & .9084 & .3651 & .9138 & .9130 & .3761 & .9590 \\
\cline{2-9}
 & (30,50,50) & .9721 & .9873 & .4274 & .9774 & .9778 & .4210 & .9920 \\
\cline{2-9}
 & (50,50,50) & .9985 & .9985 & .4117 & .9979 & .9969 & .4215 & .9994\\
\cline{2-9}
 & (50,50,100) & .9964 & .9966 & .7611 & .9953 & .9964 & .7199 & .9998 \\
\cline{2-9}
 & (50,100,100) & .9997 & 1.0000 & .8479 & .9997 & 1.000 & .8220 & 1.000 \\
\cline{2-9}
 & (100,100,100) & 1.000 & 1.000 & .8026 & 1.000 & 1.000 & .7737 & 1.000 \\
\cline{2-9}

\hline
\end{tabular}
\caption{
\label{tab:emp-power-seg-class-emp-3Cl}
The empirical power estimates for the tests under the segregation alternatives,
$H^1_S$, $H^2_S$, and $H^3_S$ for the three-class case
with $N_{mc}=10000$, for some combinations of $n_1,n_2 \in
\{10,30,50,100\}$ at the nominal level of $\alpha=0.05$.}
\end{table}
}


\subsection{Empirical Power Analysis under Association Alternatives in the Three-Class Case}
\label{sec:power-comp-assoc-3Cl}
For the association alternatives, we also consider three cases.
First, we generate $X_i \stackrel{iid}{\sim} \U((0,1)\times(0,1))$ for $i=1,2,\ldots,n_1$.
Then we generate $Y_j$ and $Z_{\ell}$ for $j=1,2,\ldots,n_2$ and $\ell=1,2,\ldots,n_3$
as follows.
For each $j$, we pick an $i$ randomly, then generate
$R^Y_j \stackrel{iid}{\sim} \U(0,r_y)$ with $r_y \in (0,1)$
and
$T_j \stackrel{iid}{\sim} \U(0,2\,\pi)$
set
$Y_j:=X_i+R^Y_j\,(\cos T_j, \sin T_j)'$.
Similarly,
for each $\ell$, we pick an $i$ randomly, then generate
$R^Z_{\ell}\stackrel{iid}{\sim} \U(0,r_z)$ with $r_z \in (0,1)$
and
$U_{\ell}\stackrel{iid}{\sim} \U(0,2\,\pi)$ and set
$Z_{\ell}:=X_i+R^Z_{\ell}\,(\cos U_{\ell}, \sin U_{\ell})'$.

In the pattern generated, appropriate choices of
$r_y$ (and $r_z$) values will imply association between classes $X$ and $Y$ (and $X$ and $Z$).
The three association alternatives are as
\begin{equation}
\label{eqn:assoc-alt-3Cl}
H^1_A: r_y=1/7,\,r_z=1/10,\;\;\; H^2_A: r_y=1/10,\,r_z=1/20,\;\;\; H^3_A: r_y=1/13,\,r_z=1/30.
\end{equation}
Observe that, from $H^1_A$ to $H^3_A$ (i.e., as $r_y$ decreases),
the association between $X$ and $Y$ gets stronger in the sense that
$Y$ points tend to be found more and more frequently around the $X$ points.
The same holds for $X$ and $Z$ points as $r_z$ decreases.
Furthermore, by construction,
classes $X$ and $Z$ are more associated compared to classes $X$ and $Y$.
On the other hand, classes $Y$ and $Z$ are not associated,
but perhaps are mildly segregated.


We present the empirical power estimates using the asymptotic and Monte Carlo
critical values in Tables \ref{tab:emp-power-assoc-class-asy-3Cl} and
\ref{tab:emp-power-assoc-class-emp-3Cl}, respectively.
Observe that for each sample size combination,
as the association gets stronger, the power estimates tend to be higher.
Based on Remark \ref{rem:MC-emp-size},
we use the power estimates based on the Monte Carlo critical values
when at least one sample is small (i.e., $n_i \leq 10$).
For larger samples, both asymptotic and Monte Carlo based empirical power
estimates are reliable for comparative purposes.
For all sample size combinations,
NN-class-specific tests have better performance in terms of power
in the sense that the highest power estimate for the three
NN-class-specific tests tends to be larger than
the highest power estimate for the three base-class-specific test
at each sample size combination.
Moreover, among the base-class-specific tests,
the one for class 3 (i.e., class $Z$) has the highest power,
while among the NN-class-specific tests, the one for class 1 (i.e., class $X$)
has the highest power estimates.

\begin{table}[ht]
\centering
\begin{tabular}{|c|c||c|c|c||c|c|c||c|}
\hline
\multicolumn{9}{|c|}{Empirical power estimates under the association alternatives} \\
\hline
& sample sizes & \multicolumn{7}{|c|}{with asymptotic critical values} \\
\hline
& $(n_1,n_2,n_3)$ & $\bh_1^B$ & $\bh_2^B$ & $\bh_3^B$ & $\bh_1^{NN}$ & $\bh_2^{NN}$ & $\bh_3^{NN}$ & $\bh_D$\\
\hline
& (10,10,10)  & .1075 & .1635 & .3187 & .3939 & .0927 & .1002 & .2755\\
\cline{2-9}
& (10,10,30)  & .0176 & .1041 & .6104 & .6145 & .0531 & .1908 & .4680\\
\cline{2-9}
& (10,10,50)  & .0022 & .0751 & .6730 & .6085 & .0561 & .1863 &.4772\\
\cline{2-9}
& (10,30,30)  & .0266 & .1987 & .5647 & .5859 & .1616 & .0798 & .4878\\
\cline{2-9}
& (10,30,50)  & .0236 & .1156 & .6011 & .5575 & .1310 & .0712 & .4674\\
\cline{2-9}
& (10,50,50)  & .0324 & .1405 & .5380 & .5528 & .1737 & .0703 & .4640\\
\cline{2-9}\raisebox{0.ex}[0pt] {$H^1_A$}
 & (30,30,30) & .4294 & .2871 & .5801 & .7758 & .2619 & .3288 & .6461\\
\cline{2-9}
 & (30,30,50) & .4728 & .2254 & .7775 & .8698 & .2002 & .4536 & .7646\\
\cline{2-9}
 & (30,50,50) & .4058 & .3470 & .7643 & .8811 & .2904 & .3052 & .7898\\
\cline{2-9}
 & (50,50,50) & .5992 & .3400 & .6916 & .8477 & .3119 & .4296 & .7494\\
\cline{2-9}
 & (50,50,100) & .6042 & .2576 & .8869 & .9434 & .2050 & .6215 & .8788 \\
\cline{2-9}
 & (50,100,100) & .5266 & .3728 & .8552 & .9518 & .3286 & .4001 & .8953 \\
\cline{2-9}
 & (100,100,100) & .6708 & .3892 & .7638 & .9072 & .3589 & .5315 & .8236 \\
\hline
\hline

& (10,10,10)  & .1554 & .1508 & .4872 & .5060 & .1688 & .1434 & .4040\\
\cline{2-9}
& (10,10,30)  & .0186 & .0744 & .8030 & .7467 & .1478 & .2016 & .6497\\
\cline{2-9}
& (10,10,50)  & .0006 & .0473 & .8423 & .7297 & .1580 & .1449 & .6569\\
\cline{2-9}
& (10,30,30)  & .0770 & .1345 & .8239 & .7580 & .4405 & .1488 & .7403\\
\cline{2-9}
& (10,30,50)  & .0587 & .0618 & .8633 & .7108 & .4402 & .1105 & .7238\\
\cline{2-9}
& (10,50,50)  & .0915 & .0686 & .8527 & .7299 & .5473 & .1901 & .7535\\
\cline{2-9}\raisebox{0.ex}[0pt] {$H^2_A$}
 & (30,30,30) & .7549 & .3372 & .8981 & .9642 & .5955 & .5970 & .9311\\
\cline{2-9}
 & (30,30,50) & .7875 & .2403 & .9757 & .9886 & .5751 & .7241 & .9744\\
\cline{2-9}
 & (30,50,50) & .7710 & .3693 & .9798 & .9921 & .7534 & .6040 & .9882\\
\cline{2-9}
 & (50,50,50) & .9374 & .4464 & .9776 & .9952 & .7669 & .8283 & .9890\\
\cline{2-9}
 & (50,50,100) & .9527 & .2707 & .9991 & .9997 & .7342 & .9372 & .9992 \\
\cline{2-9}
 & (50,100,100) & .9264 & .3771 & .9991 & .9997 & .9018 & .8063& .9994 \\
\cline{2-9}
 & (100,100,100) & .9914 & .5407 & .9984 & .9999 & .8889 & .9656 & .9999\\
\hline
\hline

& (10,10,10)  & .1835 & .1501 & .5555 & .5528 & .2170 & .1612 & .4638 \\
\cline{2-9}
 & (10,10,30) & .0212 & .0649 & .8570 & .7858 & .2177 & .1992 & .7211 \\
\cline{2-9}
 & (10,10,50) & .0004 & .0403 & .8880 & .7643 & .2301 & .1302 & .7263 \\
\cline{2-9}
 & (10,30,30) & .1155 & .1102 & .8951 & .8131 & .5891 & .1958 & .82737 \\
\cline{2-9}
 & (10,30,50) & .0874 & .0485 & .9296 & .7713 & .6114 & .1580 & .8302 \\
\cline{2-9}
 & (10,50,50) & .1326 & .0516 & .9309 & .7927 & .7294 & .2855 & .8574 \\
\cline{2-9}\raisebox{0.ex}[0pt] {$H^3_A$}
 & (30,30,30) & .8559 & .3784 & .9539 & .9874 & .7591 & .6825 & .9737 \\
\cline{2-9}
 & (30,30,50) & .8881 & .2597 & .9941 & .9969 & .7627 & .8044 & .9943 \\
\cline{2-9}
 & (30,50,50) & .8845 & .4018 & .9965 & .9983 & .9072 & .7209 & .9982 \\
\cline{2-9}
 & (50,50,50) & .9864 & .5359 & .9971 & .9996 & .9276 & .9208 & .9993 \\
\cline{2-9}
 & (50,50,100) & .9888 & .3044 & 1.000 & 1.000 & .9249 & .9785 & 1.000 \\
\cline{2-9}
 & (50,100,100) & .9859 & .4183 & 1.000 & 1.000 & .9902 & .9228 & 1.000 \\
 \cline{2-9}
 & (100,100,100) & .9999 & .6672 & 1.000 & 1.000 & .9921 & .9971 & 1.000 \\
\hline

\end{tabular}
\caption{
\label{tab:emp-power-assoc-class-asy-3Cl}
The empirical power estimates for the tests under the association alternatives,
$H^1_A$, $H^2_A$, and $H^3_A$ for the three-class case
with $N_{mc}=10000$, for some combinations of $n_1,n_2 \in
\{10,30,50,100\}$ at the nominal level of $\alpha=0.05$.}
\end{table}


\begin{table}[ht]
\centering
\begin{tabular}{|c|c||c|c|c||c|c|c||c|}
\hline
\multicolumn{9}{|c|}{Empirical power estimates under the association alternatives} \\
\hline
& sample sizes & \multicolumn{7}{|c|}{with Monte Carlo critical values} \\
\hline
& $(n_1,n_2,n_3)$ & $\bh_1^B$ & $\bh_2^B$ & $\bh_3^B$ & $\bh_1^{NN}$ & $\bh_2^{NN}$ & $\bh_3^{NN}$ & $\bh_D$\\
\hline
& (10,10,10) & .1404 & .1904 & .3496 & .4479 & .1111 & .1252 & .3052\\
\cline{2-9}
& (10,10,30) & .0382 & .1166 & .6141 & .6311 & .0675 & .2103 & .4899 \\
\cline{2-9}
& (10,10,50) & .0136 & .1054 & .6850 & .6157 & .0568 & .1997 & .4712\\
\cline{2-9}
& (10,30,30) & .0455 & .1944 & .5594 & .5923 & .1772 & .0792 & .5090\\
\cline{2-9}
& (10,30,50) & .0231 & .1258 & .6112 & .5645 & .1445 & .0815 & .4844\\
\cline{2-9}
& (10,50,50) & .0338 & .1379 & .5339 & .4958 & .1731 & .0673 & .4590\\
\cline{2-9} \raisebox{0.ex}[0pt] {$H^1_A$}
 & (30,30,30) & .4356 & .3065 & .5886 & .7726 & .2641 & .3300 & .6469\\
\cline{2-9}
 & (30,30,50) & .4762 & .2316 & .7845 & .8784 & .2014 & .4598 & .7753\\
\cline{2-9}
 & (30,50,50) & .4188 & .3466 & .7636 & .8859 & .2917 & .2971 & .7938\\
\cline{2-9}
 & (50,50,50) & .5992 & .3377 & .6932 & .8454 & .3161 & .4353 & .7498\\
\cline{2-9}
 & (50,50,100) & .6000 & .2603 & .8871 & .9463 & .2049 & .6289 & .8821\\
\cline{2-9}
 & (50,100,100) & .5591 & .3736 & .8556 & .9535 & .3390 & .4034 & .8954\\
\cline{2-9}
 & (100,100,100) & .6697 & .3858 & .7642 & .9100 & .3664 & .5365 & .8338\\
\hline
\hline

& (10,10,10)  & .2017 & .1798 & .5185 & .5655 & .1936 & .1754 & .4373\\
\cline{2-9}
& (10,10,30)  & .0518 & .0838 & .8061 & .7625 & .1764 & .2237 & .6656  \\
\cline{2-9}
& (10,10,50)  & .0040 & .0654 & .8521 & .7367 & .1594 & .1580 & .6516\\
\cline{2-9}
& (10,30,30)  & .1085 & .1296 & .8201 & .7627 & .4638 & .1482 & .7563\\
\cline{2-9}
& (10,30,50)  & .0587 & .0699 & .8710 & .7156 & .4668 & .1274 & .7392\\
\cline{2-9}
& (10,50,50)  & .0923 & .0669 & .8509 & .6725 & .5457 & .1860 & .7498\\
\cline{2-9} \raisebox{0.ex}[0pt] {$H^2_A$}
 & (30,30,30) & .7581 & .3573 & .9010 & .9634 & .5976 & .5987 & .9312\\
\cline{2-9}
 & (30,30,50) & .7901 & .2440 & .9767 & .9894 & .5761 & .7304 & .9765\\
\cline{2-9}
 & (30,50,50) & .7817 & .3681 & .9798 & .9927 & .7553 & .5946 & .9884\\
\cline{2-9}
 & (50,50,50) & .9372 & .4443 & .9783 & .9950 & .7702 & .8318 & .9890\\
\cline{2-9}
 & (50,50,100) & .9521 & .2733 & .9991 & .9997 & .7339 & .9390 & .9993\\
\cline{2-9}
 & (50,100,100) & .9339 & .3782 & .9992 & .9997 & .9074 & .8094 & .9994 \\
\cline{2-9}
 & (100,100,100) & .9914 & .5375 & .9984 & .9999 & .8928 & .9662 & 1.000\\
\hline
\hline

& (10,10,10)  & .2350 & .1774 & .5875 & .6110 & .2453 & .1942 & .4980\\
\cline{2-9}
& (10,10,30)  & .0590 & .0730 & .8597 & .8005 & .2533 & .2253 & .7381\\
\cline{2-9}
& (10,10,50)  & .0016 & .0543 & .8941 & .7721 & .2314 & .1426 & .7223\\
\cline{2-9}
& (10,30,30)  & .1529 & .1055 & .8917 & .8171 & .6119 & .1956 & .8411\\
\cline{2-9}
& (10,30,50)  & .0872 & .0576 & .9343 & .7750 & .6337 & .1776 & .8427\\
\cline{2-9}
& (10,50,50)  & .1332 & .0505 & .9291 & .7379 & .7285 & .2814 & .8546\\
\cline{2-9} \raisebox{0.ex}[0pt] {$H^3_A$}
 & (30,30,30) & .8579 & .3995 & .9558 & .9872 & .7610 & .6843 & .9739\\
\cline{2-9}
 & (30,30,50) & .8899 & .2631 & .9945 & .9973 & .7634 & .8087 & .9947\\
\cline{2-9}
 & (30,50,50) & .8926 & .4000 & .9965 & .9984 & .9078 & .7132 & .9982\\
\cline{2-9}
 & (50,50,50) & .9864 & .5333 & .9972 & .9996 & .9285 & .9230 & .9993\\
\cline{2-9}
 & (50,50,100) & .9887 & .3094 & 1.000 & 1.000 & .9248 & .9790 & 1.000\\
\cline{2-9}
 & (50,100,100) & .9873 & .4196 & 1.000 & 1.000 & .9907 & .9242 & 1.000\\
\cline{2-9}
 & (100,100,100) & .9999 & .6639 & 1.000 & 1.000 & .9928 & .9972 & 1.000\\
\hline

\end{tabular}
\caption{
\label{tab:emp-power-assoc-class-emp-3Cl}
The empirical power estimates for the tests under the association alternatives,
$H^1_A$, $H^2_A$, and $H^3_A$ for the three-class case
with $N_{mc}=10000$, for some combinations of $n_1,n_2 \in
\{10,30,50,100\}$ at the nominal level of $\alpha=0.05$.}
\end{table}


\begin{remark}
\label{rem:MC-power}
\textbf{Main Result of Monte Carlo Power Analysis:}
Based on the recommendations made in Remark \ref{rem:MC-emp-size},
when at least one sample size is small
(in the sense that some cell count is $\leq 5$),
the power comparisons are more appropriate if the Monte Carlo critical values are used.
For large samples, the power comparisons can be made using
either the asymptotic or Monte Carlo critical values.
In Sections \ref{sec:power-comp-seg-2Cl} and \ref{sec:power-comp-assoc-2Cl},
we observe that the power estimates using
asymptotic and Monte Carlo critical values
tend to be significantly different from each other
when at least one of the samples is small,
very similar when samples are larger.
Under the segregation alternatives, for large samples with similar sizes,
we conclude that NN-class-specific tests have higher power;
when at least one sample is small, base-class-specific test
for the smaller (larger) class has the highest (lowest) power estimates.
Under the association alternatives,
we observe that when at least
one sample size is small,
base-class-specific test for the smaller class (i.e., class $X$)
has extremely poor performance,
while for the larger class it has higher power than the
NN-class-specific test.
On the other hand,
for larger samples, base-class-specific test for the
largest (smallest) class has the highest (lowest) power.

However base- and NN-class-specific tests answer different questions.
So it is not appropriate to compare the base- and NN-class-specific tests.
Thus, we recommend both of the base- and NN-class-specific tests
to get more information on different aspects of the NN structure.
$\square$
\end{remark}

\section{Examples}
\label{sec:examples}
We illustrate the tests on three example data sets:
an ecological data set (swamp tree data),
an epidemiological data set (leukemia data),
and an MRI data set (pyramidal neuron data).

\subsection{Swamp Tree Data}
\label{sec:swamp-data}
\cite{good:1982} considered the spatial patterns of tree species
along the Savannah River, South Carolina, U.S.A.
From this data, \cite{dixon:EncycEnv2002} used a single 50m $\times$ 200m rectangular plot
to illustrate NNCT methods.
All live or dead trees with 4.5 cm or more dbh (diameter at breast height)
were recorded together with their species.
Hence it is an example of a realization of a marked multi-variate point pattern.
The plot contains 13 different tree species,
four of which comprises over 90 \% of the 734 tree stems.
The remaining tree stems were categorized as ``other trees".
The plot consists of 215 water tupelo (\emph{Nyssa aquatica}),
205 black gum (\emph{Nyssa sylvatica}), 156 Carolina ash (\emph{Fraxinus caroliniana}),
98 bald cypress (\emph{Taxodium distichum}), and 60 stems of 8 additional species (i.e., other species).
A $5 \times 5$ NNCT-analysis is conducted for this data set.
If segregation among the less frequent species were important,
a more detailed $13 \times 13$ NNCT-analysis should be performed.
The locations of these trees in the study region are plotted in Figure \ref{fig:SwampTrees}
and the corresponding $5 \times 5$ NNCT together with percentages
based on row and column sums
are provided in Table \ref{tab:NNCT-swamp}.
For example, for black gum as the base species and Carolina ash as the NN species,
the cell count is 26 which is 13 \% of the 205 black gums (which is 28 \% of all trees),
and 15 \% of the 171 times Carolina ashes serves as NN (which is 23 \% of all trees).
Observe that the percentages and Figure \ref{fig:SwampTrees} are suggestive of segregation for
all tree species, especially for Carolina ashes, water tupelos, black gums, and the ``other" trees
since the observed percentage of species with themselves as the NN is much larger than the marginal
(row or column) percentages.

\begin{figure}[ht]
\rotatebox{-90}{ \resizebox{3. in}{!}{\includegraphics{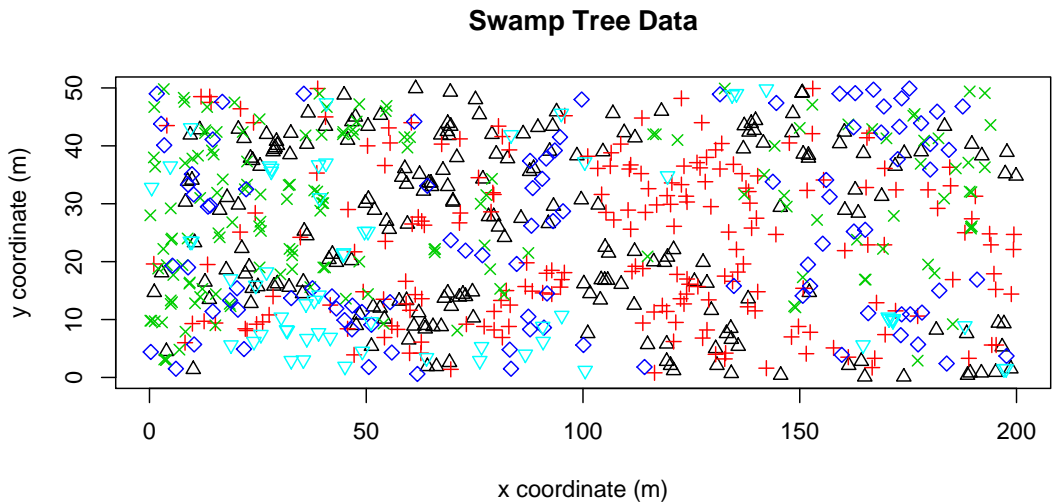} }}
 \caption{
\label{fig:SwampTrees}
The scatter plot of the locations of water tupelos (triangles $\triangle$),
black gum trees (pluses $+$), Carolina ashes (crosses $\times$),
bald cypress trees (diamonds $\diamond$), and other trees (inverse triangles $\triangledown$).}
\end{figure}

\begin{table}[ht]
\centering
\begin{tabular}{cc|ccccc|c}
\multicolumn{2}{c}{}& \multicolumn{5}{c}{NN}& \\
\multicolumn{2}{c}{}& W.T. & B.G. & C.A. &  B.C. & O.T.   &   sum  \\
\hline
& W.T. &    112  &   40  &  29  &  20 &  14 &  215 \\
& B.G. &    38   &  117  &  26  &  16 &   8 &  205 \\
& C.A. &    23   &   23  &  82  &  22 &   6 &  156 \\
\raisebox{2.5ex}[0pt]{base}
& B.C. &    19   &   29  &  29  &  14 &   7 &  98  \\
& O.T. &    7    &   8   &  5   &  7  &  33 &  60  \\
\hline
& sum  &   199   &  217  & 171  &  79 &  68 &  734 \\
\end{tabular}
\begin{tabular}{cc|ccccc|c}
\multicolumn{2}{c}{}& \multicolumn{5}{c}{NN}& \\
\multicolumn{2}{c}{}& W.T. & B.G. & C.A. &  B.C. & O.T.   &   sum  \\
\hline
& W.T. &    52 \% &   19 \% &  13 \%  &  9 \%  &  7 \%  &  29 \% \\
& B.G. &    19 \% &   57 \% &  13 \%  &  8 \%  &  4 \%  &  28 \% \\
& C.A. &    15 \% &   15 \% &  53 \%  &  14 \% &  4 \%  &  21 \% \\
\raisebox{2.5ex}[0pt]{base}
& B.C. &    19 \% &   30 \% &  30 \% &  14 \% &   7 \%  &  13 \% \\
& O.T. &    12 \% &   13 \% &  8 \%  &  12 \% &  55 \%  &  8 \%  \\
\hline
\end{tabular}
\begin{tabular}{cc|ccccc|c}
\multicolumn{2}{c}{}& \multicolumn{5}{c}{NN}& \\
\multicolumn{2}{c}{}& W.T. & B.G. & C.A. &  B.C. & O.T.   \\
\hline
& W.T. &    56 \% &   18 \% &  17 \%  &  25 \%  &  21 \%  \\
& B.G. &    19 \% &   54 \% &  15 \%  &  20 \%  &  12 \%  \\
& C.A. &    12 \% &   11 \% &  48 \%  &  28 \% &  9 \%  \\
\raisebox{2.5ex}[0pt]{base}
& B.C. &    10 \% &   13 \% &  17 \% &  18 \% &   10 \%  \\
& O.T. &    4 \% &    4 \% &  3 \%  &  9 \% &  49 \%  \\
\hline
& sum  &   27 \% &  30 \% & 23 \% &  11 \% &  9 \%  \\
\end{tabular}
\caption{\label{tab:NNCT-swamp}
The NNCT for swamp tree data (top),
the percentages of each row, i.e.,
the percentages of tree species serving as NN to each tree species (middle),
and the percentages of each column, i.e,
the percentages of tree species serving as base to each tree species (bottom).
W.T. = water tupelo, B.G. = black gum, C.A. = Carolina ash,
B.C.  = bald cypress, and O.T. = other tree species.}
\end{table}

\begin{table}[ht]
\centering
\begin{tabular}{|cc||c|c|c|c|}
\hline
\multicolumn{6}{|c|}{Test statistics and $p$-values for segregation tests} \\
\hline
&  & $\chi^2$-statistic & $p_{asy}$ & $p_{mc}$ & $p_{rand}$ \\
\hline
\multicolumn{2}{|c||}{overall} & 275.64 & $<.0001$ & $<.0001$ & $<.0001$ \\
\hline
     & W.T. & 42.27 & $<.0001$ & $<.0001$ & $<.0001$    \\
     & B.G. & 65.13 & $<.0001$ & $<.0001$ & $<.0001$    \\
base & C.A. & 70.99 & $<.0001$ & $<.0001$ & $<.0001$    \\
     & B.C. &  7.09 & .1313  & .1315 & .1291 \\
     & O.T. &117.48 & $<.0001$ & $<.0001$ & $<.0001$ \\
\hline
     & W.T. & 61.37 & $<.0001$ & $<.0001$ & $<.0001$    \\
     & B.G. & 75.96 & $<.0001$ & $<.0001$ & $<.0001$    \\
NN   & C.A. & 81.06 & $<.0001$ & $<.0001$ & $<.0001$    \\
     & B.C. & 10.73 & .0571  & .0611 & .0518 \\
     & O.T. &118.23 & $<.0001$ & $<.0001$ & $<.0001$ \\
\hline
\end{tabular}
\caption{
\label{tab:pval-swamp-class}
The  $\chi^2$-statistics for the segregation tests
and the corresponding $p$-values.
$p_{asy}$ stands for the $p$-value based on the asymptotic approximation,
$p_{mc}$ is the $p$-value based on $10000$ Monte Carlo replication of the CSR independence
pattern in the same region, and $p_{rand}$ is based on Monte Carlo
randomization of the labels on the given locations of the trees 10000 times.
W.T. = water tupelo, B.G. = black gum, C.A. = Carolina ash,
B.C.  = bald cypress, and O.T. = other tree species.}
\end{table}

The locations of the tree species can be viewed a priori resulting
from different processes so the more appropriate null hypothesis is the CSR independence pattern.
Hence our inference will be a conditional one (see Remark \ref{rem:QandR}).
We calculate $Q=472$ and $R=454$ for this data set.
We present the overall test of segregation,
class-specific test statistics and the associated $p$-values
in Table \ref{tab:pval-swamp-class},
where $p_{asy}$ stands for the $p$-value based on the asymptotic approximation
(i.e., the corresponding $\chi^2$ distribution),
$p_{mc}$ is the $p$-value based on $10000$ Monte Carlo replication of the CSR independence
pattern in the same plot, and $p_{rand}$ is based on Monte Carlo
randomization of the labels on the given locations of the trees 10000 times.
Observe that $p_{asy}$, $p_{mc}$, and $p_{rand}$ are similar for each test.
Overall test of segregation is significant implying
significant deviation from the CSR independence pattern
for at least one pair of tree species.
Base-class-specific tests are all significant for all species but bald cypresses
implying significant deviation in rows than expected under CSR independence, except for bald cypress trees.
These findings are in agreement with the results of (\cite{dixon:EncycEnv2002}).
NN-class-specific tests are significant for all species but bald cypresses,
implying significant deviation in columns than expected under CSR independence
for all species except for bald cypress trees.
Hence except for bald cypresses, each tree species seem
to result from a (perhaps) different first order inhomogeneous Poisson process.

\begin{figure}[t]
\centering
\rotatebox{-90}{ \resizebox{2 in}{!}{\includegraphics{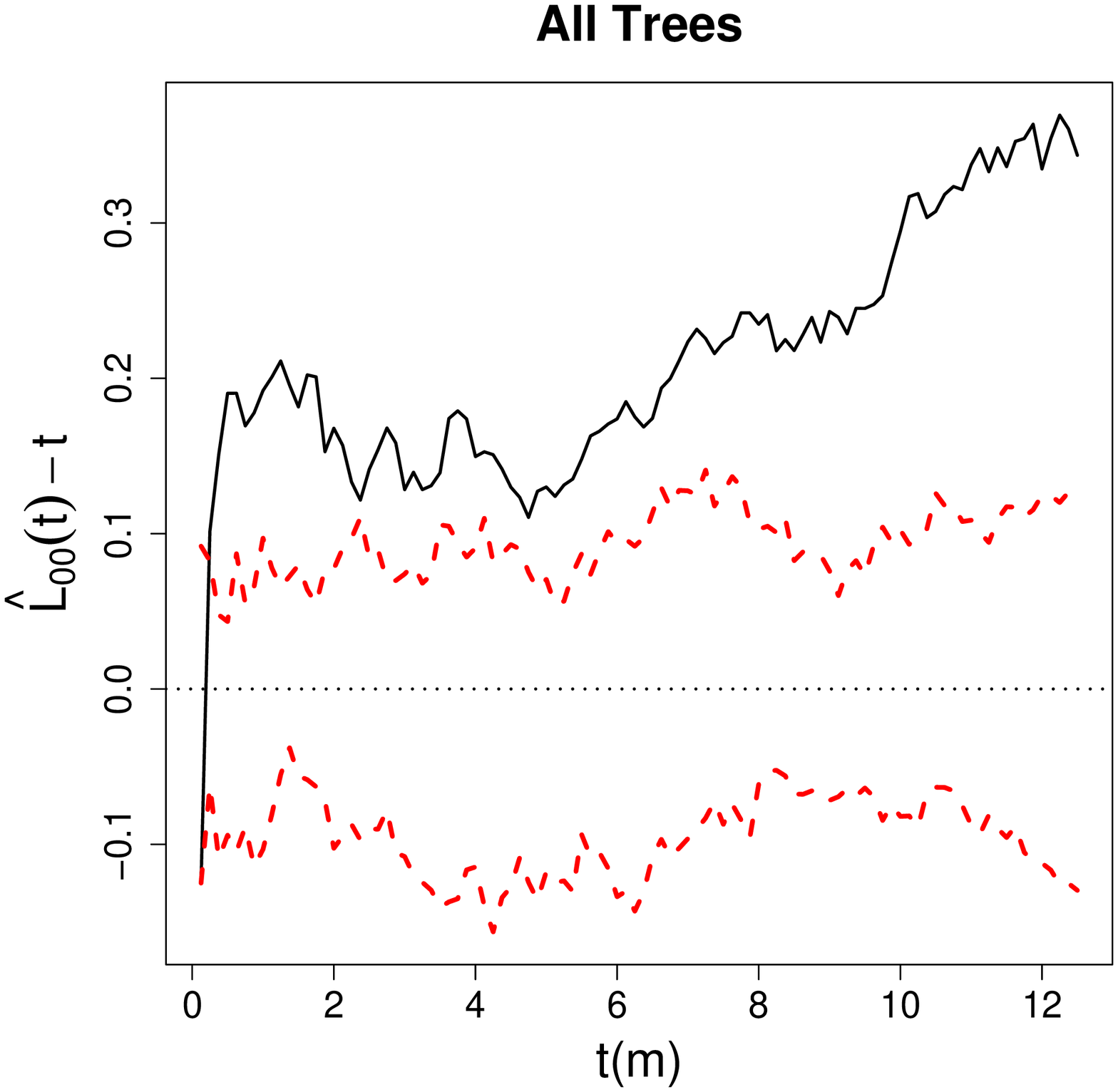} }}
\rotatebox{-90}{ \resizebox{2 in}{!}{\includegraphics{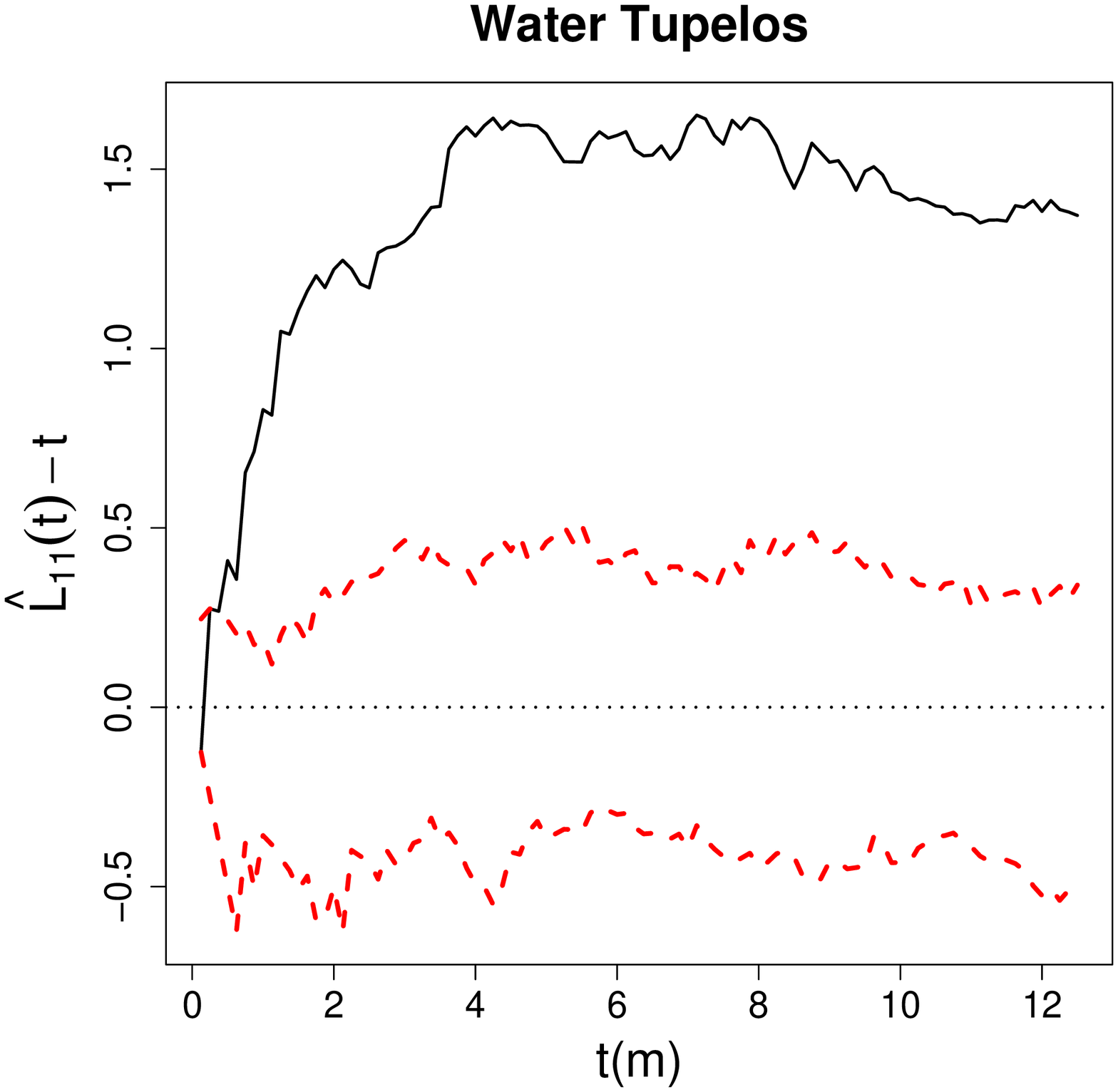} }}
\rotatebox{-90}{ \resizebox{2 in}{!}{\includegraphics{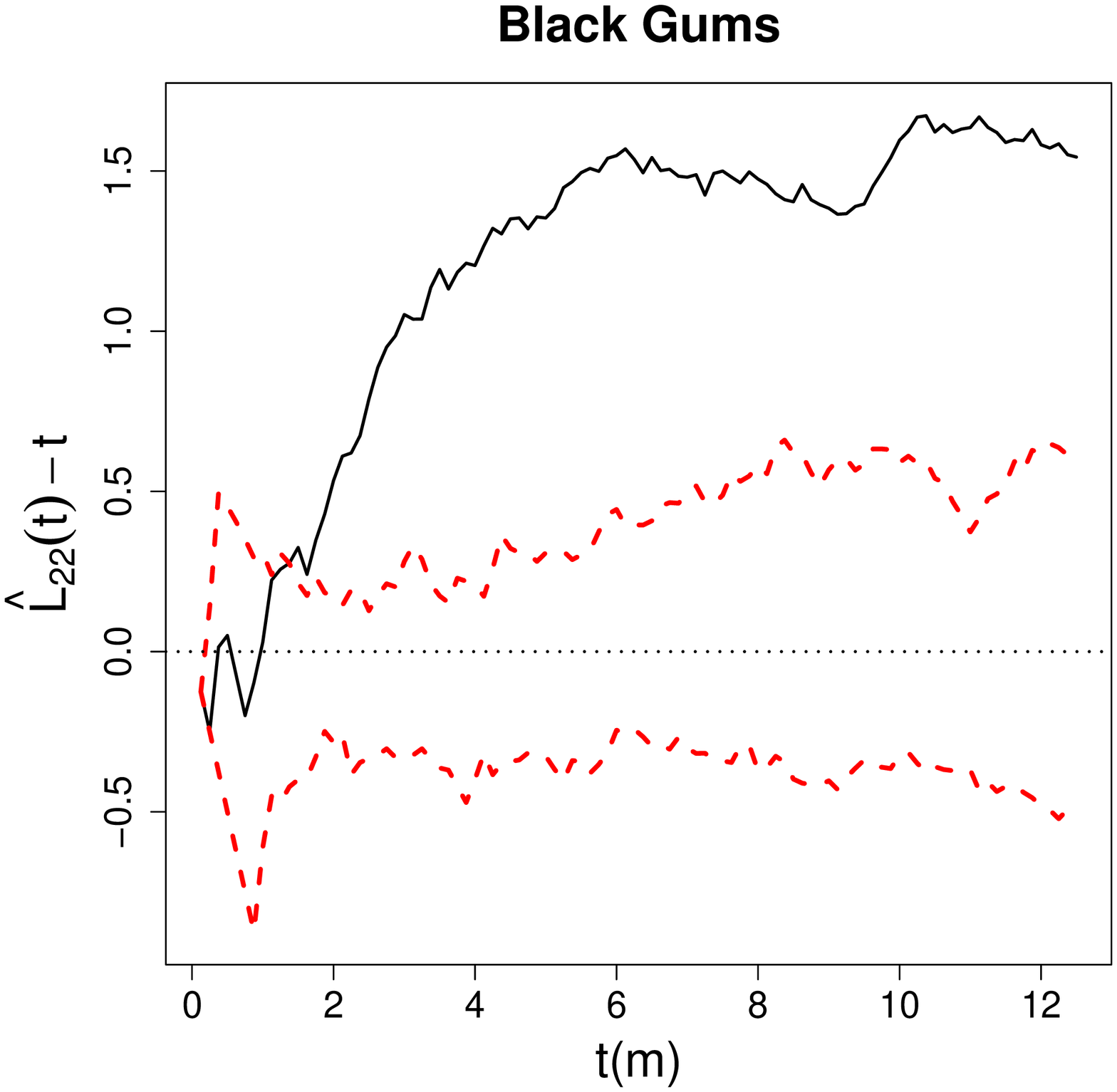} }}
\rotatebox{-90}{ \resizebox{2 in}{!}{\includegraphics{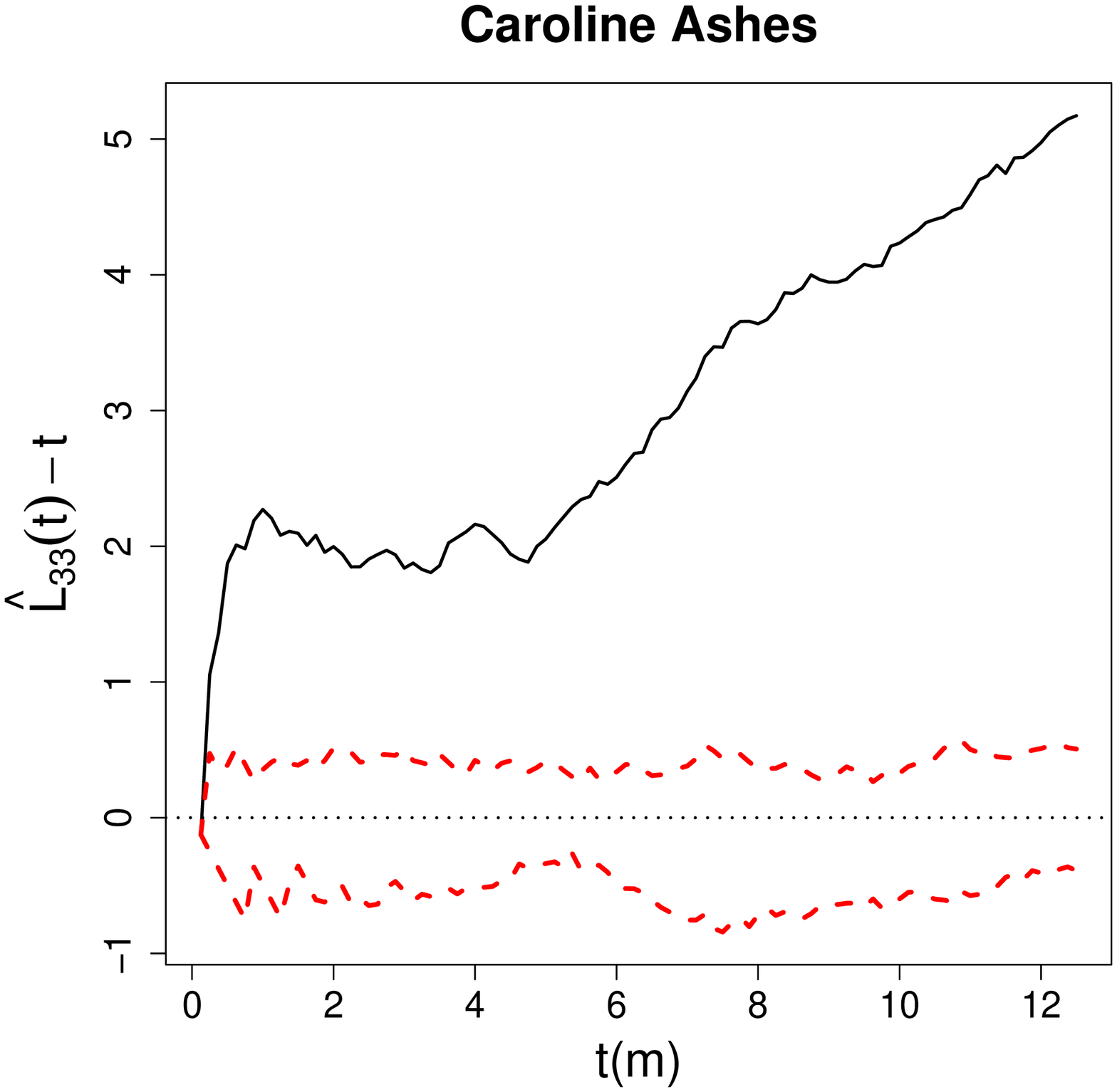} }}
\rotatebox{-90}{ \resizebox{2 in}{!}{\includegraphics{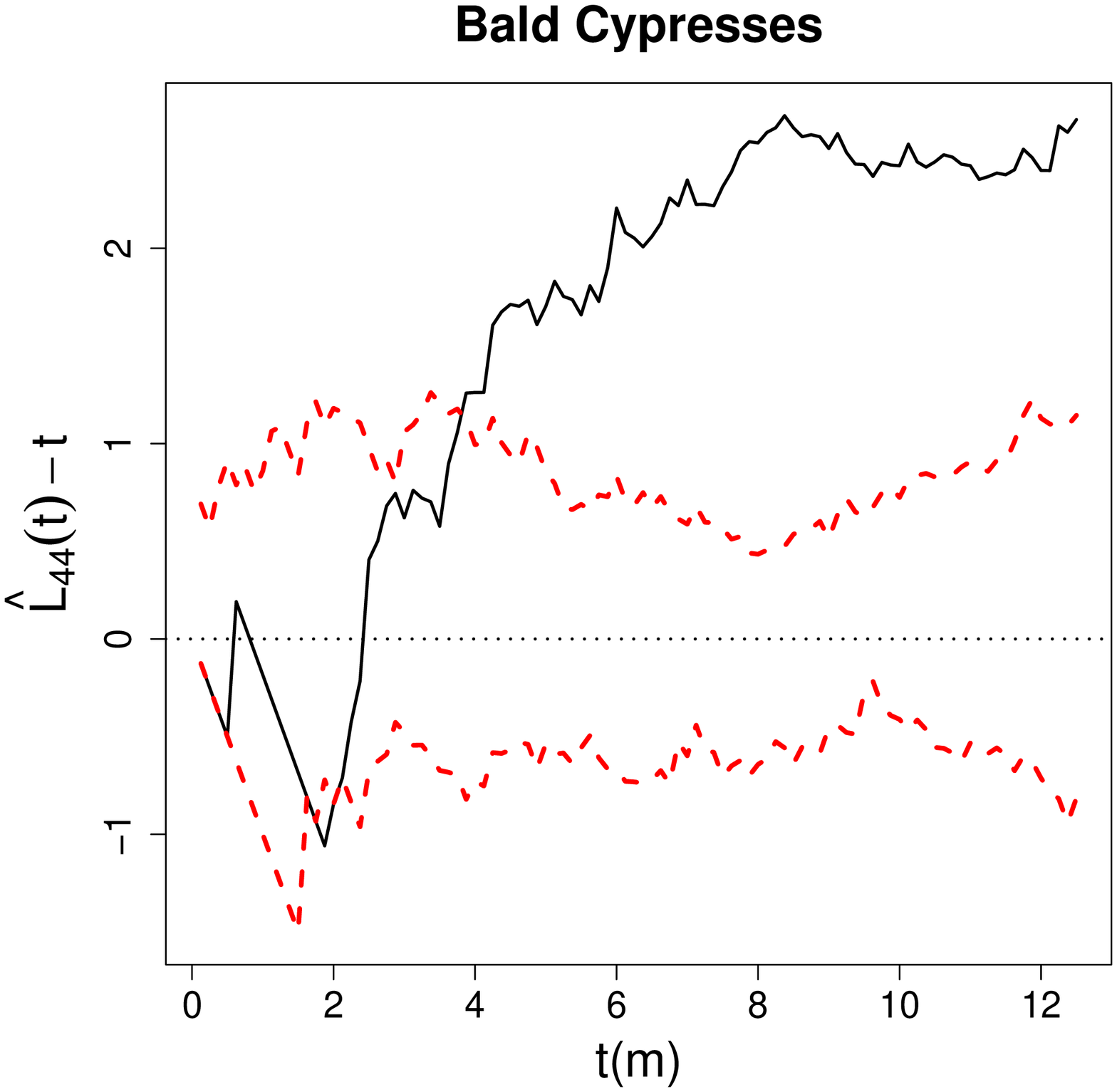} }}
\rotatebox{-90}{ \resizebox{2 in}{!}{\includegraphics{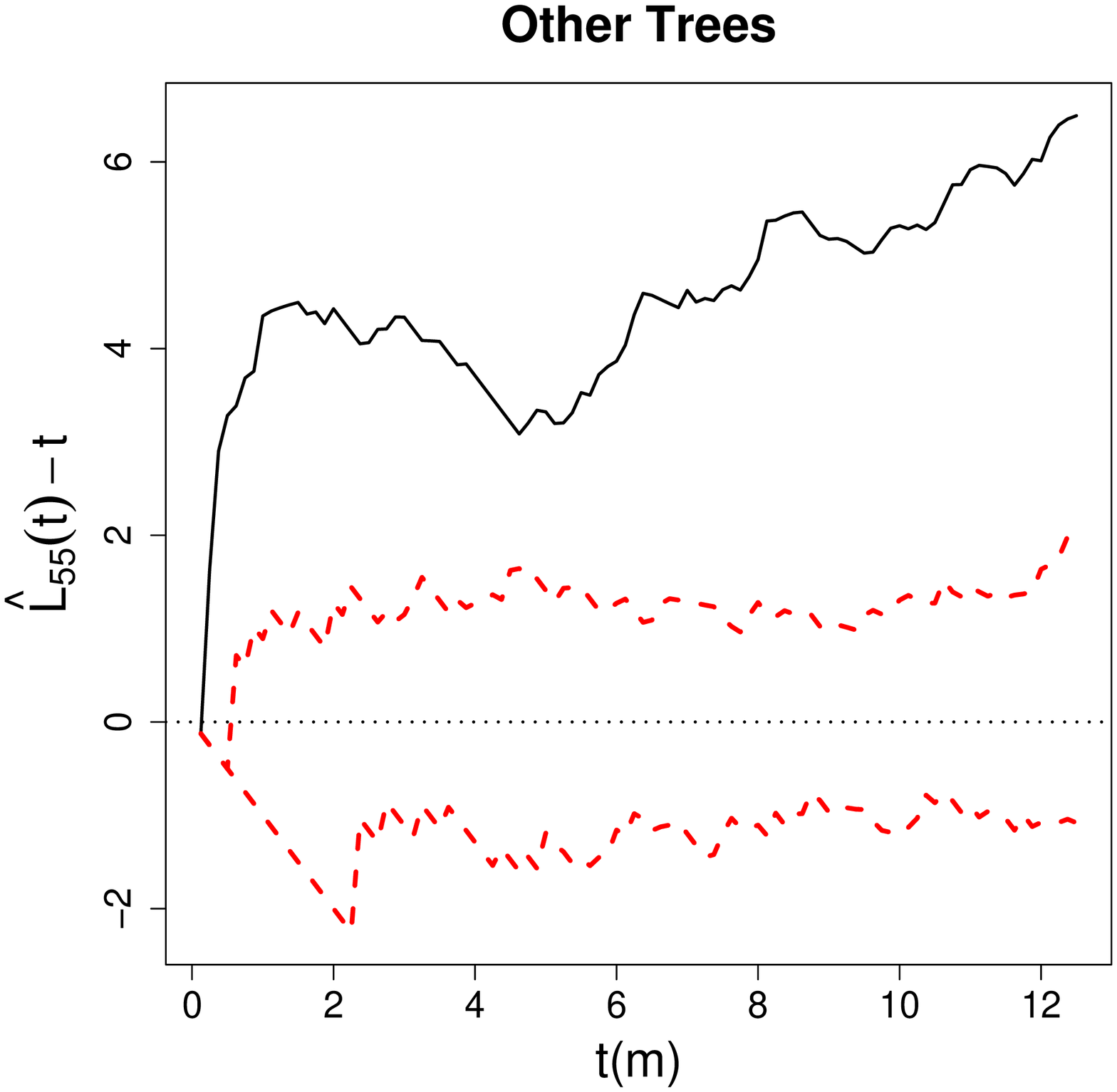} }}
\caption{
\label{fig:swamp-Liihat}
Second-order properties of swamp tree data.
Functions plotted are Ripley's univariate $L$-functions
$\widehat{L}_{ii}(t)-t$ for $i=0,1,\ldots,5$,
where $i=0$ stands for all data combined, $i=1$ for water tupelos, $i=2$ for black gums,
$i=3$ for Carolina ashes, $i=4$ for bald cypresses, and $i=5$ for other trees.
Wide dashed lines around 0 are the upper and lower 95 \% confidence bounds for the
$L$-functions based on Monte Carlo simulation under the CSR independence pattern.}
\end{figure}

\begin{figure}[t]
\centering
\rotatebox{-90}{ \resizebox{2 in}{!}{\includegraphics{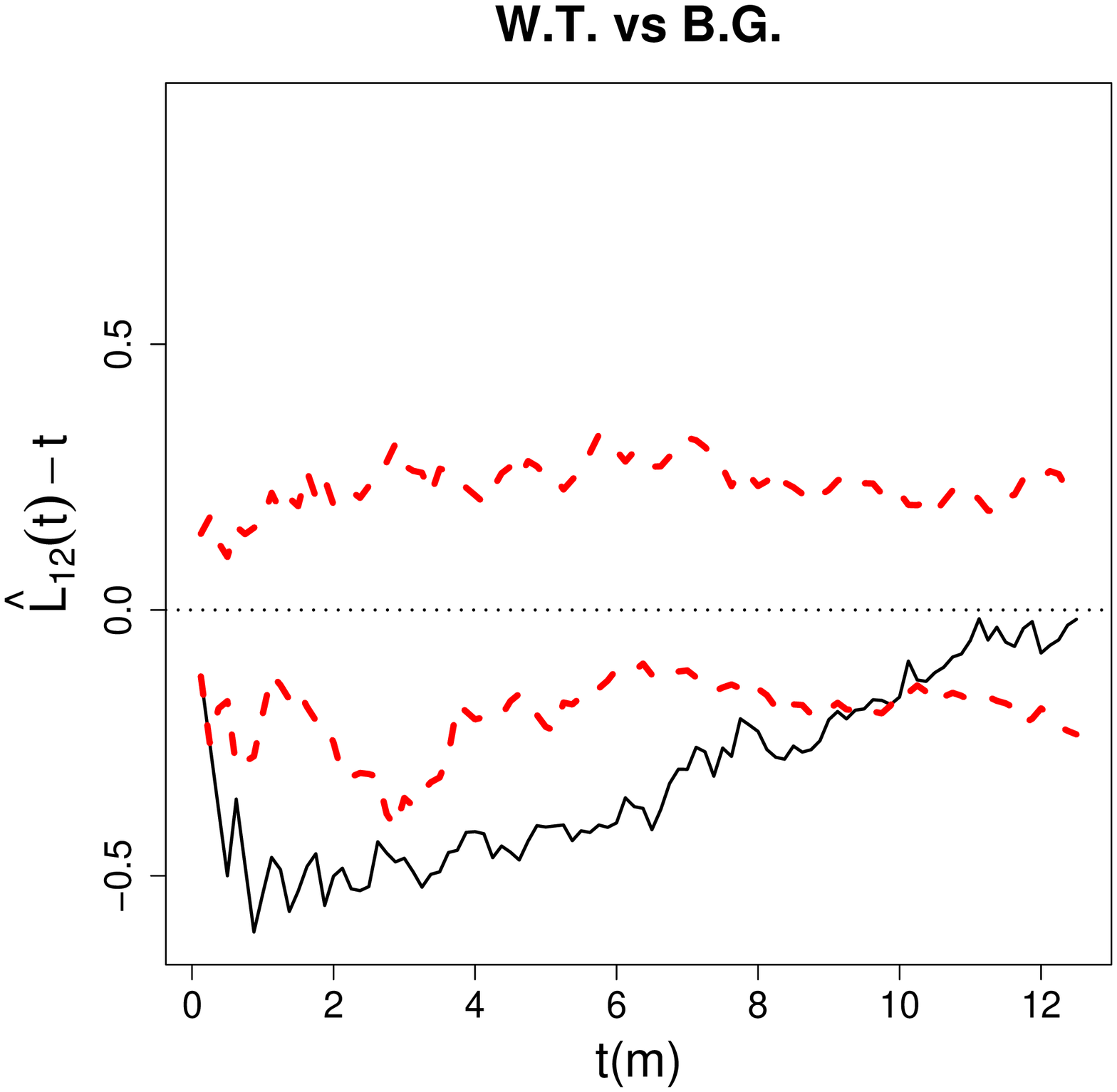} }}
\rotatebox{-90}{ \resizebox{2 in}{!}{\includegraphics{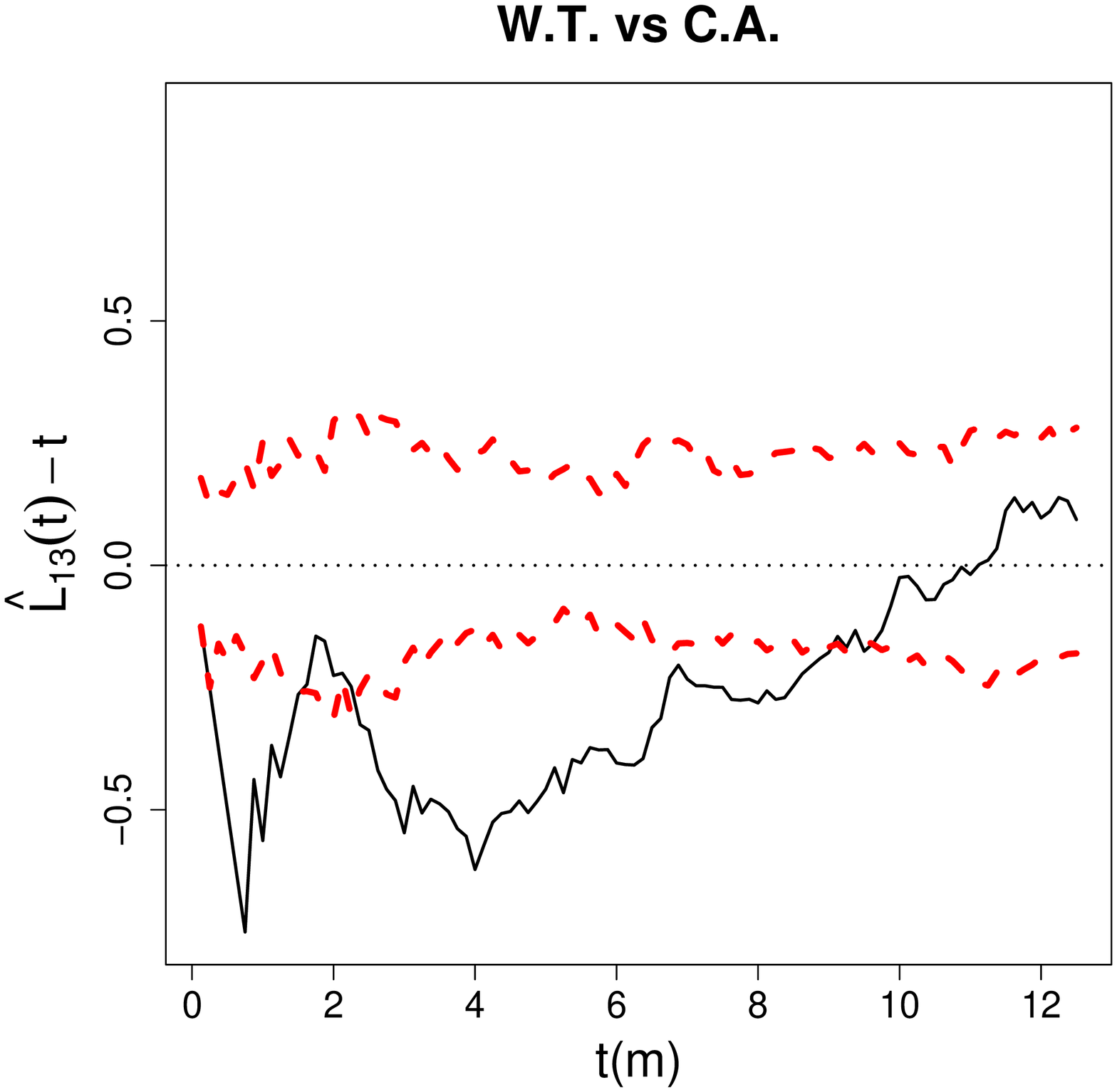} }}
\rotatebox{-90}{ \resizebox{2 in}{!}{\includegraphics{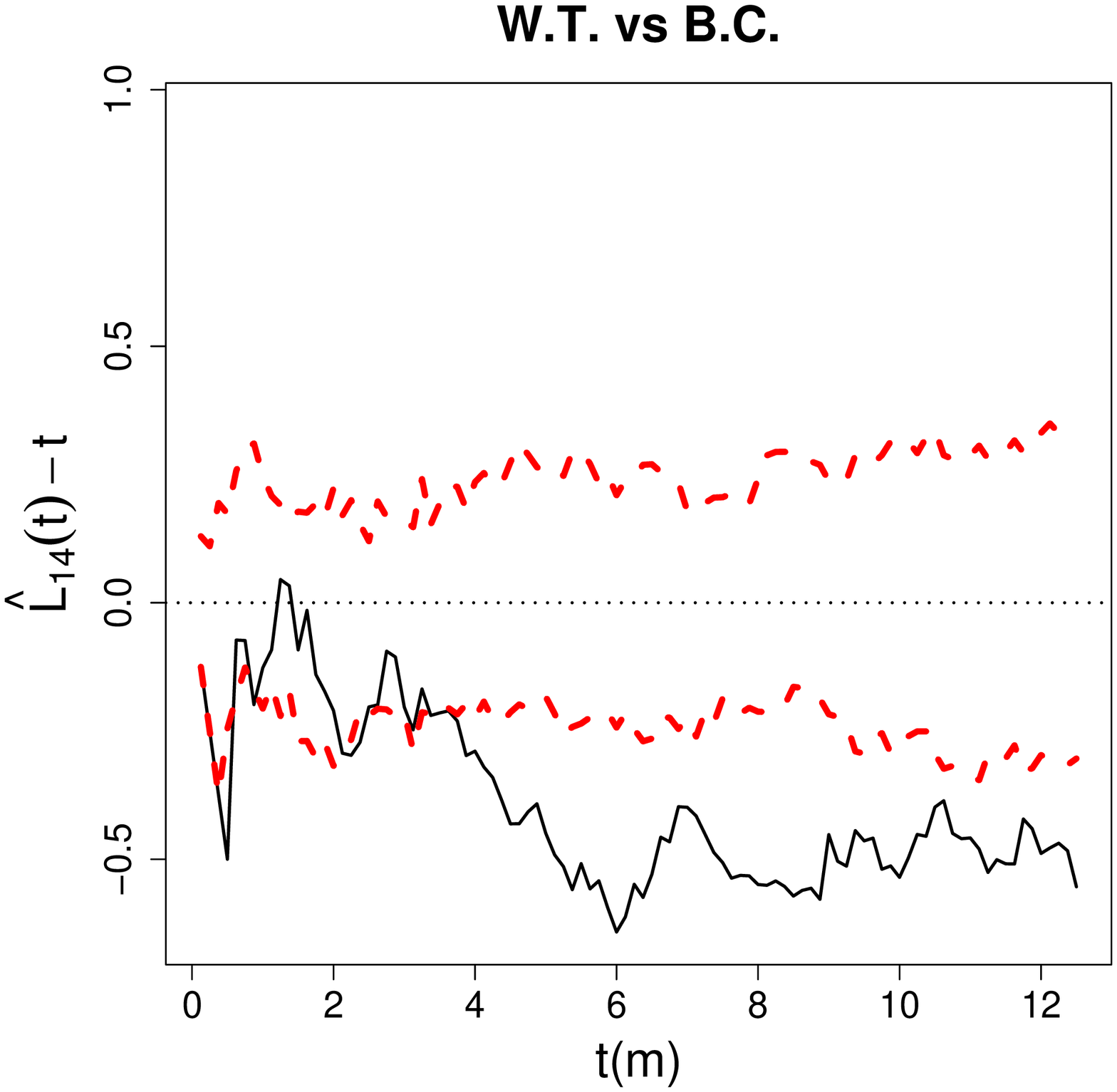} }}
\rotatebox{-90}{ \resizebox{2 in}{!}{\includegraphics{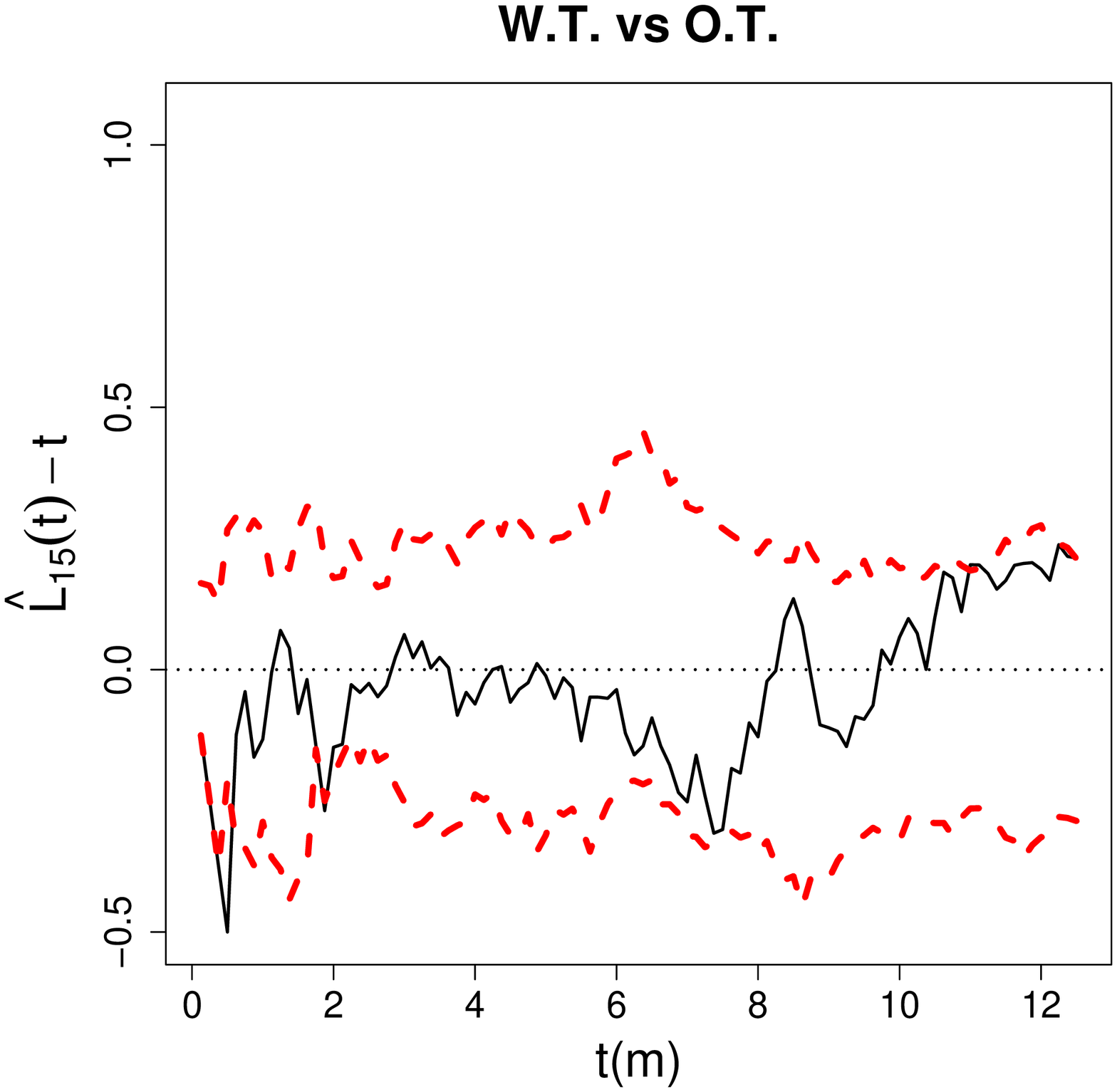} }}
\rotatebox{-90}{ \resizebox{2 in}{!}{\includegraphics{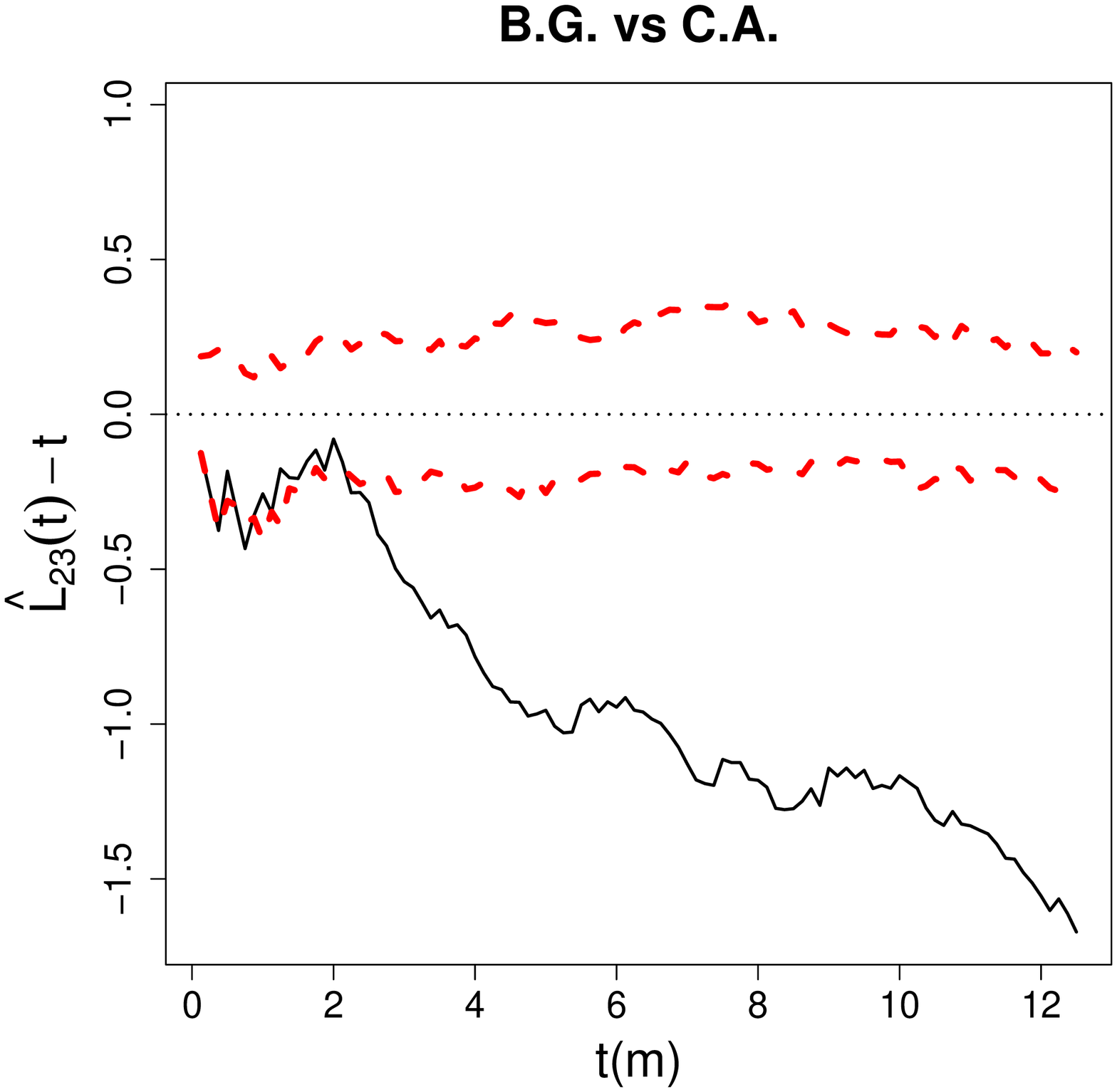} }}
\rotatebox{-90}{ \resizebox{2 in}{!}{\includegraphics{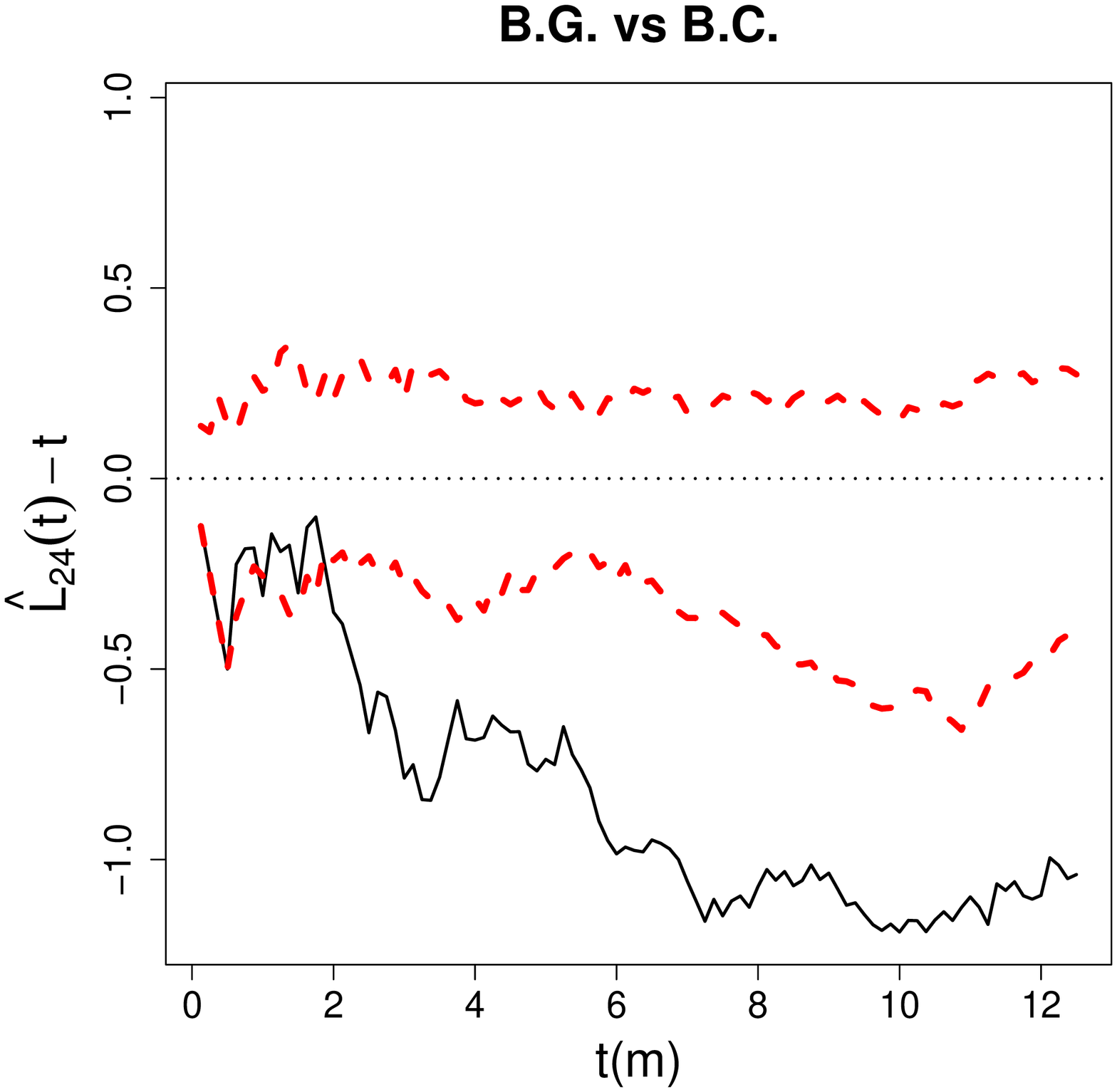} }}
\rotatebox{-90}{ \resizebox{2 in}{!}{\includegraphics{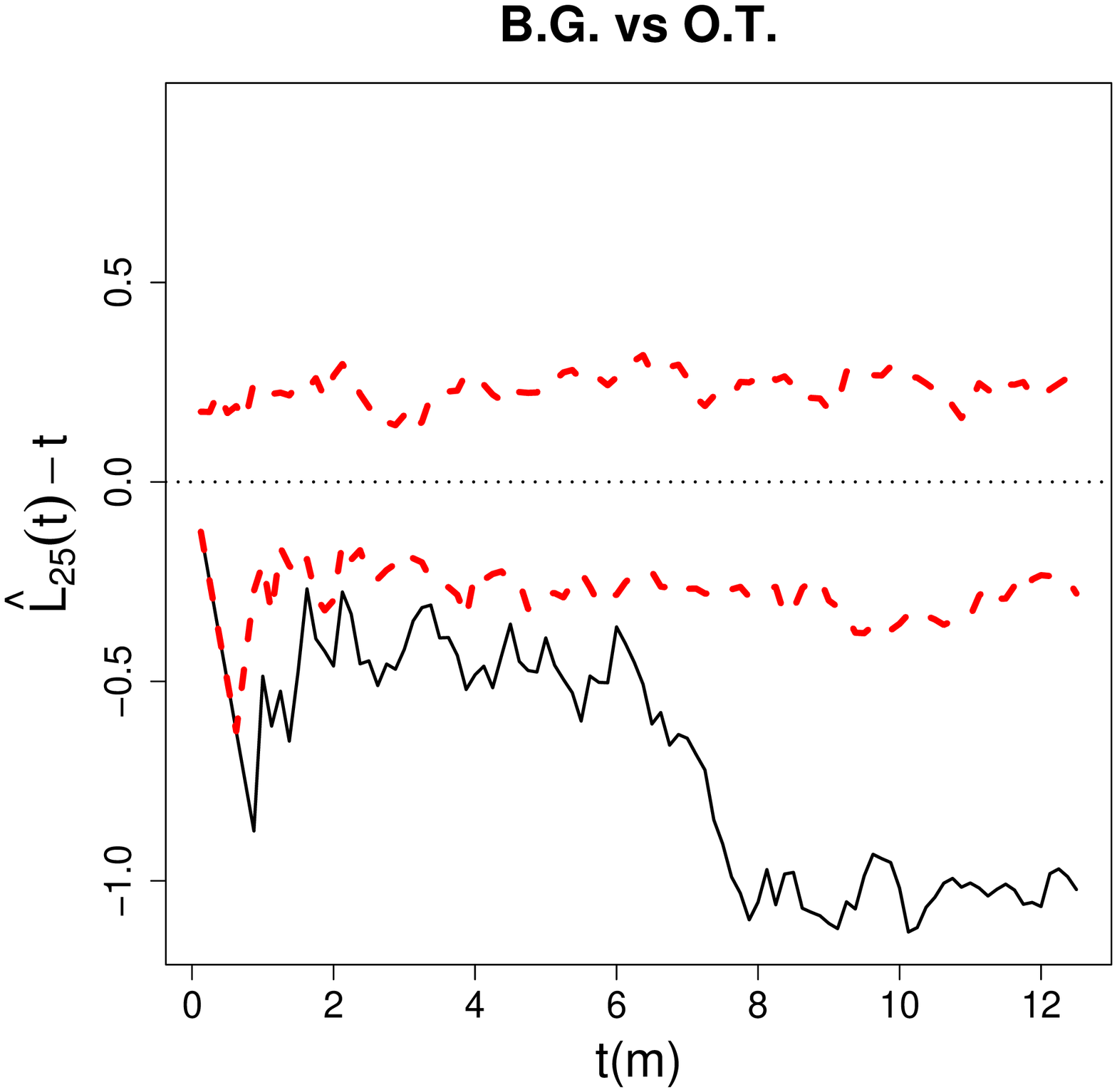} }}
\rotatebox{-90}{ \resizebox{2 in}{!}{\includegraphics{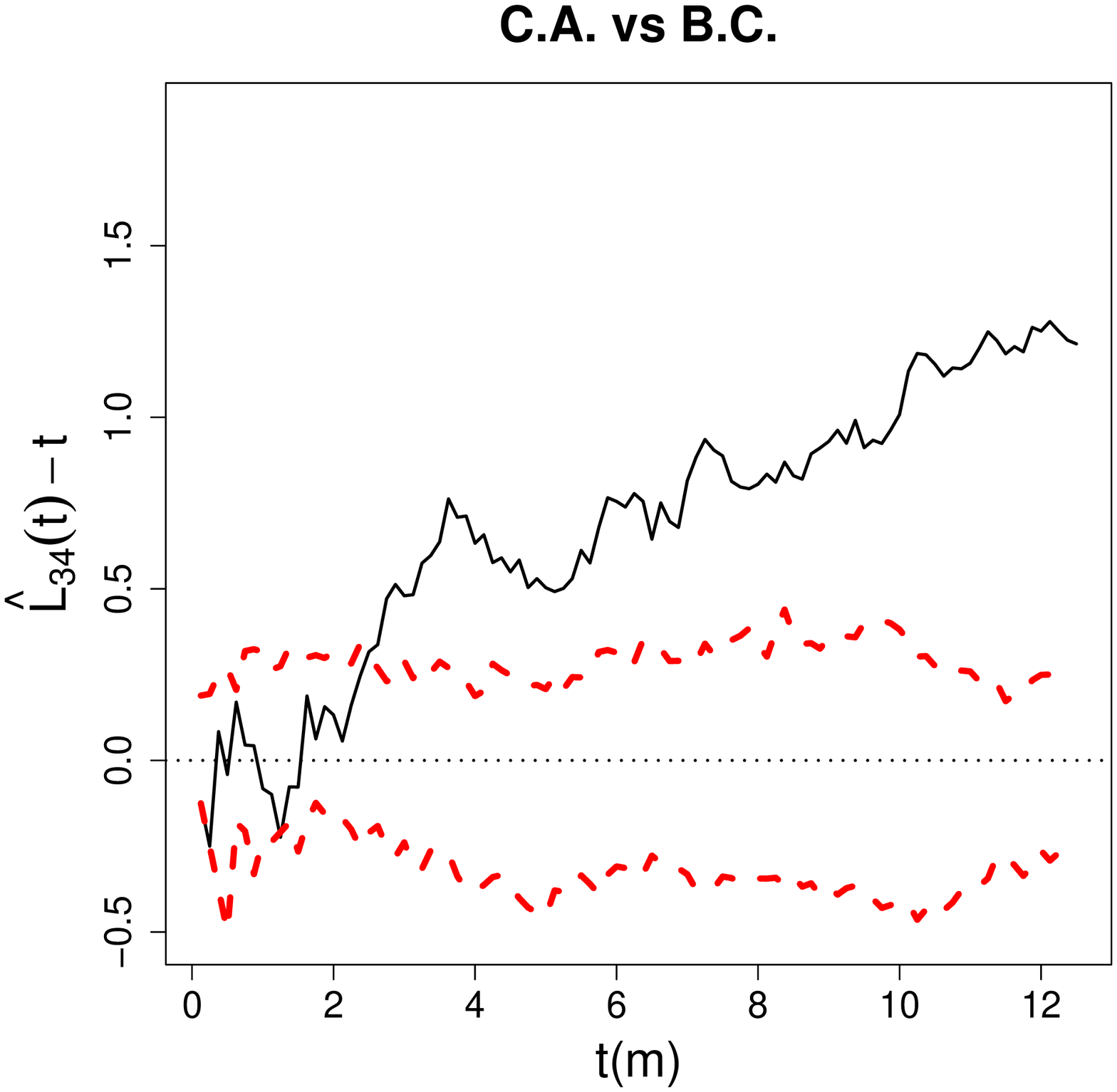} }}
\rotatebox{-90}{ \resizebox{2 in}{!}{\includegraphics{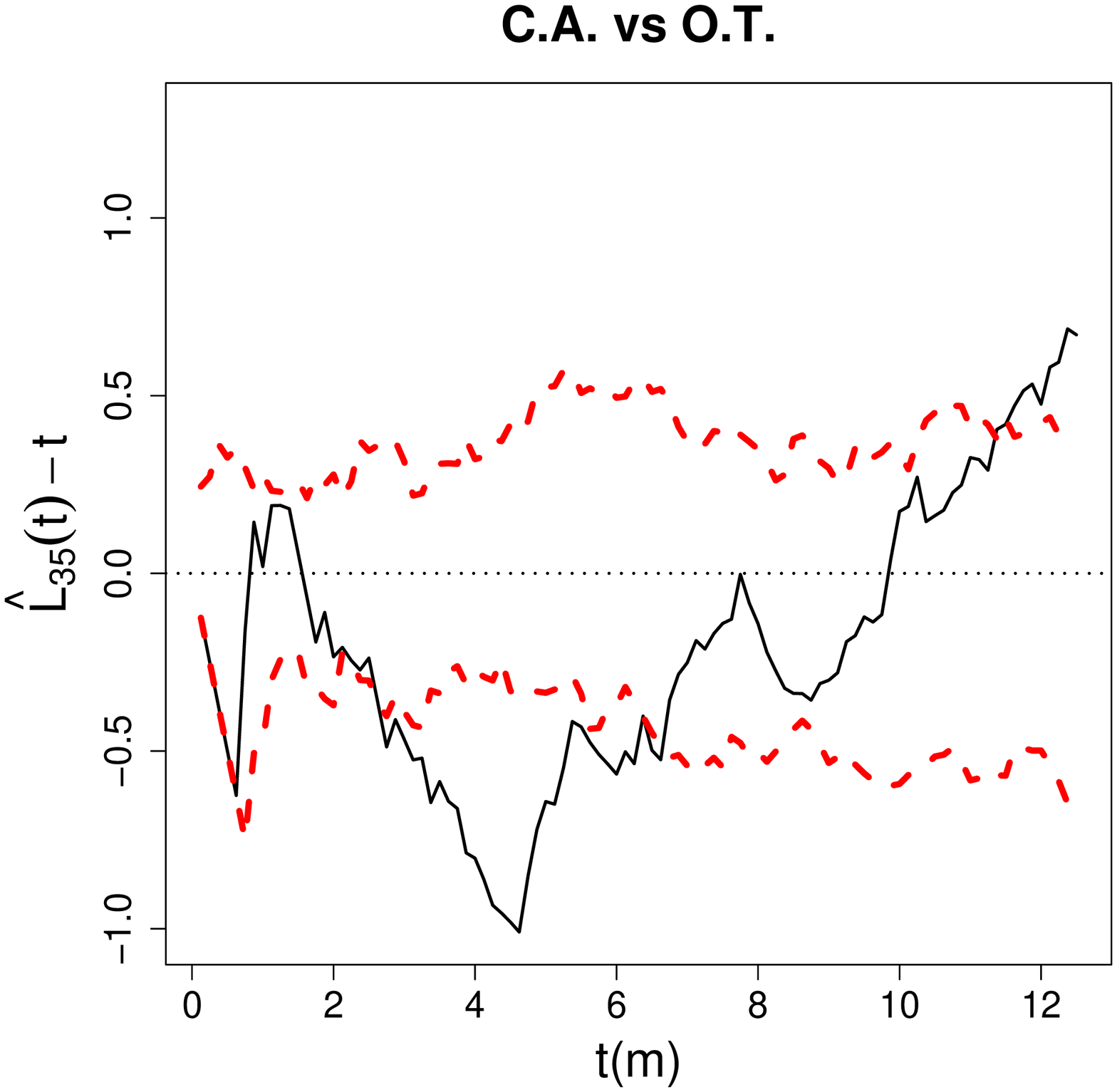} }}
\rotatebox{-90}{ \resizebox{2 in}{!}{\includegraphics{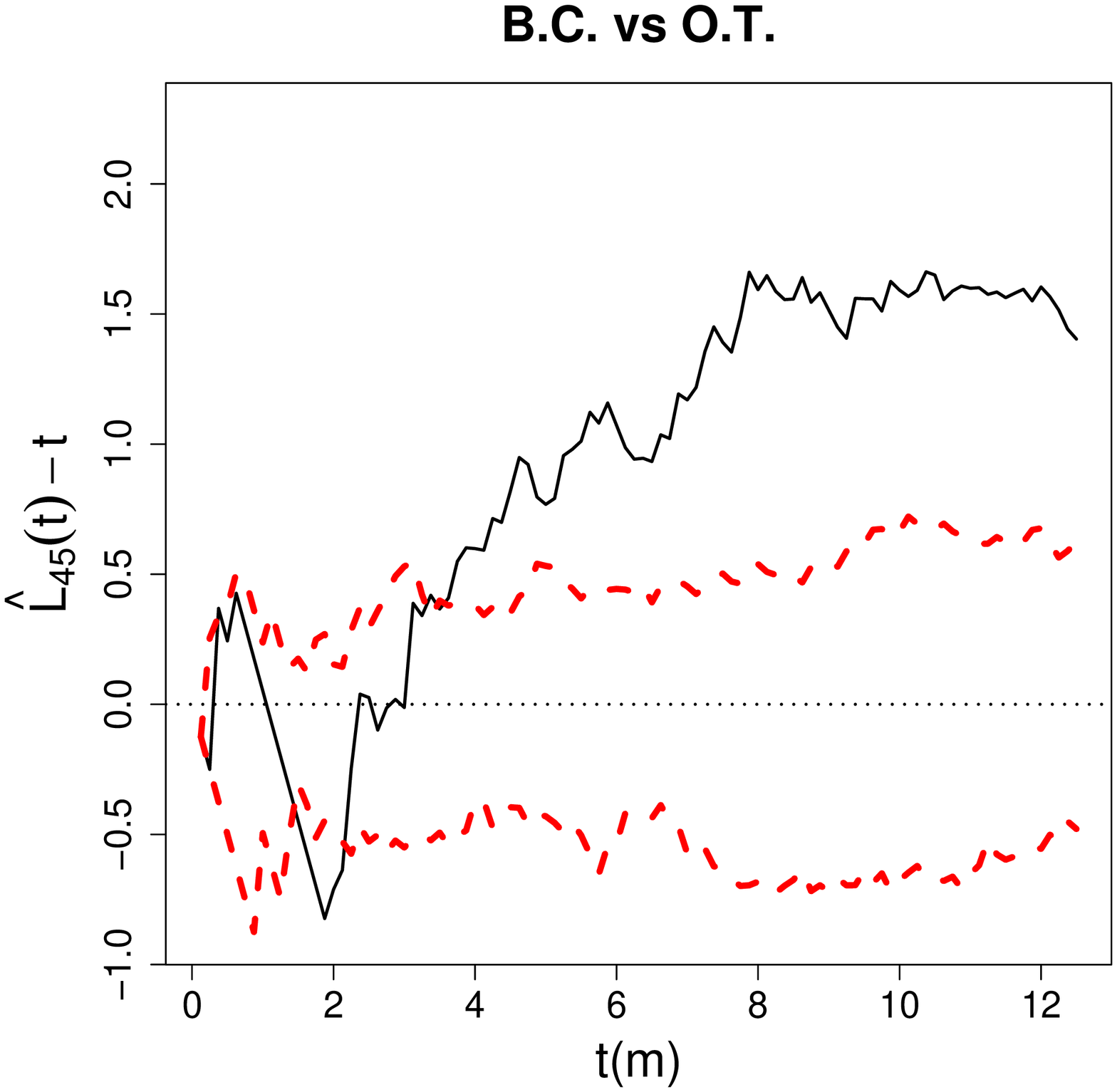} }}
\caption{
\label{fig:swamp-Lijhat}
Second-order properties of swamp tree data.
Functions plotted are Ripley's bivariate $L$-functions
$\widehat{L}_{ij}(t)-t$ for $i,j=1,2,\ldots,5$ and $i \not= j$
where $i=0$ stands for all data combined, $i=1$ for water tupelos, $i=2$ for black gums,
$i=3$ for Carolina ashes, $i=4$ for bald cypresses, and $i=5$ for other trees.
Wide dashed lines around 0 are the upper and lower 95 \% confidence bounds for the
$L$-functions based on Monte Carlo simulations under the CSR independence pattern.
W.T. = water tupelo, B.G.
= black gum, C.A. = Carolina ash, B.C.  = bald cypress, and O.T. =
other tree species.}
\end{figure}

\begin{figure}[t]
\centering
\rotatebox{-90}{ \resizebox{2 in}{!}{\includegraphics{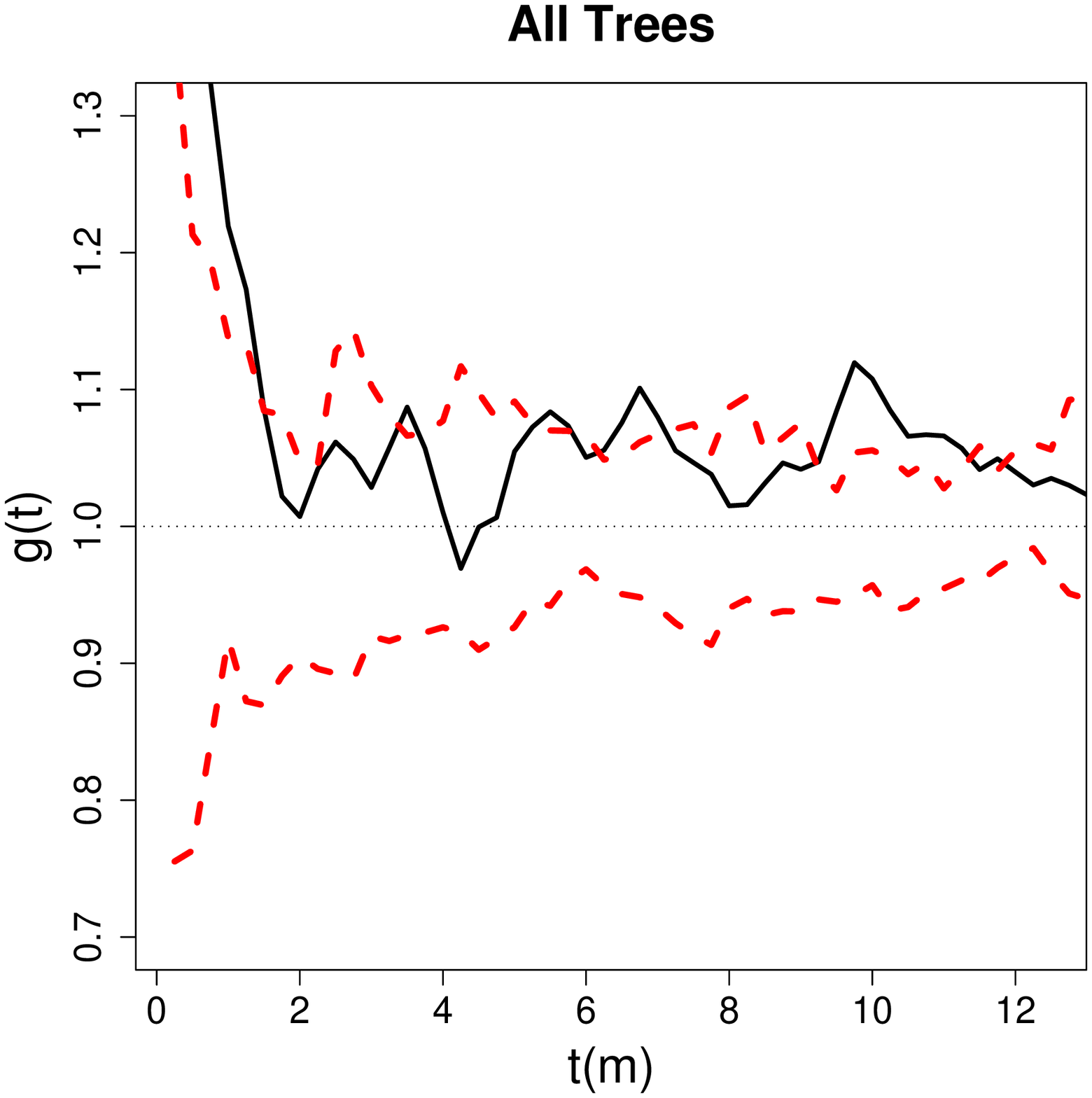} }}
\rotatebox{-90}{ \resizebox{2 in}{!}{\includegraphics{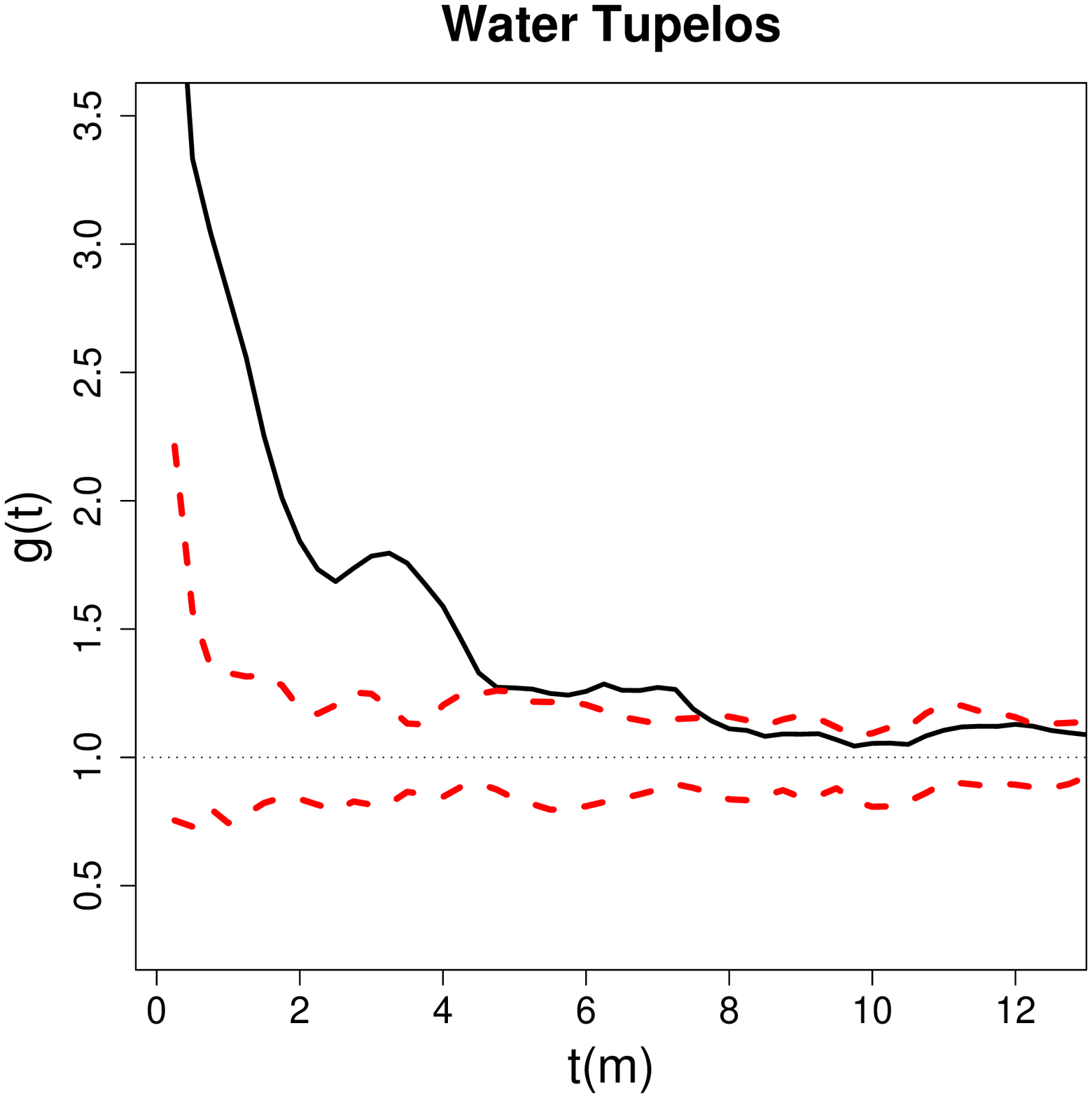} }}
\rotatebox{-90}{ \resizebox{2 in}{!}{\includegraphics{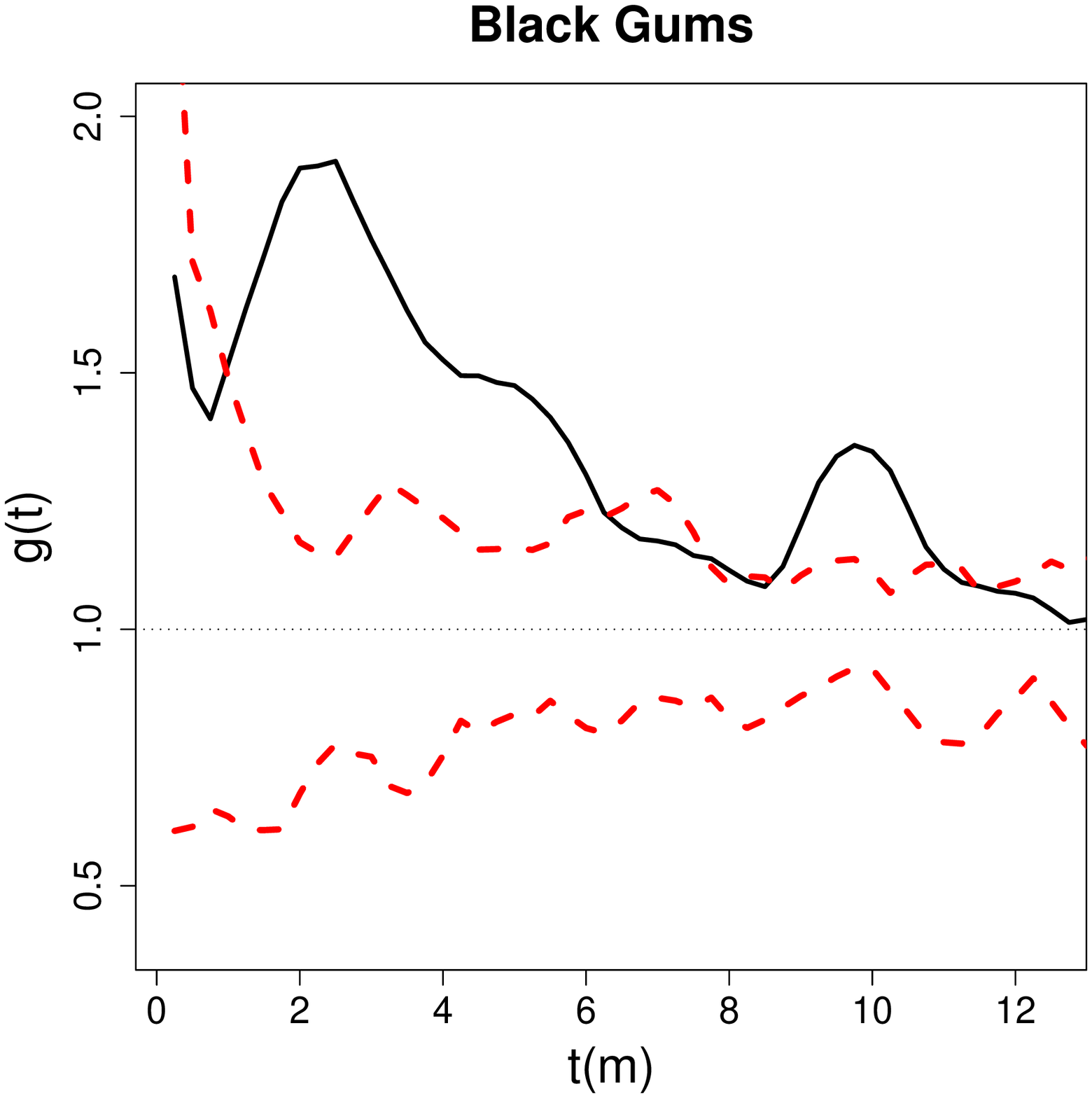} }}
\rotatebox{-90}{ \resizebox{2 in}{!}{\includegraphics{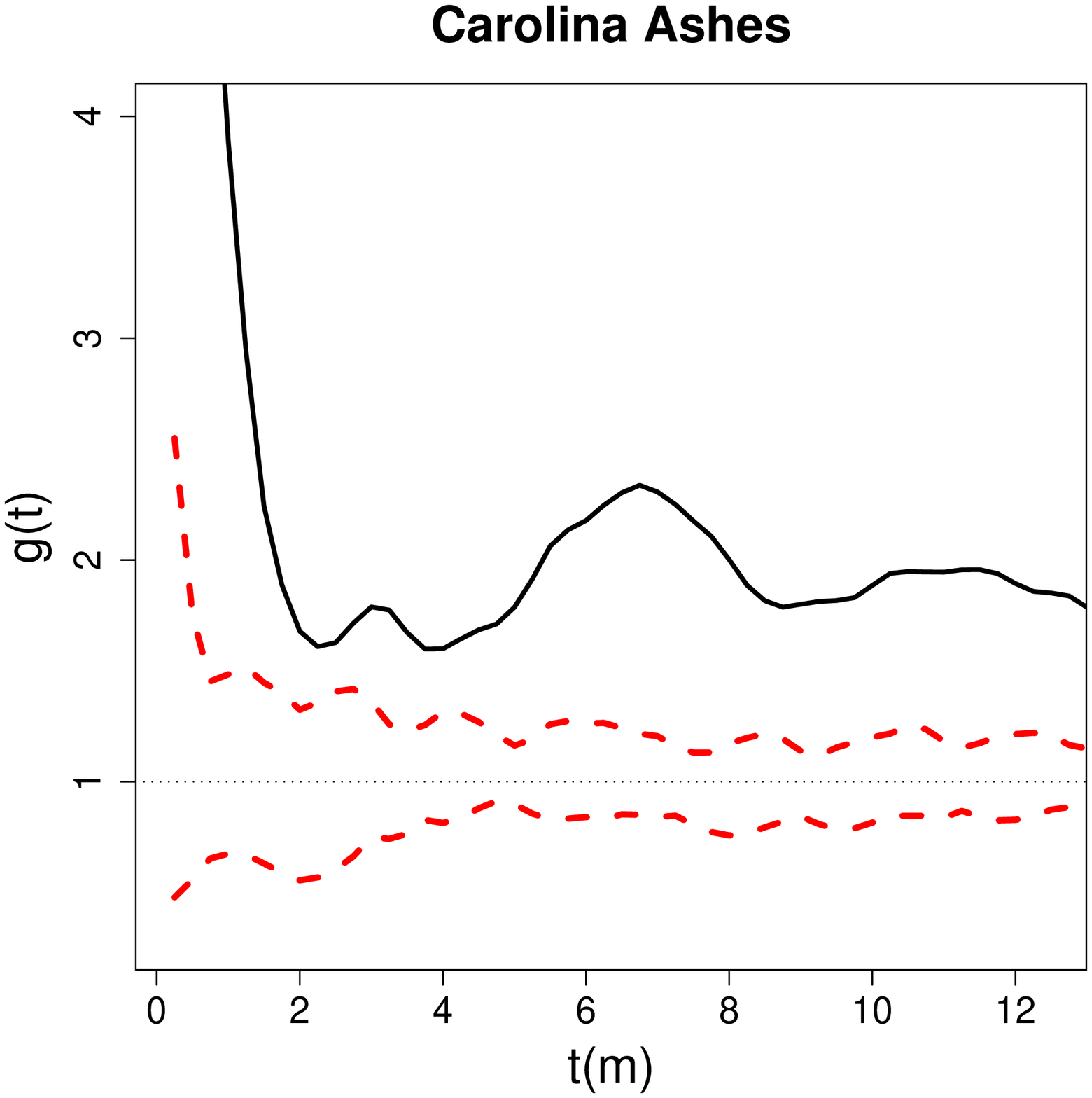} }}
\rotatebox{-90}{ \resizebox{2 in}{!}{\includegraphics{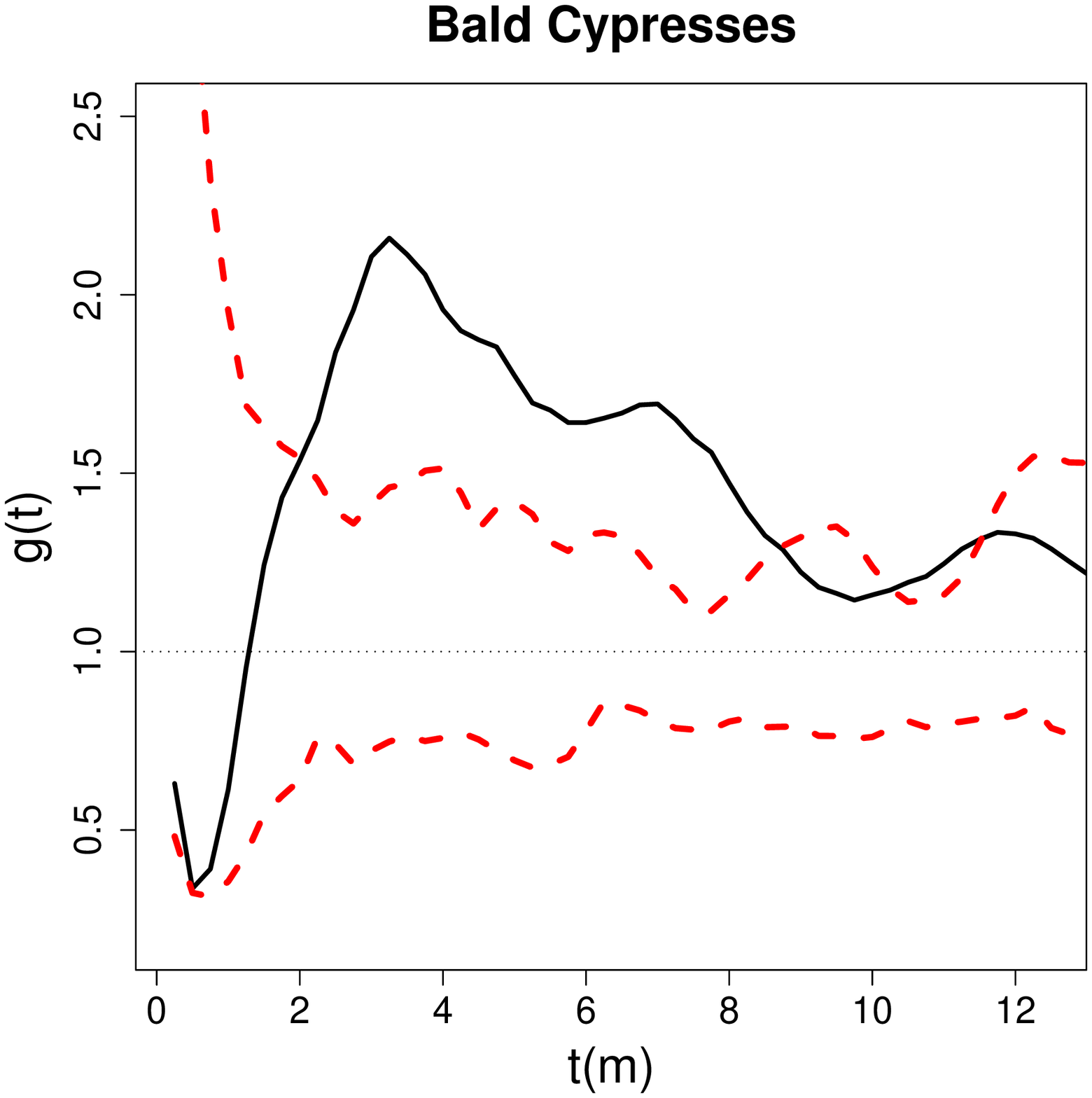} }}
\rotatebox{-90}{ \resizebox{2 in}{!}{\includegraphics{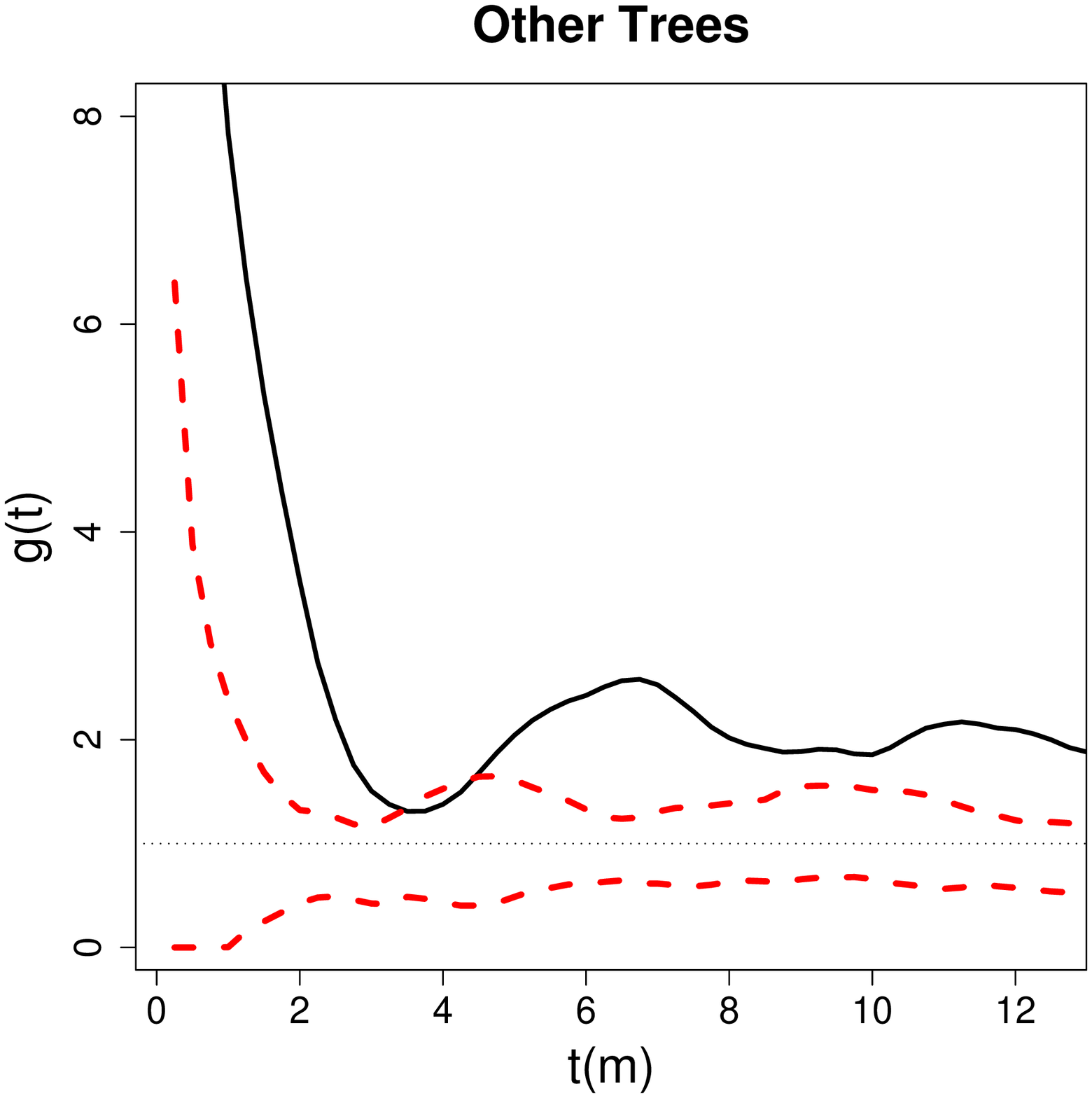} }}
\caption{
\label{fig:swamp-PCFii}
Pair correlation functions for all trees combined and for each species in the swamp tree data.
Wide dashed lines around 1 (which is the theoretical value)
are the upper and lower (pointwise) 95 \% confidence bounds for the
$L$-functions based on Monte Carlo simulation under the CSR independence pattern.}
\end{figure}

\begin{figure}[t]
\centering
\rotatebox{-90}{ \resizebox{2 in}{!}{\includegraphics{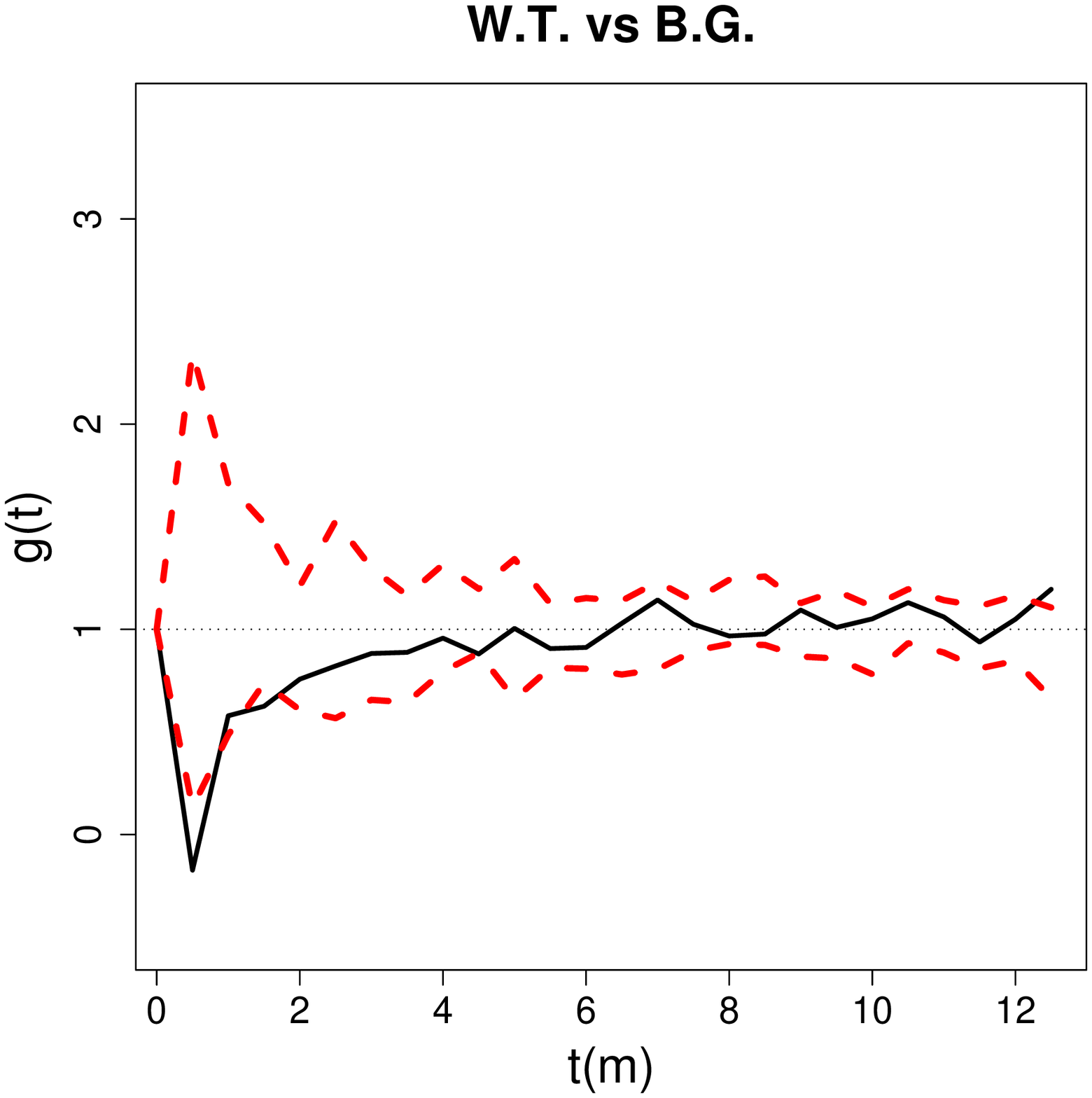} }}
\rotatebox{-90}{ \resizebox{2 in}{!}{\includegraphics{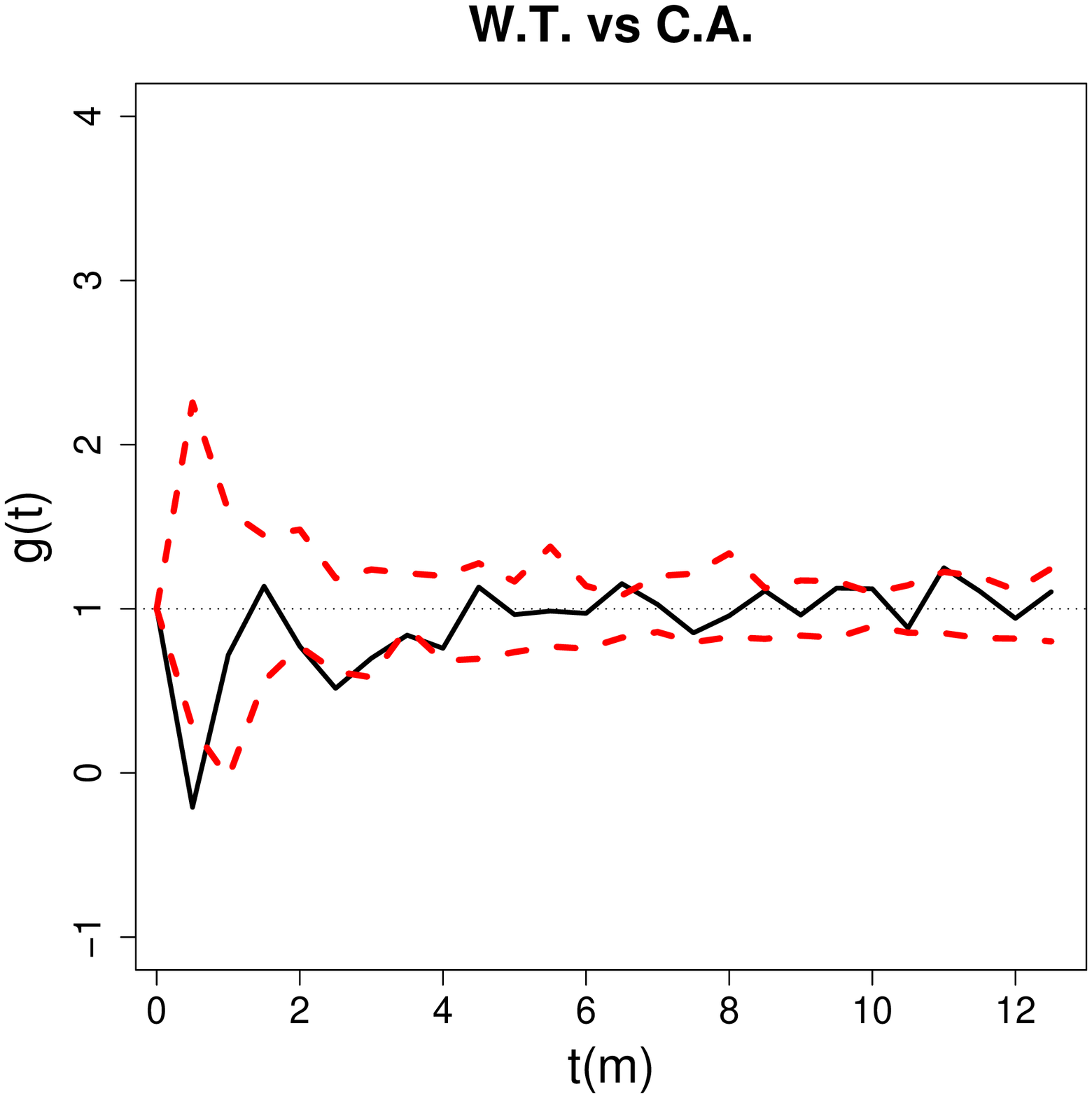} }}
\rotatebox{-90}{ \resizebox{2 in}{!}{\includegraphics{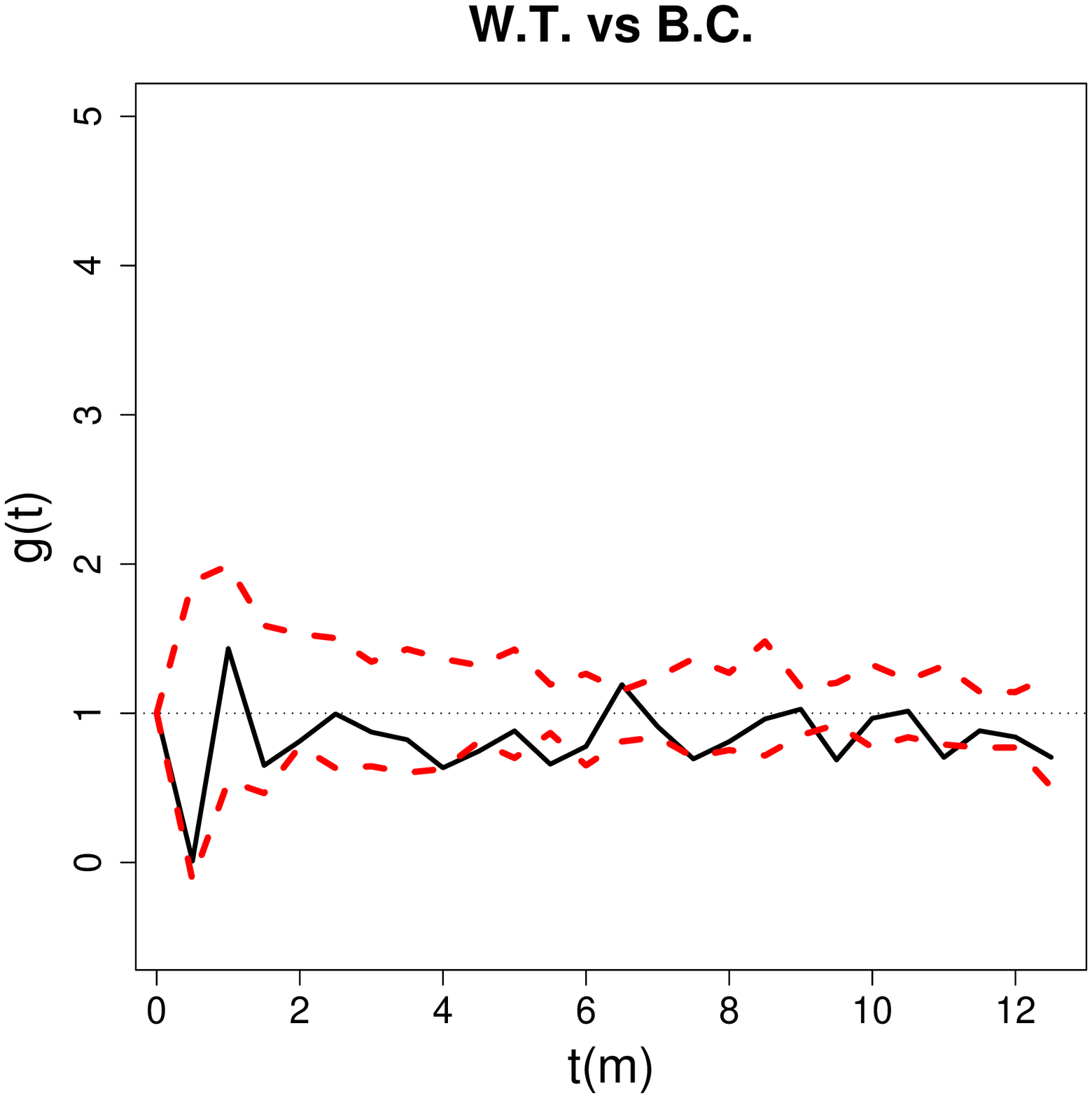} }}
\rotatebox{-90}{ \resizebox{2 in}{!}{\includegraphics{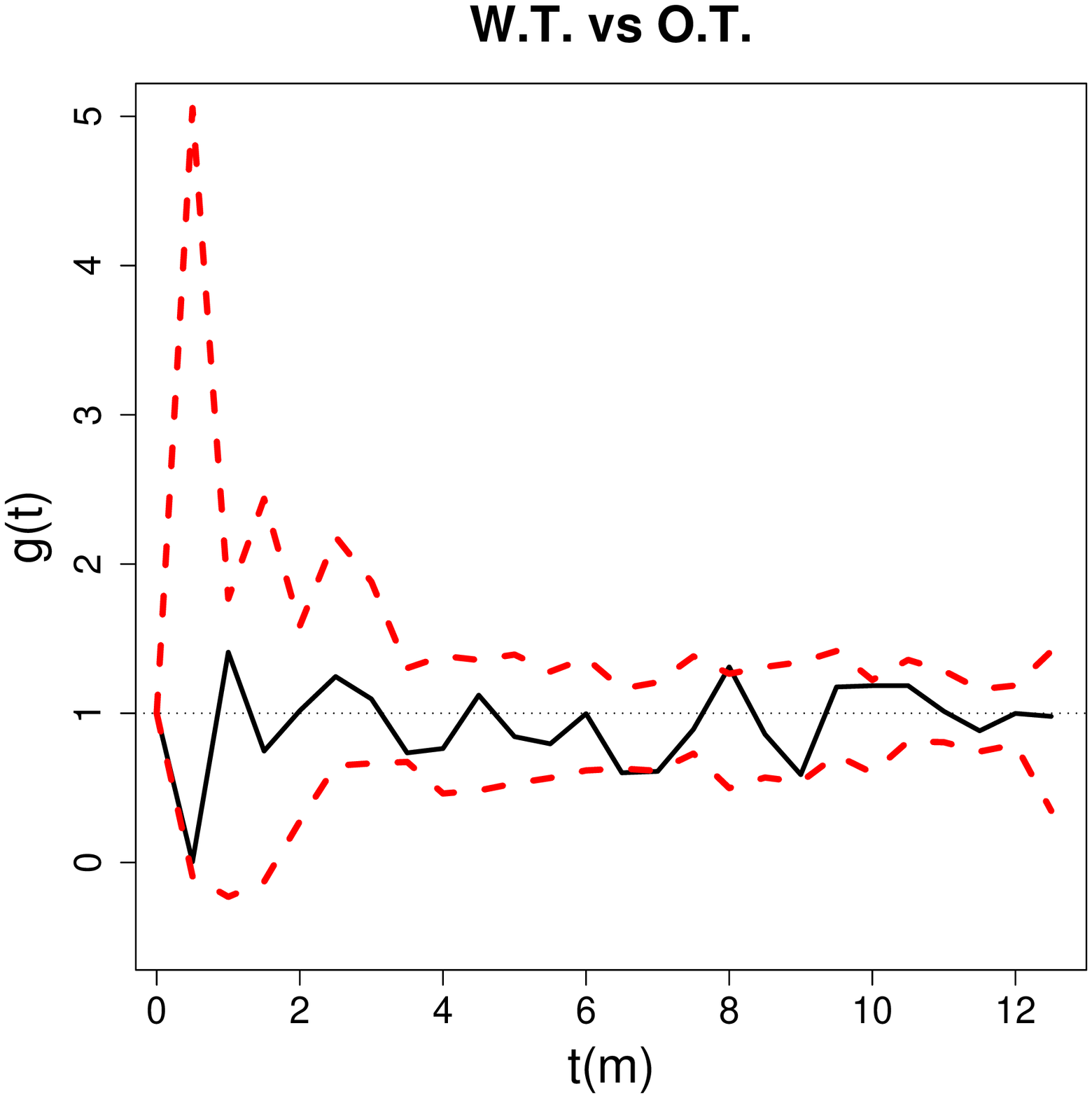} }}
\rotatebox{-90}{ \resizebox{2 in}{!}{\includegraphics{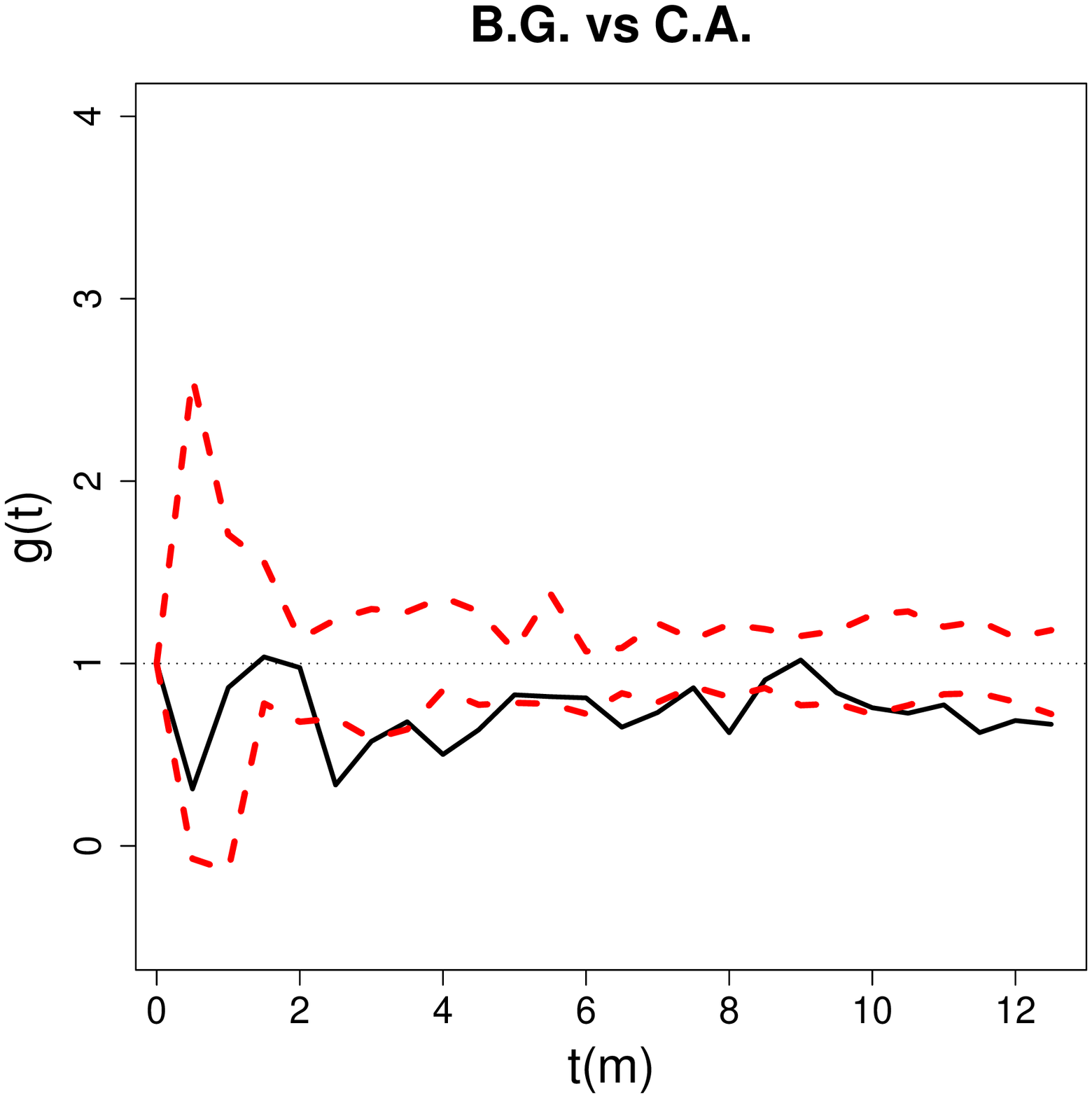} }}
\rotatebox{-90}{ \resizebox{2 in}{!}{\includegraphics{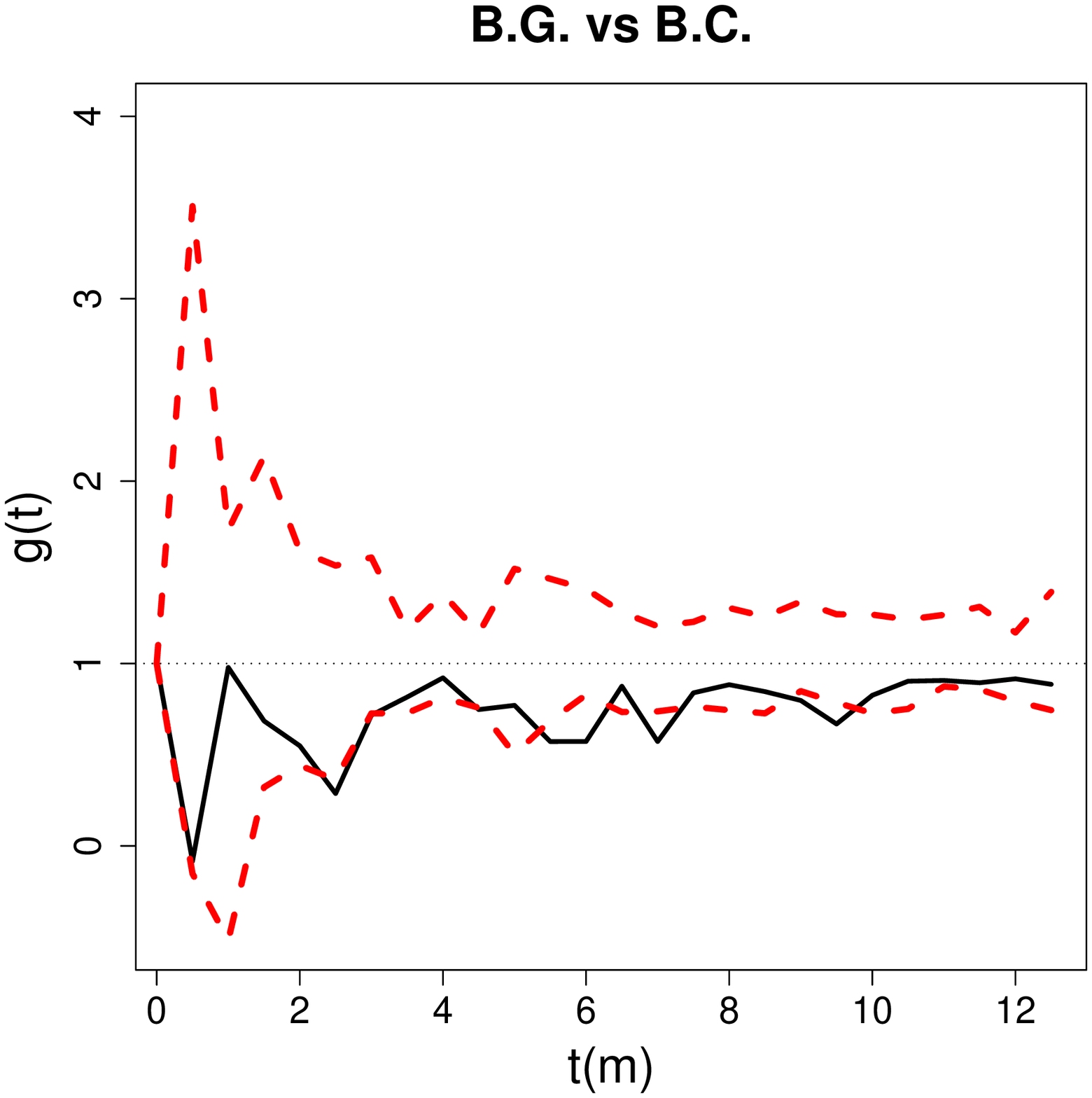} }}
\rotatebox{-90}{ \resizebox{2 in}{!}{\includegraphics{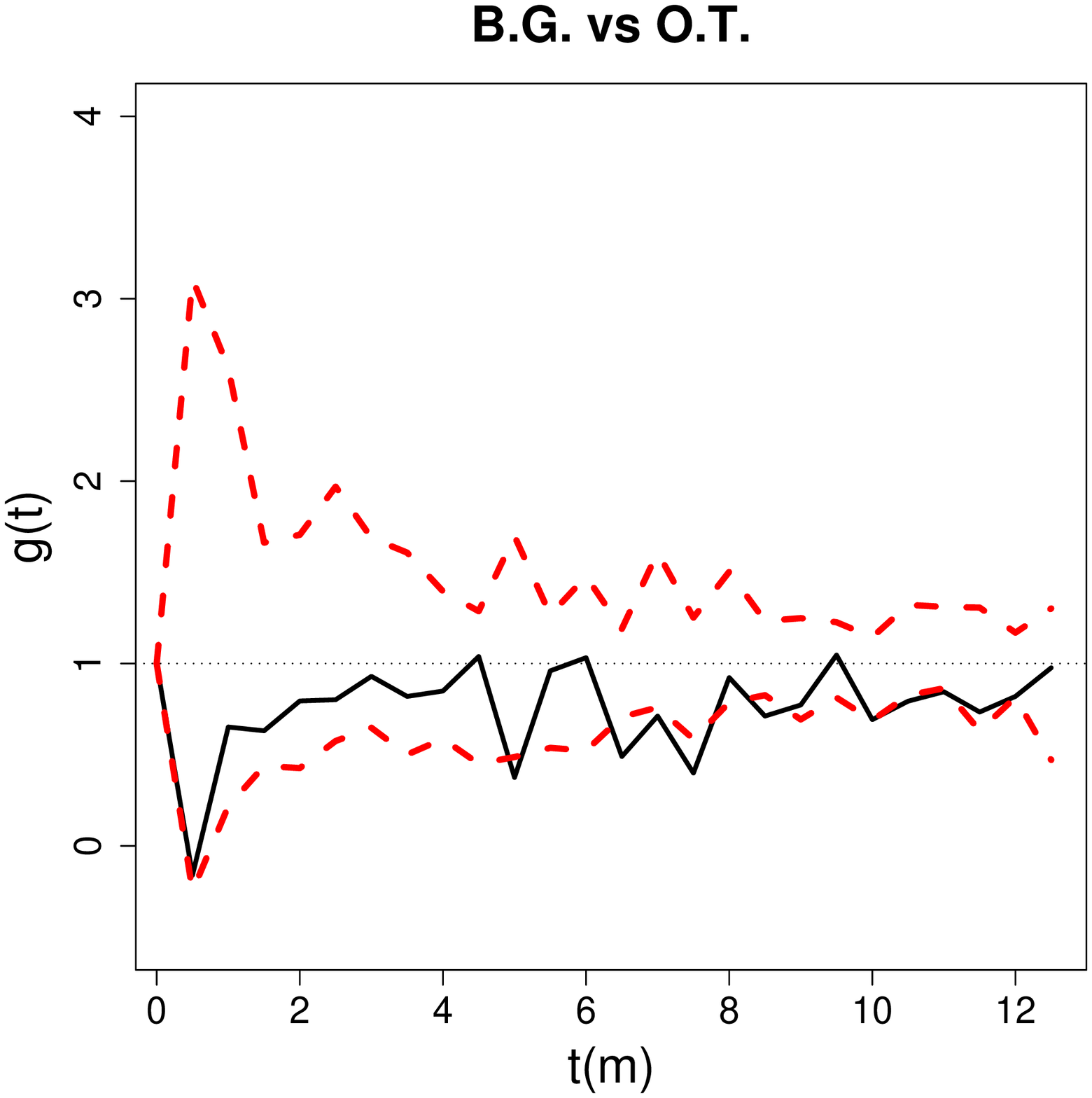} }}
\rotatebox{-90}{ \resizebox{2 in}{!}{\includegraphics{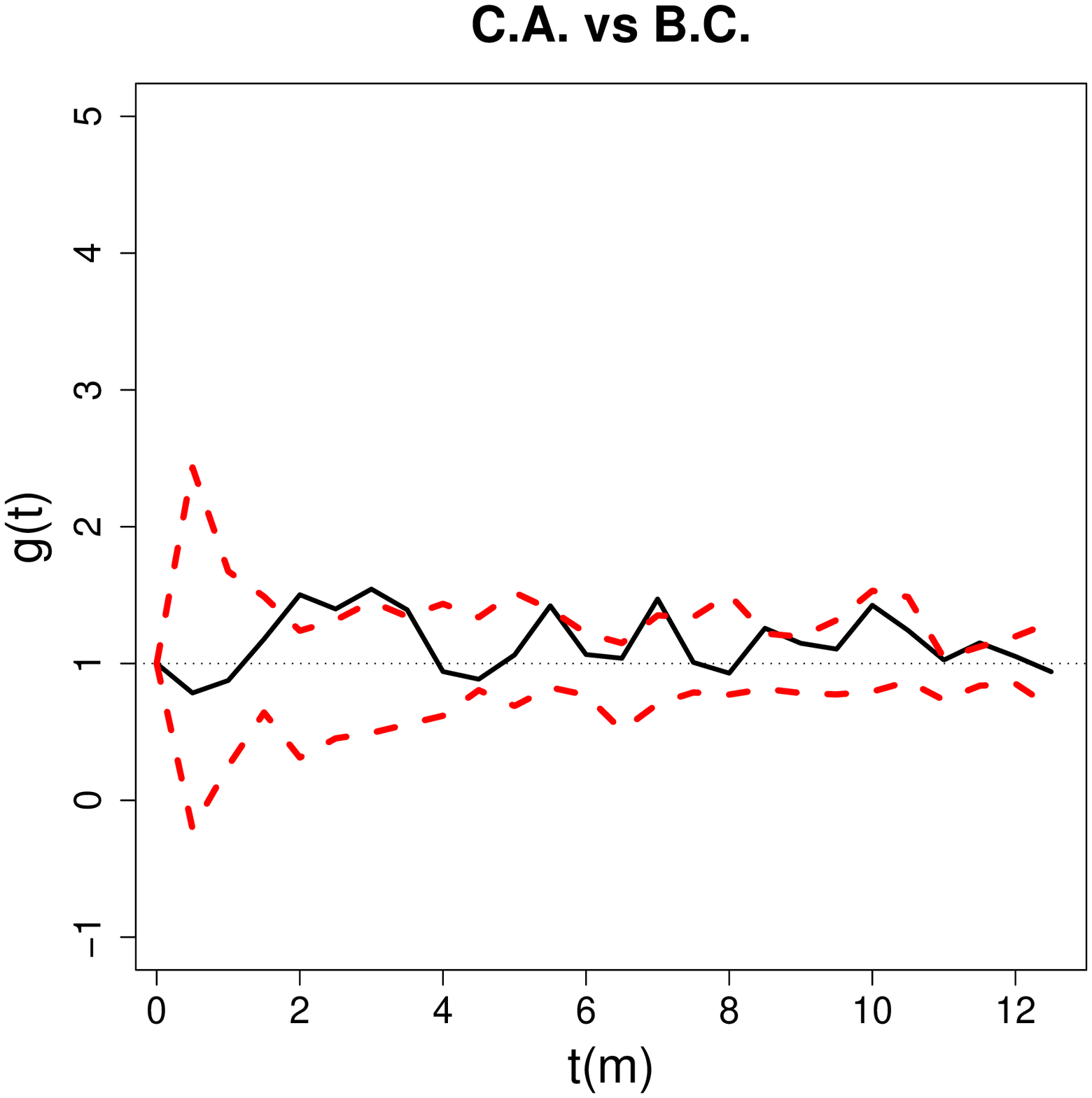} }}
\rotatebox{-90}{ \resizebox{2 in}{!}{\includegraphics{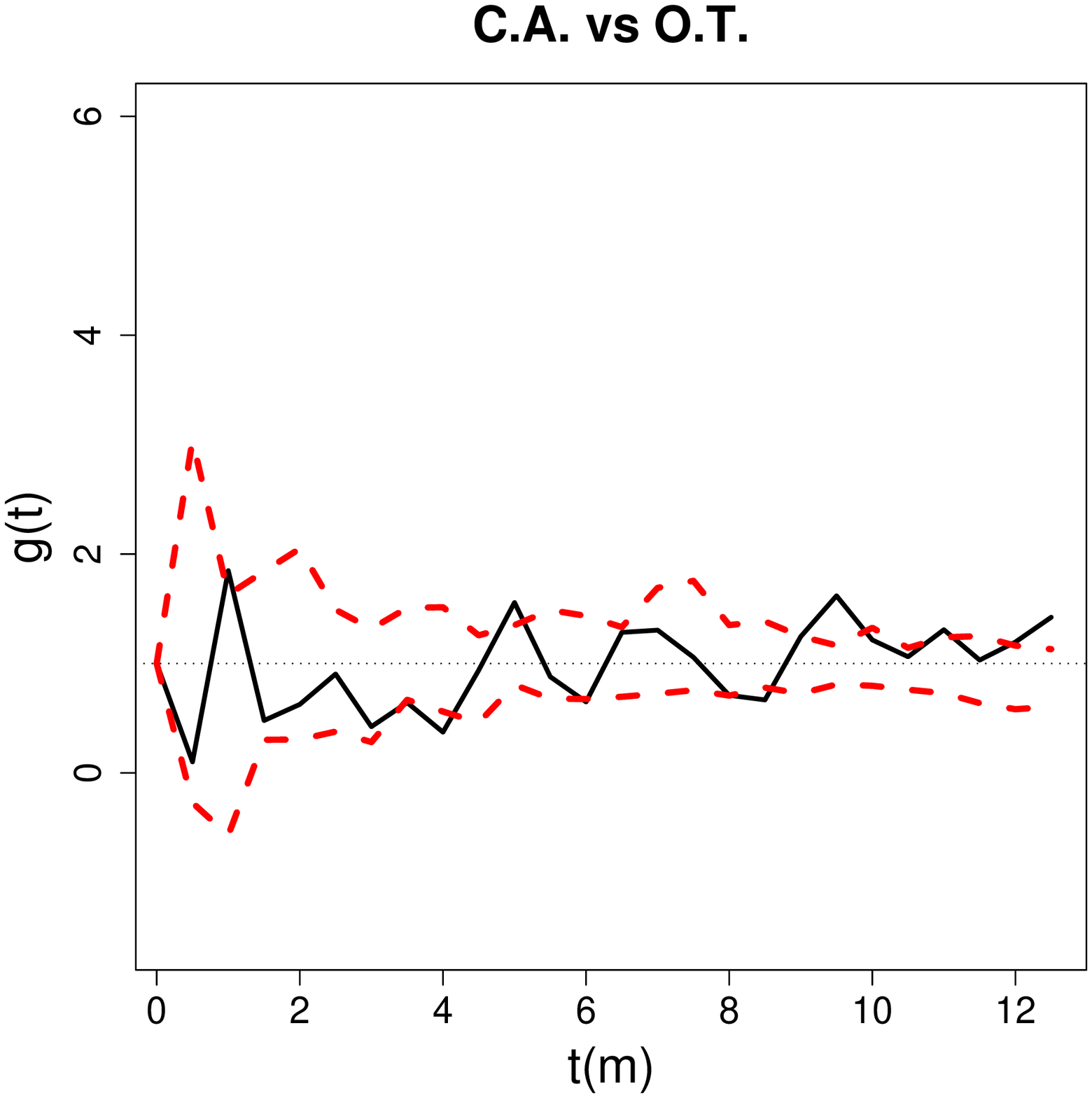} }}
\rotatebox{-90}{ \resizebox{2 in}{!}{\includegraphics{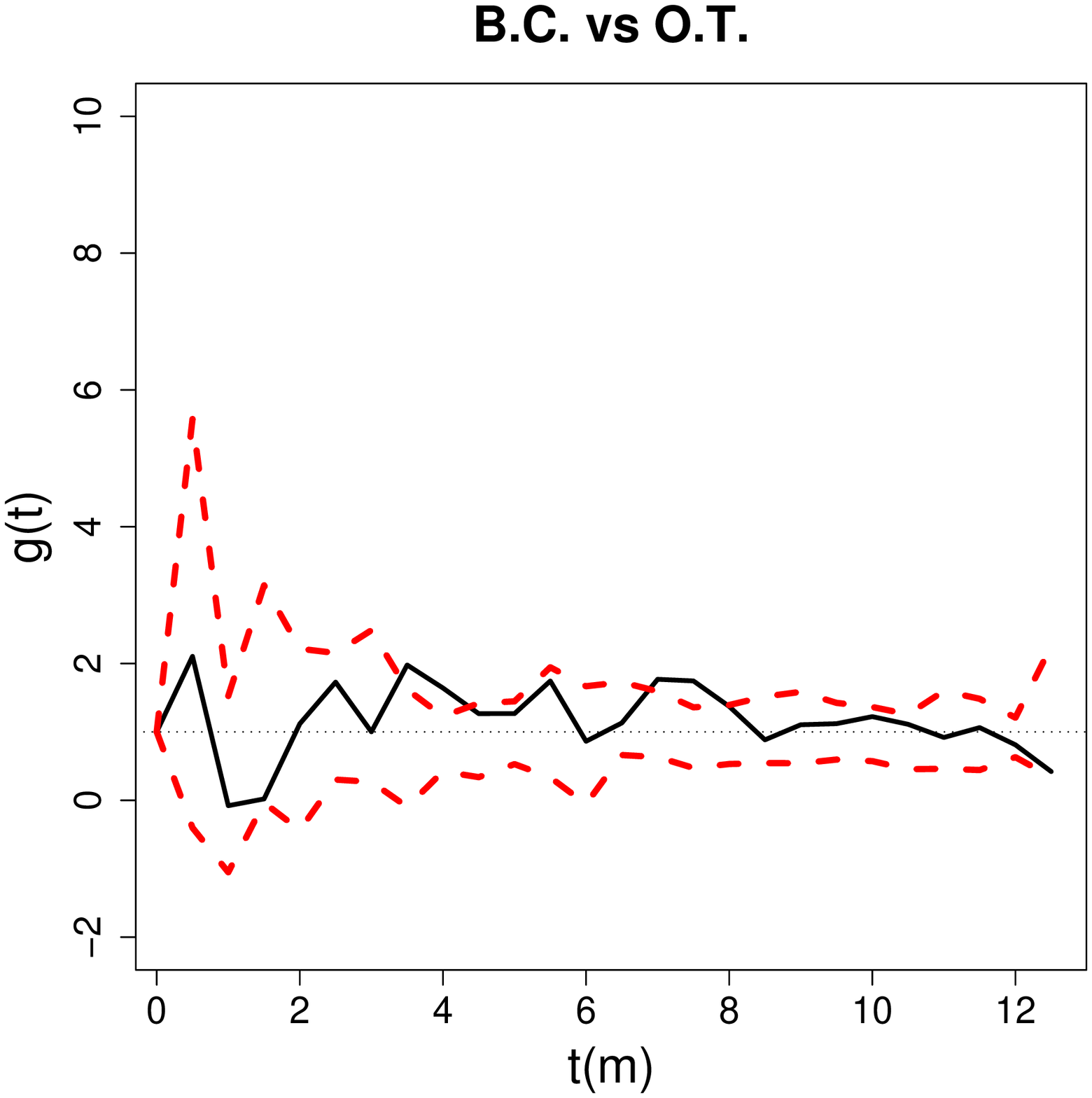} }}
\caption{
\label{fig:swamp-PCFij}
Pair correlation functions for each pair of species in the swamp tree data.
Wide dashed lines around 1 (which is the theoretical value)
are the upper and lower (pointwise) 95 \% confidence bounds for the
$L$-functions based on Monte Carlo simulations under the CSR independence pattern.
W.T. = water tupelos, B.G. = black gums, C.A. = Carolina ashes,
B.C.  = bald cypresses, and O.T. = other tree species.}
\end{figure}

Based on the NNCT-tests above, we conclude that tree species exhibit
significant deviation from the CSR independence pattern, except for bald cypresses.
Considering Figure \ref{fig:SwampTrees} and the corresponding NNCT in Table \ref{tab:NNCT-swamp},
this deviation is toward the segregation of the species.
Then, we might also be interested in the causes of the segregation
and the type and level of interaction between the tree species
at different scales (i.e., distances between the trees).
To answer such questions, we also present the second-order analysis of the swamp tree data.
We calculate Ripley's (univariate) $L$-function which is the modified version of $K$ function
as $\widehat{L}_{ii}(t)=\sqrt{\left( \widehat{K}_{ii}(t)/\pi \right)}$
where $t$ is the distance from a randomly chosen event (i.e., location of a tree),
$\widehat{K}_{ii}(t)$ is an estimator of
\begin{equation}
\label{eqn:Kii}
K(t)=\lambda^{-1}\E[\text{\# of extra events within distance $t$ of a randomly chosen event}]
\end{equation}
with $\lambda$ being the density (number per unit area) of events
and is calculated as
\begin{equation}
\label{eqn:Kiihat}
\widehat{K}_{ii}(t)=\widehat{\lambda}^{-1}\sum_i\sum_{j \not= i}w(l_i,l_j)\I(d_{ij}<t)/N
\end{equation}
where $\widehat{\lambda}=N/A$ is an estimate of density ($N$ is the observed number of points
and $A$ is the area of the study region),
$d_{ij}$ is the distance between points $i$ and $j$,
$\I(\cdot)$ is the indicator function,
$w(l_i,l_j)$ is the proportion of the circumference of the circle
centered at $l_i$ with radius $d_{ij}$ that falls in the study area,
which corrects for the boundary effects.
Under CSR, $L(t)-t=0$ holds.
If the univariate pattern exhibits aggregation,
then $L(t)-t$ tends to be positive,
if it exhibits regularity then $L(t)-t$ tends to be negative.
The estimator $\widehat{K}(t)$ is approximately unbiased for $K(t)$ at each fixed $t$.
Bias depends on the geometry of the study area and increases with $t$.
For a rectangular region
it is recommended to use $t$ values up to 1/4 of the smaller side
length of the rectangle.
See (\cite{diggle:2003}) for more detail.
So we take the values $t \in [0,12.5]$ in our analysis,
since the smaller side of the rectangular region of swamp tree data is 50 m.
In Figure \ref{fig:swamp-Liihat},
we present the plots of $\widehat{L}_{ii}(t)-t$ functions for each species
as well as the plot of all trees combined.
We also present the upper and lower (pointwise)
95 \% confidence bounds for each $L_{ii}(t)-t$.
Observe that for all trees combined there is significant aggregation of trees
(the $L_{00}(t)-t$ curve is above the upper confidence bound) at all scales (i.e., distances).
Water tupelos exhibit significant aggregation for the range of the plotted distances;
black gums exhibit significant aggregation for distances $t>1$ m;
Carolina ashes exhibit significant aggregation for the range of plotted distances;
bald cypresses exhibit no deviation from CSR independence for $t \lesssim 5$ m,
then they exhibit significant spatial aggregation for $t>4$ m;
other trees exhibit significant aggregation for the range of plotted distances.
Hence, segregation of the species might be due to different
levels and types of aggregation of the species in the study region.

We also calculate Ripley's bivariate $L$-function
as $\widehat{L}_{ij}(t)=\sqrt{\left( \widehat{K}_{ij}(t)/\pi \right)}$
where $\widehat{K}_{ij}(t)$ is an estimator of
$$K_{ij}(t)=\lambda_j^{-1}\E[\text{\# of extra type $j$ events within distance $t$ of a randomly chosen type $i$ event}]$$
with $\lambda_j$ being the density of type $j$ events
and is calculated as
\begin{equation}
\label{eqn:Kijhat}
\widehat{K}_{ij}(t)=\left( \widehat{\lambda}_i\widehat{\lambda}_j A \right)^{-1}\sum_i\sum_{j}w(i_k,j_l)\I(d_{i_k,j_l}<t)
\end{equation}
where $d_{i_k,j_l}$ is the distance between $k^{th}$ type $i$ and $l^{th}$ type $j$ points,
$w(i_k,j_l)$ is the proportion of the circumference of the circle
centered at $k^{th}$ type $i$ point with radius $d_{i_k,j_l}$ that falls in the study area,
which is used for edge correction.
Under CSR independence, $L_{ij}(t)-t=0$ holds.
If the bivariate pattern is segregation,
then $L_{ij}(t)-t$ tends to be negative,
if it is association then $L_{ij}(t)-t$ tends to be positive.
See (\cite{diggle:2003}) for more detail.
In Figure \ref{fig:swamp-Lijhat},
we present the bivariate plots of $\widehat{L}_{ij}(t)-t$ functions
together with the upper and lower (pointwise) 95 \% confidence bounds
for each pair of species (due to the symmetry of $L_{ij}(t)$
there are only 10 different pairs).
Observe that for distances up to $t \approx 10$ m,
water tupelos and black gums exhibit significant segregation
($\widehat{L}_{12}(t)-t$ is below the lower confidence bound)
and for the rest of the plotted distances their interaction is not significantly
different from the CSR independence pattern;
water tupelos and Carolina ashes are significantly segregated up to about $t \approx 10$ m;
water tupelos and bald cypresses do not have significant deviation
from the CSR independence pattern for distances up to 4 m,
for larger distances they exhibit significant segregation;
water tupelos and the other trees do not deviate from CSR independence
for the range of the plotted distances.
Black gums and Carolina ashes are significantly segregated for $t>2$ m;
black gums and bald cypresses are significantly segregated for  $t>2$ m;
black gum and other trees are significantly segregated for all the distances plotted.
Carolina ashes and bald cypresses are significantly associated for distances
larger than 3 m;
and Carolina ashes and the other trees exhibit significant segregation
for $3< t < 7$ m and for $t>11$ m they exhibit significant association.
On the other hand, bald cypresses and other trees are significantly associated
for distance larger than 4 m.

But Ripley's $K$-function is cumulative,
so interpreting the spatial interaction at larger distances
is problematic (\cite{wiegand:2007}).
The (accumulative) pair correlation function $g(t)$
is better for this purpose (\cite{stoyan:1994}).
The pair correlation function of a (univariate)
stationary point process is defined as
$$g(t) = \frac{K'(t)}{2\,\pi\,t}$$
where $K'(t)$ is the derivative of $K(t)$.
For a univariate stationary Poisson process, $g(t)=1$;
values of $g(t) < 1$ suggest inhibition (or regularity) between points;
and values of $g(t) > 1$ suggest clustering (or aggregation).
The pair correlation functions for all trees
and each species for the swamp tree data are plotted
in Figure \ref{fig:swamp-PCFii}.
Observe that all trees are aggregated around
distance values of 0-1,3,4,5,7,9-10 m;
water tupelos are aggregated for distance values of 0-4 and 5-7 m;
black gums are aggregated for distance values of 1-6 and 8-11 m;
Carolina ashes are aggregated for all the range of the plotted distances;
bald cypresses are aggregated for distance values of 2-8 and around 11 m;
and other trees are aggregated for all distance values except 3-5 m.
Comparing Figures \ref{fig:swamp-Liihat} and \ref{fig:swamp-PCFii},
we see that Ripley's $L$ and pair correlation functions
detect the same patterns but with different distance values.
That is, Ripley's $L$ implies that the particular pattern is significant
for a wider range of distance values compared to $g(t)$,
since Ripley's $L$ is cumulative, so the values of $L$ at
small scales confound the values of $L$ at larger scales (\cite{loosmore:2006}).
Hence the results based on pair correlation function $g(t)$
are more reliable.

The same definition of the pair correlation function
can be applied to Ripley's bivariate $K$ or $L$-functions as well.
The benchmark value of $K_{ij}(t) = \pi \, t^2$ corresponds to $g(t) = 1$;
$g(t) < 1$ suggests segregation of the classes;
and $g(t) > 1$ suggests association of the classes.
The bivariate pair correlation functions for the
species in swamp tree data are plotted in Figure \ref{fig:swamp-PCFij}.
Observe that
water tupelos and black gums are segregated for distance values of 0-1 m;
water tupelos and Carolina ashes are segregated for values of 0-1 and 2.5 m
and are associated for values about 6 m;
water tupelos and bald cypresses are segregated for 0-1, 5.5, 9.5, and 11 m
and are associated for 6.5 m;
water tupelos and other trees are segregated for 0-0.5 and 7 m
and are associated for 8 m;
black gums and Carolina ashes are segregated for 2-2.5, 3.5-4.5, 6-8.5, and 9.5-12 m;
black gums and bald cypresses are segregated for 3.5, 5.5-6.5,7, and 9.5 m;
black gums and other trees are segregated for 5 and 6-7.5 m;
Carolina ashes and bald cypresses are associated for 1.5-3, 5.5., and 7 m;
Carolina ashes and other trees are associated for 5 and 9-10 m;
and bald cypresses and other trees are segregated for 4 m
and are associated for 3-4 and 6.5-7.5 m.

However the pair correlation function estimates might have critical behavior
for small $t$ if $g(t)>0$,
since the estimator variance and hence the bias are considerably large.
This problem gets worse especially in cluster processes (\cite{stoyan:1996}).
See for example Figures \ref{fig:swamp-PCFii} and \ref{fig:swamp-PCFij}
where the confidence bands for smaller $t$ values are much wider compared
to those for larger $t$ values.
So pair correlation function analysis is more reliable for larger distances
and it is safer to use $g(t)$ for distances larger than the average NN distance in the data set.
Comparing Figure \ref{fig:swamp-Liihat} with Figure \ref{fig:swamp-PCFii}
and Figure \ref{fig:swamp-Lijhat} with Figure \ref{fig:swamp-PCFij}
we see that Ripley's $L$ and pair correlation functions
usually detect the same large-scale pattern but at different ranges of distance values.
Ripley's $L$ suggests that the particular pattern is significant
for a wider range of distance values compared to $g(t)$,
since 
values of $L$ at small scales confound the values of $L$
at larger scales where $g(t)$ is more reliable to use (\cite{loosmore:2006}).

While second order analysis (using Ripley's $K$ and $L$-functions or pair correlation function)
provides information on the univariate and bivariate patterns
at all scales (i.e., for all distances),
NNCT-tests summarize the spatial interaction for the smaller scales
(for distances about the average NN distance in the data set).
In particular, for the swamp tree data average NN distance ($\pm$ standard deviation)
is about 1.8 ($\pm$ 1.04) meters and notice that Ripley's $L$-function and
NNCT-tests yield similar results for distances about 2 meters.
It is interesting to observe that both methods indicate that
bald cypresses have a very different pattern compared to the others.

\subsection{Leukemia Data}
\label{sec:leukemia-data}
\cite{cuzick:1990} considered the spatial locations of 62 cases of childhood
leukemia in the North Humberside region of the UK between the years 1974 to 1982 (inclusive).
A sample of 143 controls are selected using the completely randomized design
from the same region.
We analyze the spatial distribution of leukemia cases and controls
in this data using a $2 \times 2$ NNCT.
We plot the locations of these points in the study region in
Figure \ref{fig:leukemia} and provide the corresponding NNCT
together with percentages based on row and column sums in Table \ref{tab:NNCT-leukemia}.
Observe that the percentages in the diagonal cells are about the same as the
marginal (row or column) percentages of the subjects in the study,
which might be interpreted as the lack of any deviation from the null case for both classes.
Figure \ref{fig:SwampTrees} is also supportive of this observation.

\begin{figure}[t]
\centering
\rotatebox{-90}{ \resizebox{3. in}{!}{\includegraphics{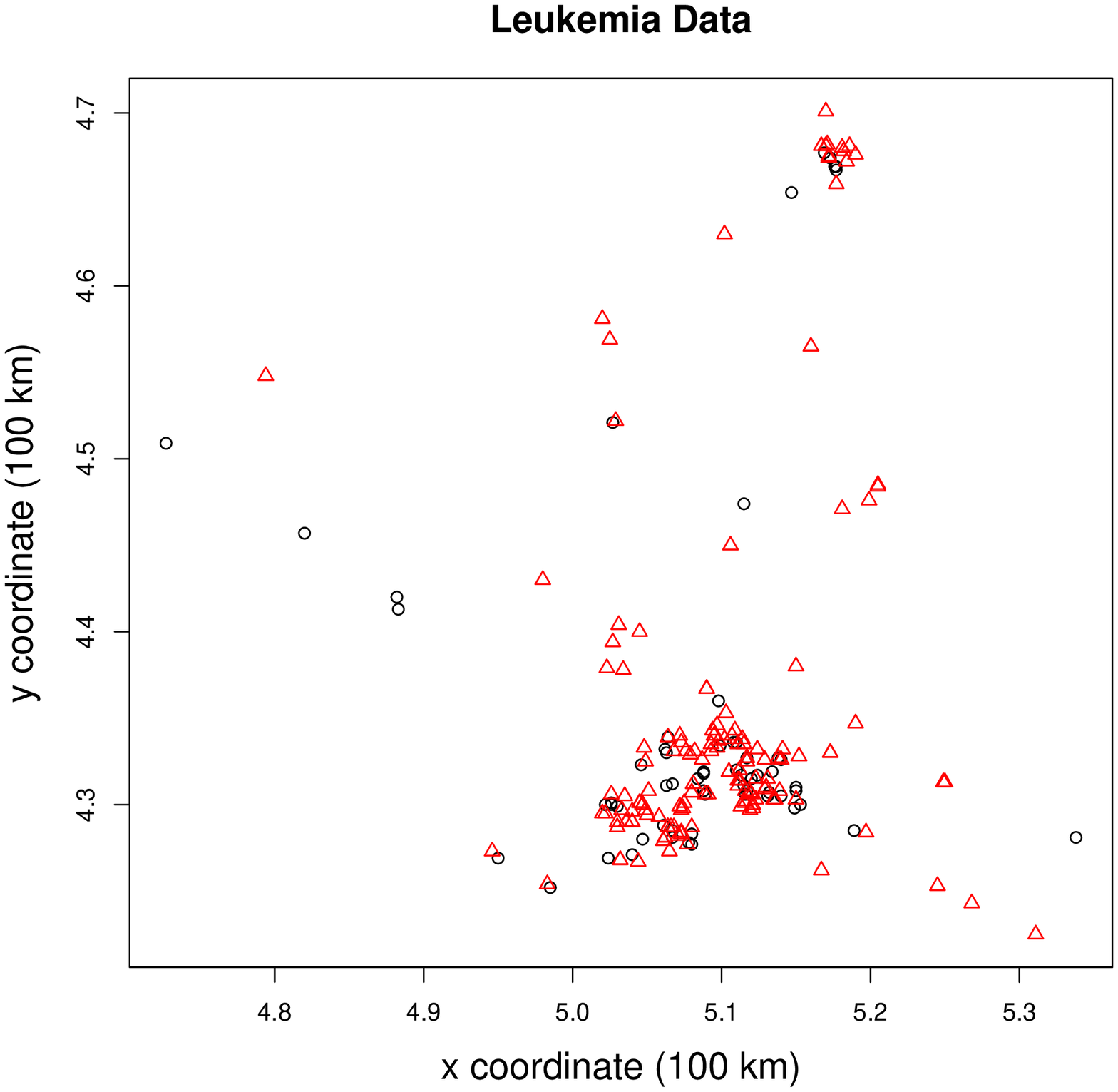} }}
 \caption{
\label{fig:leukemia}
The scatter plot of the locations of controls (circles $\circ$) and childhood leukemia cases
(triangles $\triangle$) in North Humberside, UK.}
\end{figure}

\begin{table}[]
\centering
\begin{tabular}{c}

\begin{tabular}{cc|cc|c}
\multicolumn{2}{c}{}& \multicolumn{2}{c}{NN}& \\
\multicolumn{2}{c}{}&    case &  control   &   sum  \\
\hline
& case &    25  &   41    &   66  \\
\raisebox{1.5ex}[0pt]{base}
& control &    39  &   113    &  152  \\
\hline
&sum  &    64  &   154    &  218  \\
\end{tabular}
\hspace{.25 cm}
\begin{tabular}{c|cc|c}
& \multicolumn{2}{c}{NN}& \\
&    case &  control   & sum \\
\hline
 case &  38 \%  &  62 \%    &   30 \%  \\
 control &  26 \%  &  74 \%    &   70 \%  \\
\hline
  &    &      &    \\
\end{tabular}
\hspace{.25 cm}
\begin{tabular}{c|cc|c}
& \multicolumn{2}{c}{NN}& \\
&   case &  control  &  \\
\hline
 case &  39 \%  &  27 \%    &    \\
 control &  61 \%  &  73 \%    &    \\
\hline
sum &  29 \%  &  71 \%    &   \\
\end{tabular}

\end{tabular}
\caption{\label{tab:NNCT-leukemia}
The NNCT for the leukemia data (left) and the corresponding percentages (right).}
\end{table}

\begin{table}[]
\centering
\begin{tabular}{|cc||c|c|c|c|}
\hline
\multicolumn{6}{|c|}{Test statistics and $p$-values for segregation tests} \\
\hline
&  & $\chi^2$-statistic & $p_{asy}$ & $p_{mc}$ & $p_{rand}$ \\
\hline
\multicolumn{2}{|c||}{overall} & 2.25 & .3249 & .3303 & .3519 \\
\hline
 & case    & 1.44 & .2293 & .2266 & .2389    \\
\raisebox{1.5ex}[0pt]{base}
     & control & 1.65 & .1995 & .2041 & .2237    \\
\hline
  & case    & 2.25 & .3249 & .3303 & .3519    \\
\raisebox{1.5ex}[0pt]{NN}
     & control & 2.25 & .3249 & .3303 & .3519   \\
\hline
\end{tabular}
\caption{
\label{tab:pval-leukemia-class}
The  $\chi^2$-statistics for the segregation tests
and the corresponding $p$-values.
}
\end{table}

We can assume that some processes affect a posteriori
the population of North Humberside region
so that some of the individuals get to be cases,
while others continue to be healthy (i.e., they are controls).
So the appropriate null hypothesis is the RL pattern.
We calculate $Q=152$ and $R=142$ for this data set.
In Table \ref{tab:pval-leukemia-class},
we present the overall test of segregation,
class-specific test statistics and the associated $p$-values
(i.e., $p$-values based on the asymptotic approximation,
Monte Carlo simulation, and Monte Carlo randomization methods).
Observe that $p_{asy}$, $p_{mc}$, and $p_{rand}$ are similar for each test
and none of the tests yields a significant result.

\clearpage
\begin{figure}[ht]
\centering
\rotatebox{-90}{ \resizebox{3 in}{!}{\includegraphics{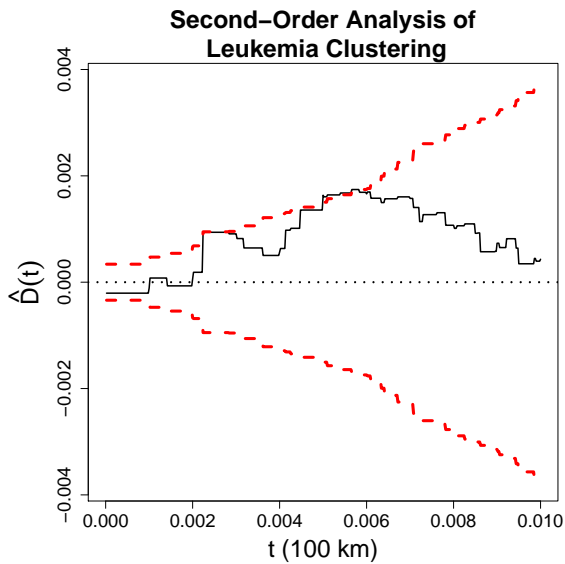} }}
 \caption{
\label{fig:leukemia-Dhat}
Second-order analysis of North Humberside childhood leukemia data:
Function plotted is Diggle's modified bivariate $K$-function
$\widehat{D}(t)=\widehat{K}_{11}(t)-\widehat{K}_{22}(t)$ with
$i=1$ for controls and $i=2$ for leukemia cases.
Wide dashed lines around 0 are plus and minus two standard errors
of $\widehat{D}(t)$ under RL of cases and controls.}
\end{figure}

Based on the NNCT-tests above, we conclude that the cases and controls
do not exhibit significant clustering (i.e., segregation).
However, NNCT-methods only provide information on spatial interaction
for distances about expected NN distance in the data set,
so it might be the case that the type and level of interaction
might still be different at larger scales (i.e., distances between the subjects' locations).
However, the locations of the subjects in this population (cases and controls together)
seem to be from an inhomogeneous Poisson process
(see also Figure \ref{fig:leukemia}).
Hence Ripley's $K$- or $L$-functions in the general form are not appropriate
to test for the spatial clustering of the cases (\cite{kulldorff:2006}).
However, \cite{diggle:2003} suggests a version based on Ripley's univariate $K$-functions
as $D(t)=K_{11}(t)-K_{22}(t)$ where $K_{ii}(t)$ is defined in Equation \eqref{eqn:Kii}.
In this setup, ``no spatial clustering" is equivalent to RL of cases and controls
on the locations in the sample, which implies $D(t)=0$,
since $K_{22}(t)$ measures the degree of spatial aggregation of the controls
(i.e., the population at risk), while $K_{11}(t)$ measures
this same spatial aggregation plus any additional clustering due to the disease.
The test statistic $D(t)$ is estimated by
$\widehat D(t)=\widehat K_{11}(t)-\widehat K_{22}(t)$,
where $\widehat K_{ii}(t)$ is as in Equation \eqref{eqn:Kiihat}.
Figure \ref{fig:leukemia-Dhat} shows the plot of $\widehat D(t)$
plus and minus two standard errors under RL.
Observe that at distances about 200 and 600 meters,
there is evidence for mild clustering of diseases
(i.e., segregation of cases from controls)
since the empirical function $\widehat D(t)$ gets close or a little
above of the upper limit.
At smaller scales, plot in Figure \ref{fig:leukemia-Dhat} is
consistent with the results of the NNCT analysis.
In particular average NN distance for leukemia data is 700 ($\pm$ 1400) m,
and NNCT analysis summarizes the pattern for about $t=1000$ m
which is depicted in Figure \ref{fig:leukemia-Dhat}.
This same data set was also analyzed
by (\cite{diggle:2003} pp 131-132) and similar plots and results were obtained.

\subsection{Pyramidal Neuron Data}
\label{sec:benes-data}
This data set gives the $(x,y)$-coordinates of pyramidal neurons in area 24, layer 2
of the cingulate cortex.
The data are taken from a unit square region
(unit of measurement unknown) in each of 31 subjects, grouped as
follows: controls consists of 12 subjects and correspond to cell numbers 1--655,
schizoaffectives consists of 9 subjects and correspond cell numbers 656--1061,
and schizophrenics consists of 10 subjects and correspond cell numbers 1062--1400.
Controls are the subjects with no previous history of any mental disorder,
schizoaffective disorder is a psychiatric disorder
where both the symptoms of mood disorder and psychosis occur,
and schizophrenia is a psychotic disorder
characterized by severely impaired thinking, emotions, and behavior.
\cite{diggle-benes:1991} applied several methods for the analysis of
the spatial distributions of pyramidal neurons in the cingulate cortex of human subjects
in three diagnostic groupings.
With a scaled Poisson analysis they found significant differences
between the groups in the mean numbers of neurons
in the sampled region, as well as a high degree of extra-Poisson
variation in the distribution of cell counts within these groups.
They employed two different functional descriptors of spatial
pattern for each subject to investigate departures from
completely random patterns, both between subjects and between groups,
while adjusting for cell count differences.
Since the distributions of their main functional pattern descriptor and
of their derived test statistic are unknown,
they applied a bootstrap procedure to attach $p$-values to their findings.

Since the definition of the rectangular domain for identifying neuron positions
is independent of neuronal cell density or the pattern
and this sampling domain is almost identical for each subject,
\cite{diggle-benes:1991} merged (pooled) the data for each group.
That is, the pyramidal neuron locations from control subjects
were pooled into one group, from schizoaffective into another,
and schizophrenic into another.
Although, the spatial distributions between subjects are not the same,
we think pooling the data by group might reveal more than what might be concealed.
\cite{diggle-benes:1991} computed and compared Ripley's univariate $K$ functions
to detect differences between patterns across the three groups.
Pattern analysis of the cellular arrangements demonstrated
significant departures from complete spatial randomness
in favor of spatial regularity for each group, in particular for schizophrenics.
On the pooled data, we also apply multivariate pattern tests.
In particular, we apply a $3 \times 3$ NNCT-analysis
and Ripley's bivariate $L$-functions (in addition to the univariate $L$-functions).

We plot the locations of these points in the study region
in Figure \ref{fig:benes} and provide the corresponding NNCT together with
percentages based on row and column sums in Table \ref{tab:NNCT-benes}.
Observe that the pyramidal neuron locations appear to be somewhat regularly
spaced, especially for schizophrenics.
Also the percentages are slightly smaller for the diagonal cells,
especially for the controls, compared to the marginal (row or column) percentages,
which might be interpreted as presence of a deviation from CSR independence
in favor of association.

\begin{figure}[ht]
\centering
\rotatebox{-90}{ \resizebox{3 in}{!}{\includegraphics{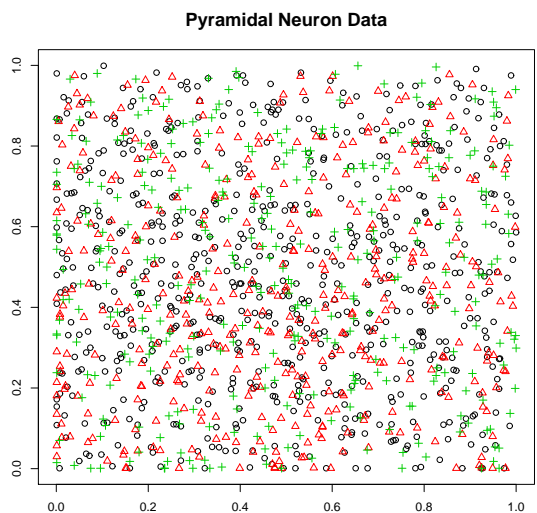} }}
 \caption{
\label{fig:benes}
The scatter plots of the locations of controls (circles $\circ$),
schizoaffectives (triangles $\triangle$), and schizophrenics (pluses $+$).}
\end{figure}

\begin{table}[ht]
\centering
\begin{tabular}{c}

\begin{tabular}{cc|ccc|c}
\multicolumn{2}{c}{}& \multicolumn{3}{c}{NN}& \\
\multicolumn{2}{c}{}&    Ctrl &  S.A. &  Sch.   &   sum  \\
\hline
& Ctrl &    271  &   216    &   171 & 658 \\
\raisebox{1.5ex}[0pt]{base}
& S.A. &    212  &   107    &  89 & 408 \\
& Sch. &    175  &   89    &  75  & 339 \\
\hline
&sum  &    658  &   412    &  335 & 1405 \\
\end{tabular}
\hspace{.25 cm}
\begin{tabular}{c|ccc|c}
& \multicolumn{3}{c}{NN}& \\
&    Ctrl &  S.A. &  Sch.   &   sum  \\
\hline
 Ctrl &  41 \%  &  33 \%    &   26 \% & 47 \% \\
 S.A. &  52 \%  &  26 \%    &   22 \%  & 29 \% \\
 Sch. &  52 \%  &  26 \%    &   22 \%  & 24 \% \\
\hline
  &    &      &    \\
\end{tabular}
\\
\begin{tabular}{c|ccc|c}
& \multicolumn{3}{c}{NN}& \\
&    Ctrl &  S.A. &  Sch.   &     \\
\hline
 Ctrl &  41 \%  &  52 \%    &   51 \% &  \\
 S.A. &  32 \%  &  26 \%    &   27 \%  & \\
 Sch. &  27 \%  &  22 \%    &   22 \%  &  \\
\hline
 sum & 47 \% & 29 \%   & 24 \%     &    \\
\end{tabular}

\end{tabular}
\caption{\label{tab:NNCT-benes}
The NNCT for the pyramidal neuron data (left)
and the corresponding percentages (middle and right).
Ctrl = Control, S.A. = Schizoaffective, and Sch. = Schizophrenics.}
\end{table}

\begin{table}[ht]
\centering
\begin{tabular}{|cc||c|c|c|c|}
\hline
\multicolumn{6}{|c|}{Test statistics and $p$-values for segregation tests} \\
\hline
&  & $\chi^2$-statistic & $p_{asy}$ & $p_{mc}$ & $p_{rand}$ \\
\hline
\multicolumn{2}{|c||}{overall} & 9.91 & .1283 & .1271 & 0.1280 \\
\hline
     & Ctrl    & 7.43 & .0243 & .0267 & .0230    \\
base & S.A.   & 4.19 & .1229 & .1208 & .1233    \\
     & Sch.     & 3.12 & .2098 & .2104 & .2101    \\
\hline
     & Ctrl    & 9.57 & .0226 & .0229 & .0241    \\
NN   & S.A.   & 6.36 & .0953 & .0968 & .0936    \\
     & Sch.     & 2.91 & .4060 & .4052 & .4058    \\
\hline
\end{tabular}
\caption{
\label{tab:pval-benes-class}
The $\chi^2$-statistics for the segregation tests and the corresponding $p$-values (in parentheses).
Ctrl = Control, S.A. = Schizoaffective, and Sch. = Schizophrenics.}
\end{table}

The locations of the pyramidal neurons can be viewed a priori resulting
from different processes, so the more appropriate null hypothesis is the CSR independence pattern.
Hence our inference will be a conditional one (see Remark \ref{rem:QandR}).
We calculate $Q=888$ and $R=892$ for this data set.
In Table \ref{tab:pval-benes-class},
we present the overall test of segregation,
class-specific test statistics and the associated $p$-values
based on the asymptotic approximation,
Monte Carlo simulation, and Monte Carlo randomization.
Observe that $p_{asy}$, $p_{mc}$, and $p_{rand}$ are similar for each test.
The overall test of segregation is not significant;
however, base- and NN-class-specific tests are significant for controls only,
implying significant deviation in row 1 and column 1
(i.e., row and column for controls) than expected under CSR independence.
Based on the NNCT for this data set,
we observe that the deviation is toward association
of controls with schizoaffectives and vice versa.
As mentioned in \cite{diggle-benes:1991},
the locations of the pyramidal neurons for controls and schizoaffectives
are more similar than those of schizophrenics.
Thus, it is possible to support to the idea that alterations
of cortical function may occur in schizophrenics,
which might agree with an earlier suggestion
that neuronal numbers are lower in the cerebral cortex of schizophrenic patients.

\begin{figure}[t]
\centering
\rotatebox{-90}{ \resizebox{2 in}{!}{\includegraphics{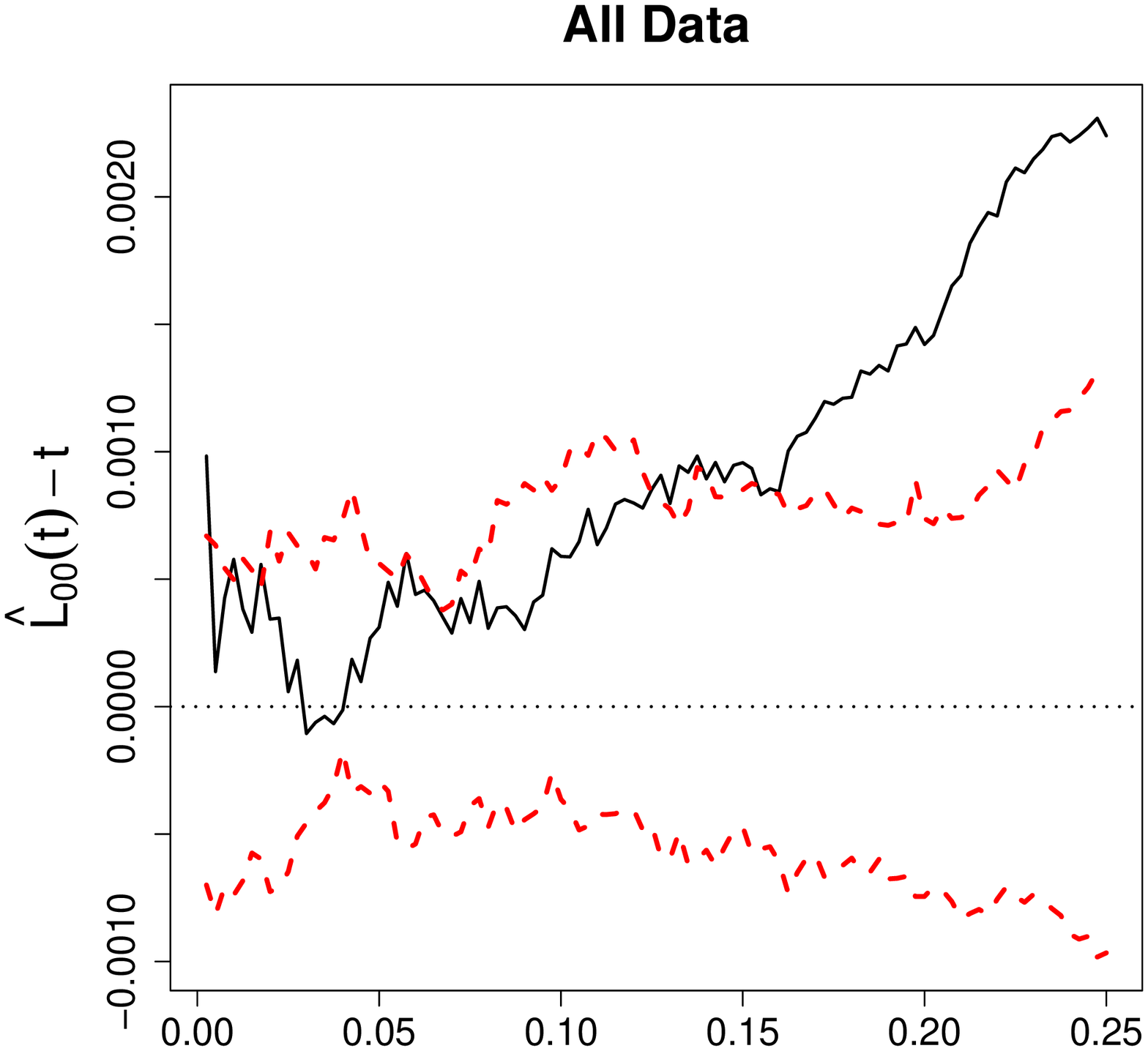} }}
\rotatebox{-90}{ \resizebox{2 in}{!}{\includegraphics{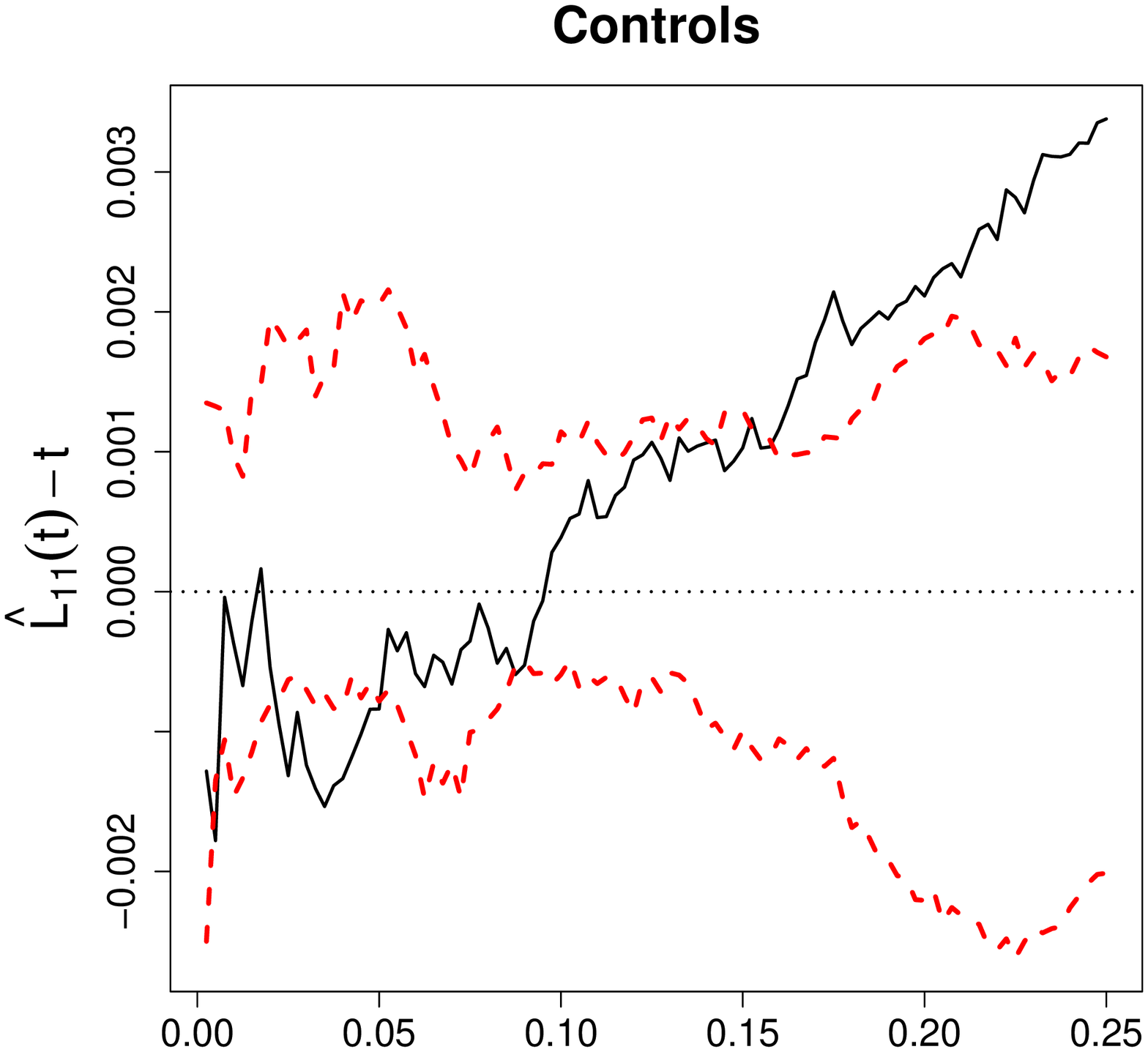} }}
\rotatebox{-90}{ \resizebox{2 in}{!}{\includegraphics{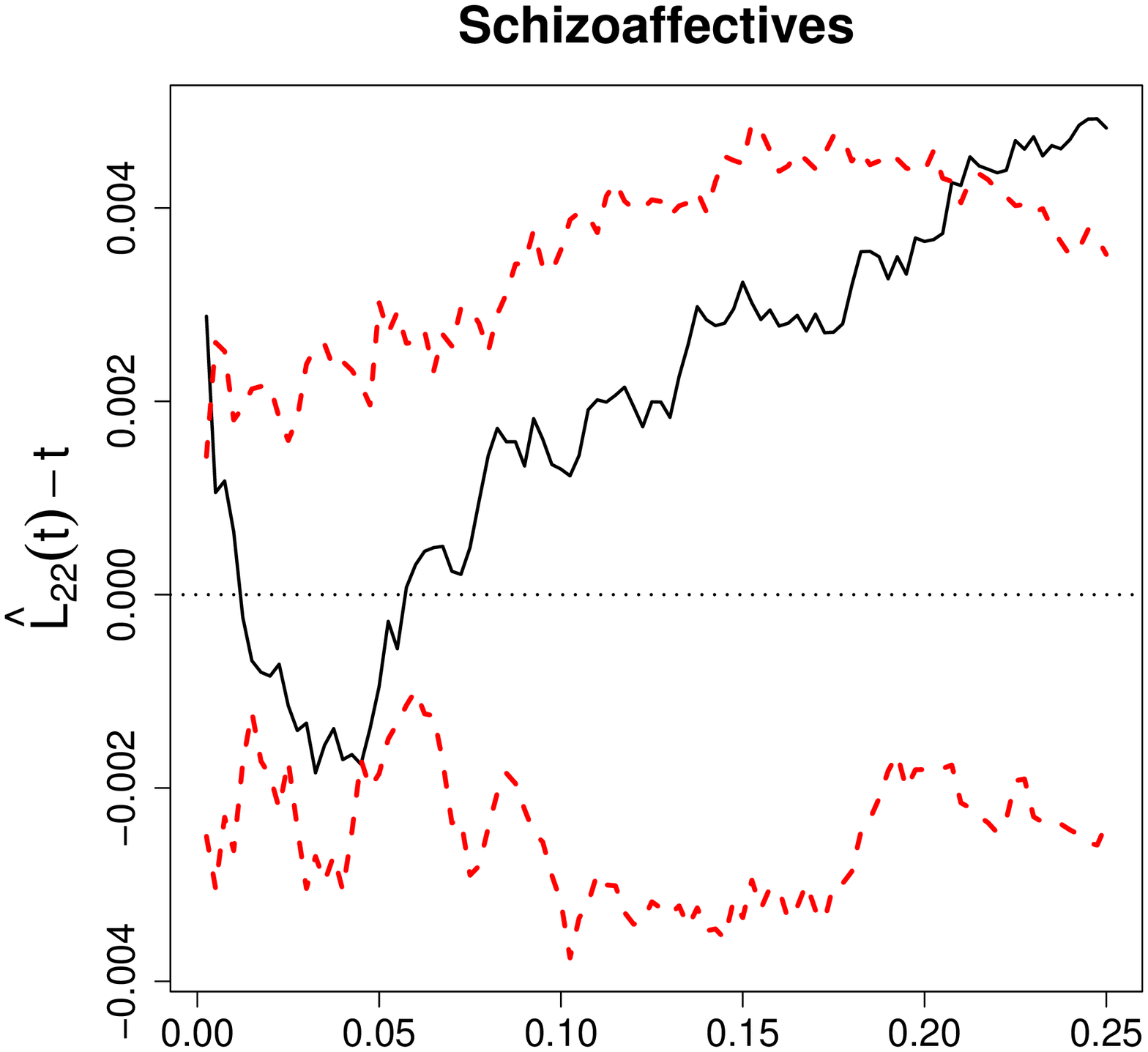} }}
\rotatebox{-90}{ \resizebox{2 in}{!}{\includegraphics{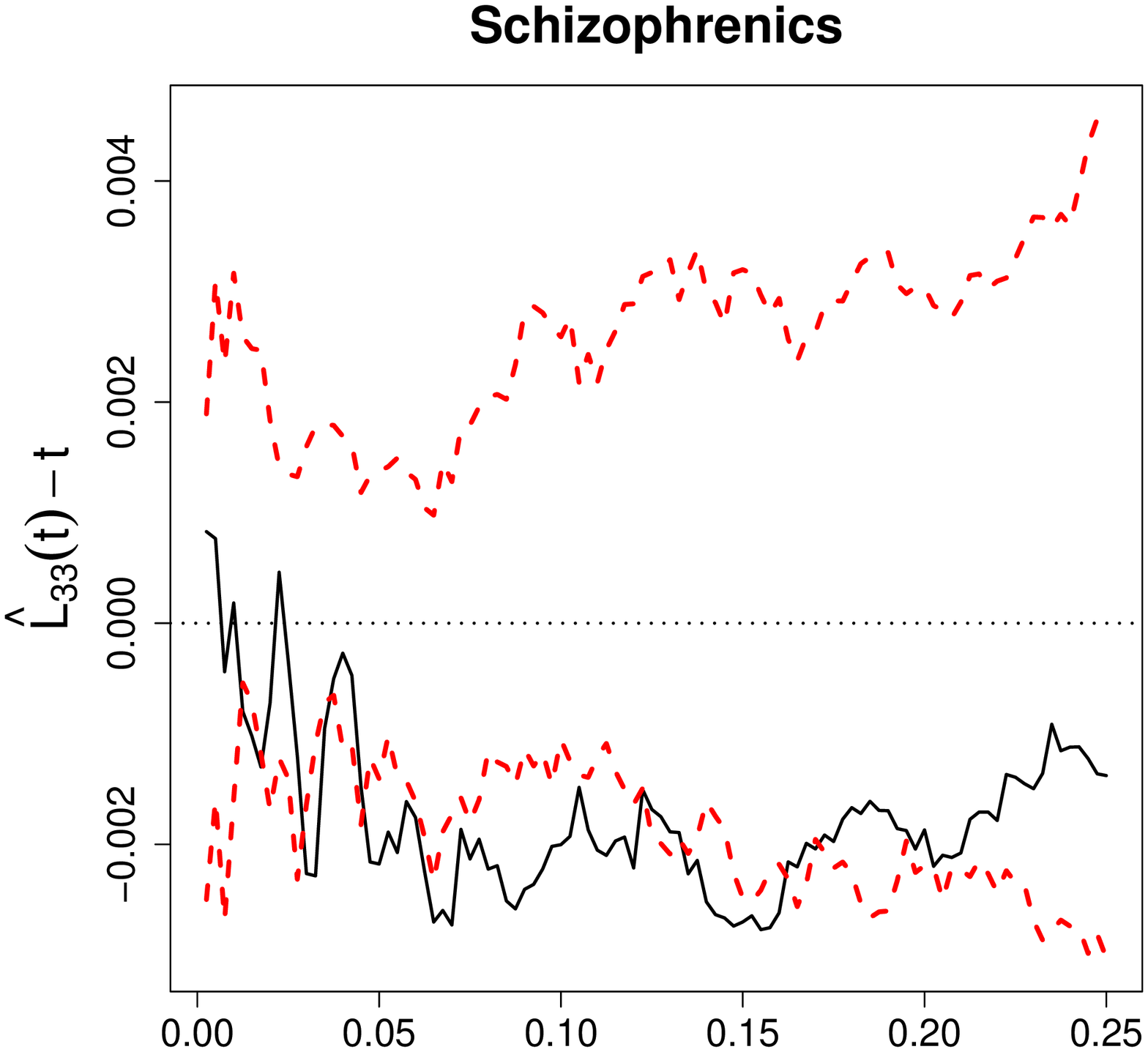} }}
\caption{
\label{fig:benes-Liihat}
Second-order properties of pyramidal neuron cells.
Functions plotted are Ripley's univariate $L$-functions
$\widehat{L}_{ii}(t)-t$ (left) for $i=0,1,2,3$
$i=0$ stands for all data combined, $i=1$ for controls, $i=2$ for schizoaffectives
and $i=3$ for schizophrenics.
Wide dashed lines around 0 are the upper and lower 95 \% confidence bounds for the
$L$-functions based on Monte Carlo simulations under the CSR independence pattern.}
\end{figure}

Based on the NNCT-tests above, we conclude that the pyramidal neurons exhibit
significant deviation from the CSR independence pattern for the control subjects
toward the association of controls and schizoaffectives.
Then, one might want to know what might be causing the association,
and what is the type and level of interaction at different scales
(i.e., distances between the neurons).
To answer such questions, we also present the second-order analysis of
the pyramidal neuron locations.
We plot Ripley's (univariate) $L$-function for all data combined
and for each (pooled) group in Figure \ref{fig:benes-Liihat}
where the upper and lower 95 \% confidence bounds are also provided.
Observe that for all neurons combined there is significant aggregation of locations
(the $L_{00}(t)-t$ curve is above the upper confidence bound) at distances $t > 0.18$.
For small scales ($t \approx 0.05$), controls and schizophrenics exhibit significant regularity.
At distances $t>0.20$, neurons of schizoaffectives tend to be aggregated,
while at distances $t>0.15$ neurons of controls tend to be aggregated,
while for distances up to 0.3, neurons of schizophrenics exhibit regularity.
This is along the lines of the NNCT analysis results, which indicate
deviation from CSR independence at smaller scales.
The less spatial regularity of the schizoaffectives compared
to the most regular pattern of schizophrenics might
explain the association of neurons of controls and schizoaffectives.

\begin{figure}[t]
\centering
\rotatebox{-90}{ \resizebox{2 in}{!}{\includegraphics{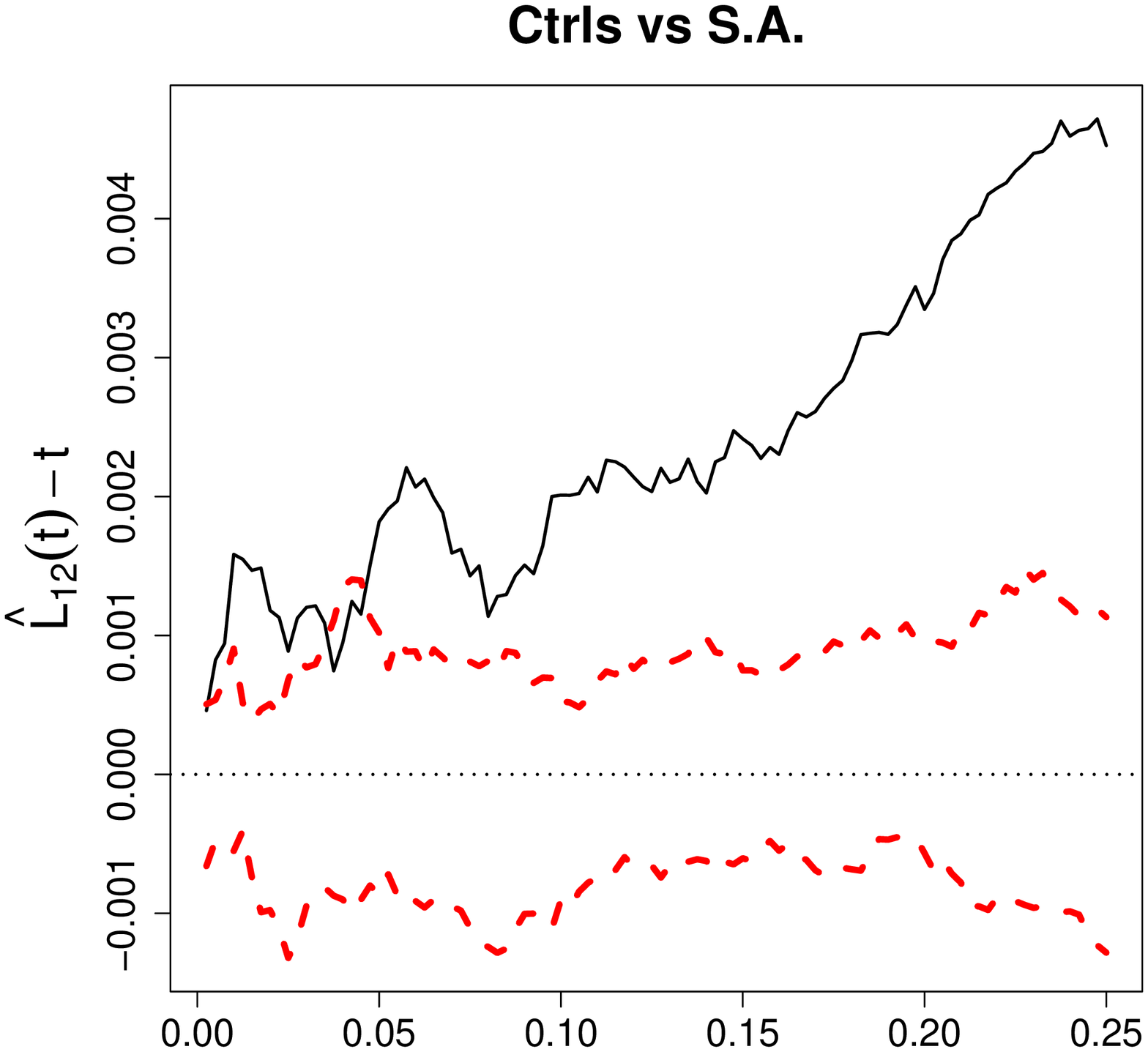} }}
\rotatebox{-90}{ \resizebox{2 in}{!}{\includegraphics{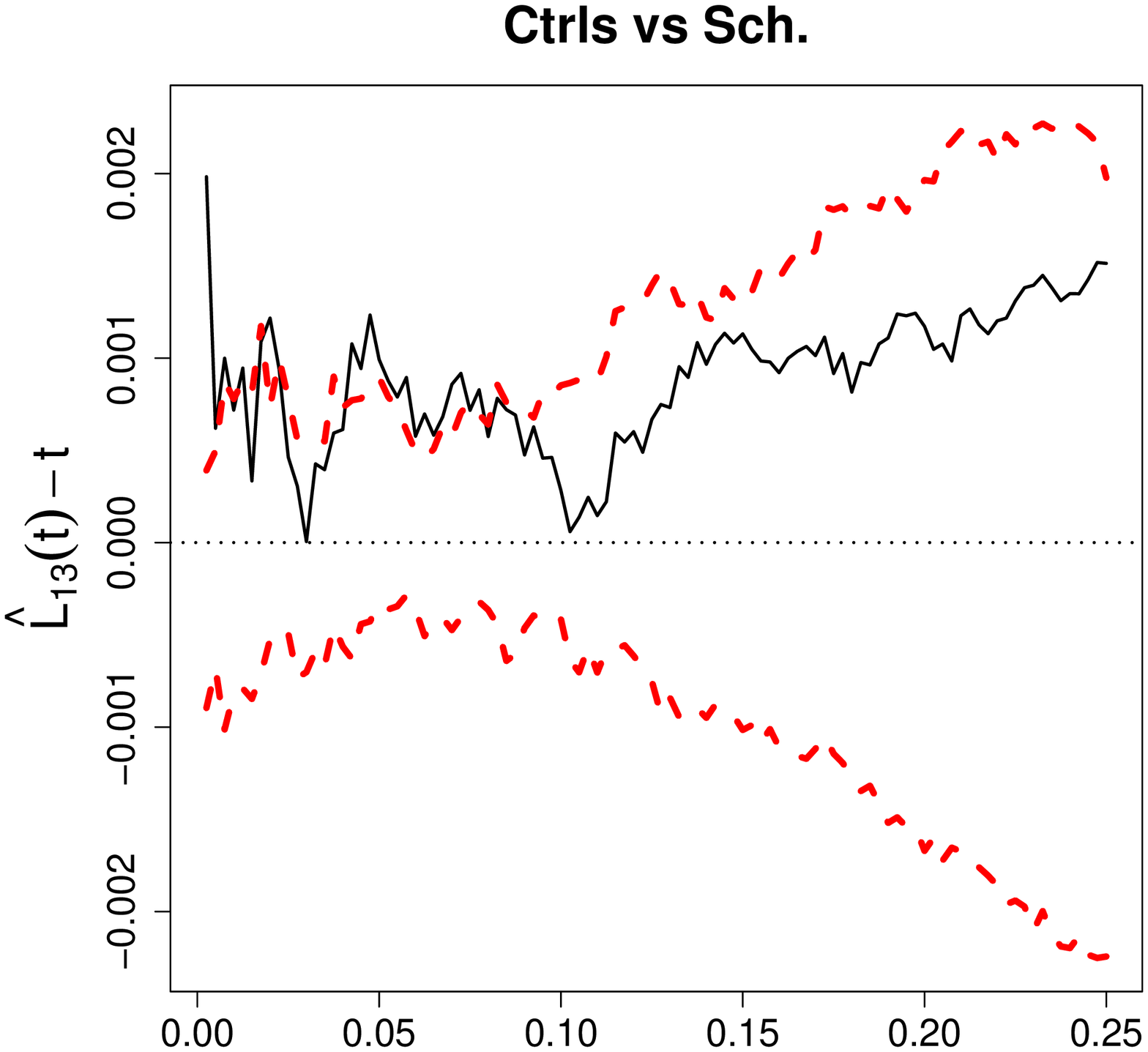} }}
\rotatebox{-90}{ \resizebox{2 in}{!}{\includegraphics{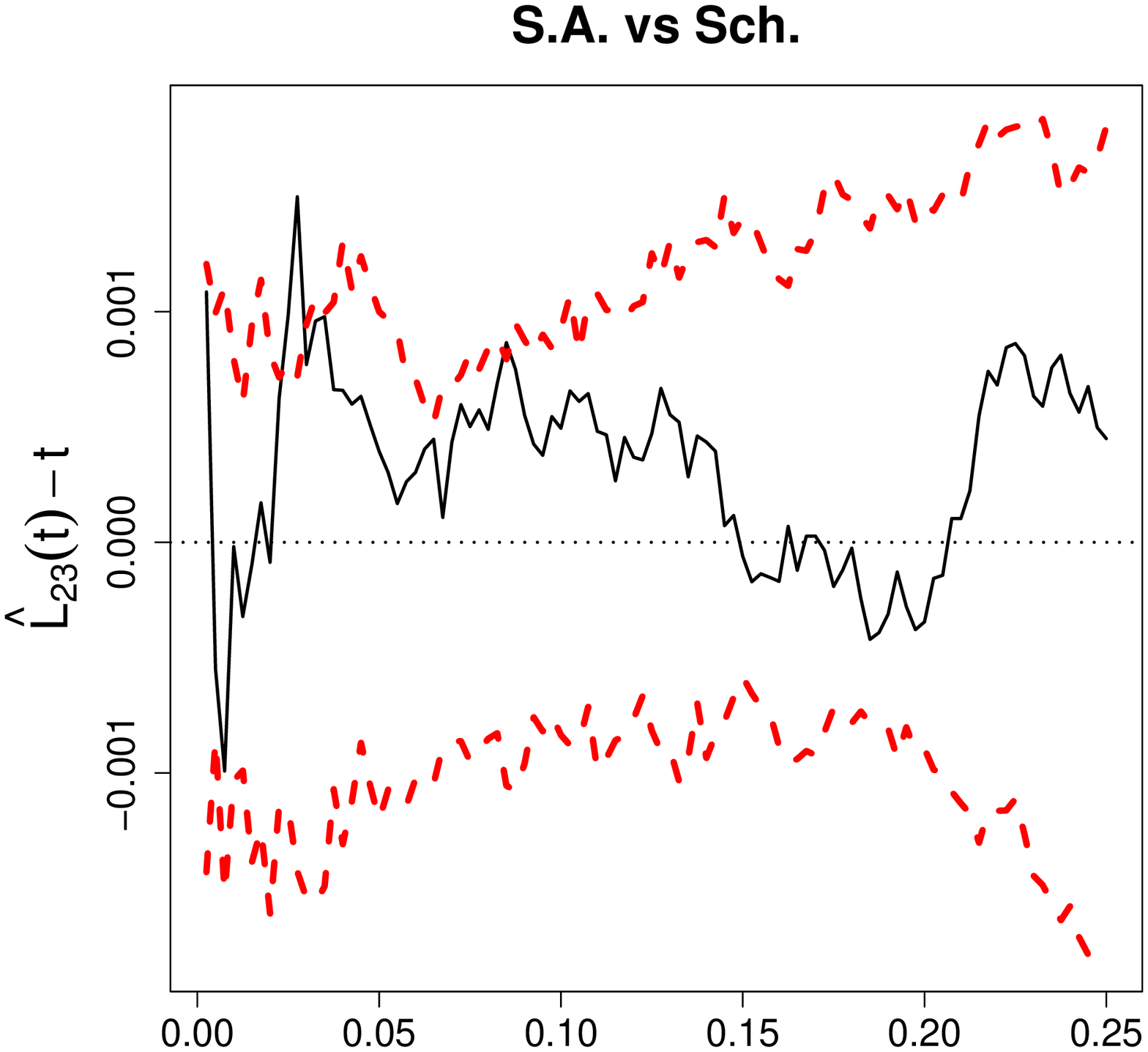} }}
\caption{
\label{fig:benes-Lijhat}
Second-order properties of pyramidal neuron cells.
Functions plotted are Ripley's bivariate $L$-functions
$\widehat{L}_{ij}(t)-t$ (right) for $i \not= j$
$i=1$ for controls, $i=2$ for schizoaffectives and $i=3$ for schizophrenics.
Wide dashed lines around 0 are the upper and lower 95 \% confidence bounds for the
$L$-functions based on Monte Carlo simulations under the CSR independence pattern.}
\end{figure}

We also plot Ripley's bivariate $L$-function
together with the upper and lower 95 \% confidence bounds
in Figure \ref{fig:benes-Lijhat}.
Observe that for almost all distances
controls and schizoaffectives exhibit significant spatial association,
while controls and schizophrenics exhibit significant spatial association
for distances $t\approx 0.04$, and schizoaffectives and schizophrenics
show significant association for distances $t\approx 0.04$ also.
The average NN distance for this data set is 0.014 ($\pm$ 0.007),
so at smaller scales (i.e., $t < 0.02$) the univariate and
bivariate $L$-functions seem to be in agreement with the NNCT results
which indicate the association of controls and schizophrenics.

Since Ripley's $K$-function is cumulative,
we also present the pair correlation functions
for all neurons and each group of neurons in Figure \ref{fig:benes-PCFii}.
Observe that all neurons are aggregated around distance values of 0.04, 0.09, 0.16, and 0.20;
control neurons are aggregated for distance values of 0.1, 0.17, and 0.25;
schizoaffective neurons are aggregated for distance values of 0.08, 0.10, 0.12, 0.16, and 0.20;
and schizophrenic neurons exhibit regularity for distance values of 0.025.
Comparing Figures \ref{fig:benes-Liihat} and \ref{fig:benes-PCFii},
we see that Ripley's $L$ and pair correlation functions
detect the same patterns but with different distance values,
so that Ripley's $L$ implies that the particular pattern is significant
for a wider range of distance values compared to $g(t)$.
But the results based on pair correlation function $g(t)$
are more reliable.

The bivariate pair correlation functions for the
groups in pyramidal neuron tree data are plotted in Figure \ref{fig:benes-PCFij}.
Observe that control and schizoaffective neurons are associated for distance values of 0.04, 0.08, and 0.15;
control and schizophrenic neurons do not significantly deviate from CSR independence
for the considered range of distances, neither do the schizoaffective and schizophrenic neurons.
Comparing Figures \ref{fig:benes-Lijhat} and \ref{fig:benes-PCFij}
we see that Ripley's $L$ and pair correlation functions
usually detect the same large-scale pattern but at narrower ranges of distance values
for the latter.

\begin{figure}[t]
\centering
\rotatebox{-90}{ \resizebox{2 in}{!}{\includegraphics{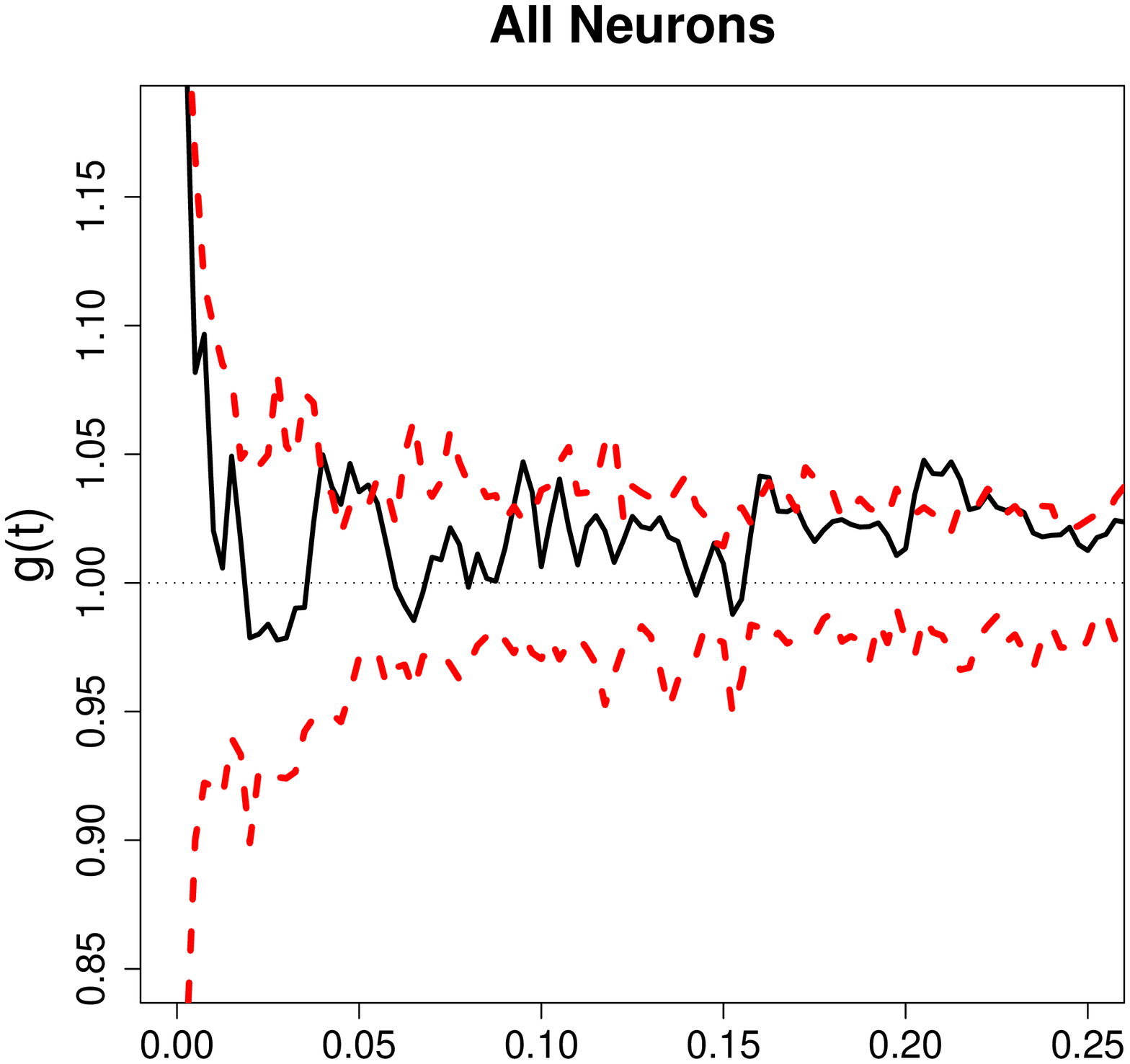} }}
\rotatebox{-90}{ \resizebox{2 in}{!}{\includegraphics{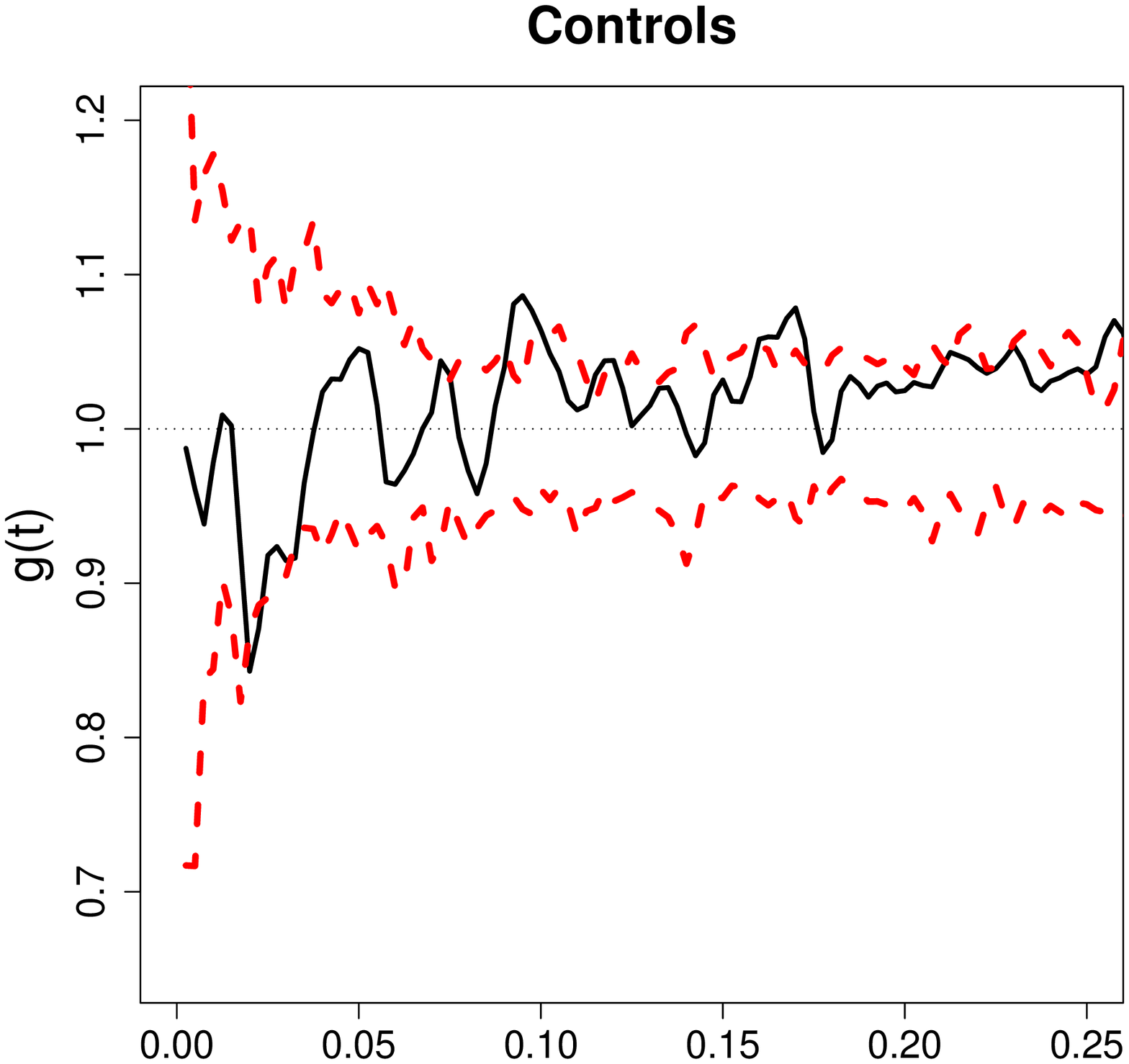} }}
\rotatebox{-90}{ \resizebox{2 in}{!}{\includegraphics{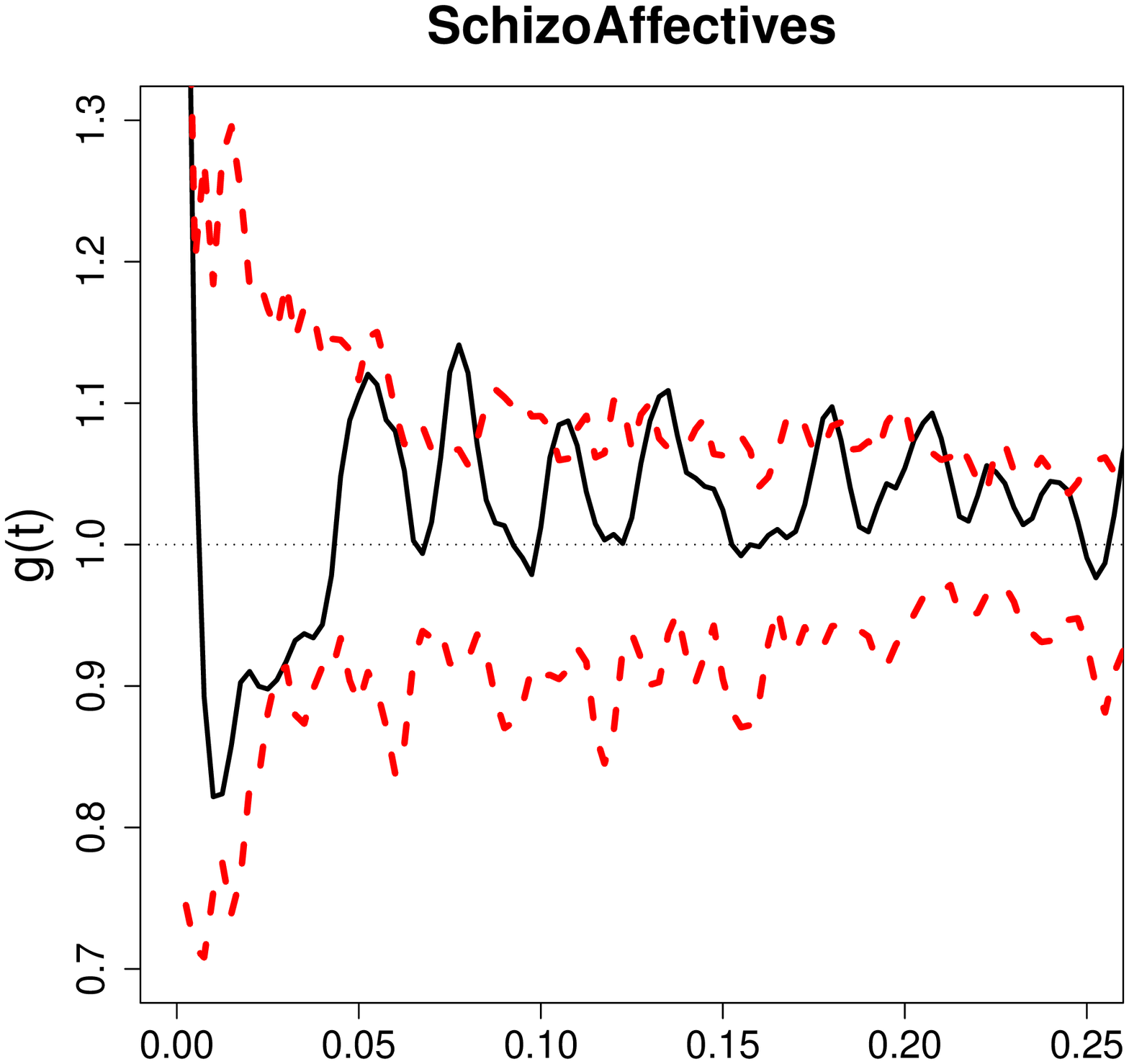} }}
\rotatebox{-90}{ \resizebox{2 in}{!}{\includegraphics{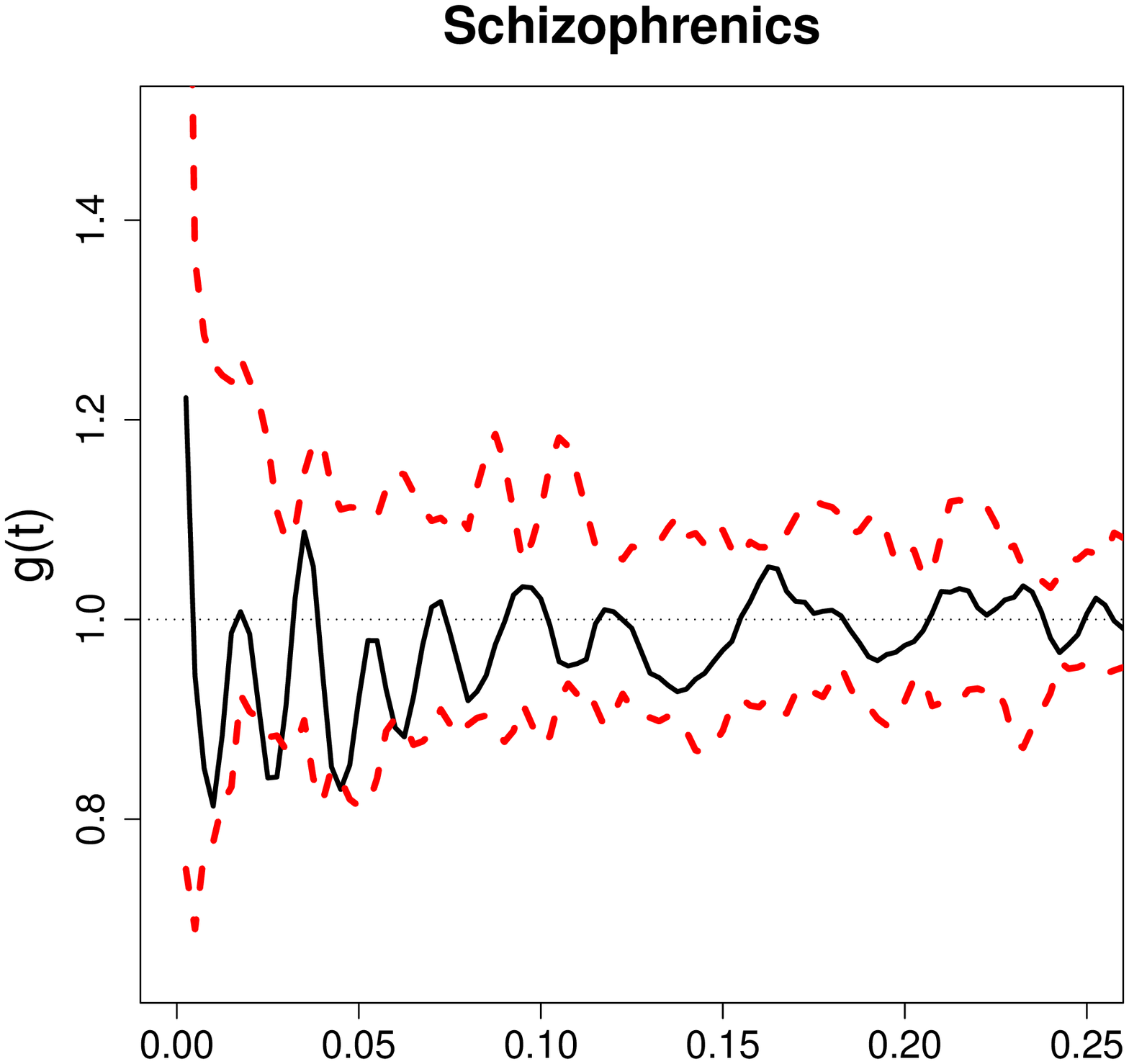} }}
\caption{
\label{fig:benes-PCFii}
Pair correlation functions for all neurons combined and for each group in the pyramidal neuron data.
Wide dashed lines around 1 (which is the theoretical value)
are the upper and lower (pointwise) 95 \% confidence bounds for the
$L$-functions based on Monte Carlo simulation under the CSR independence pattern.}
\end{figure}

\begin{figure}[t]
\centering
\rotatebox{-90}{ \resizebox{2 in}{!}{\includegraphics{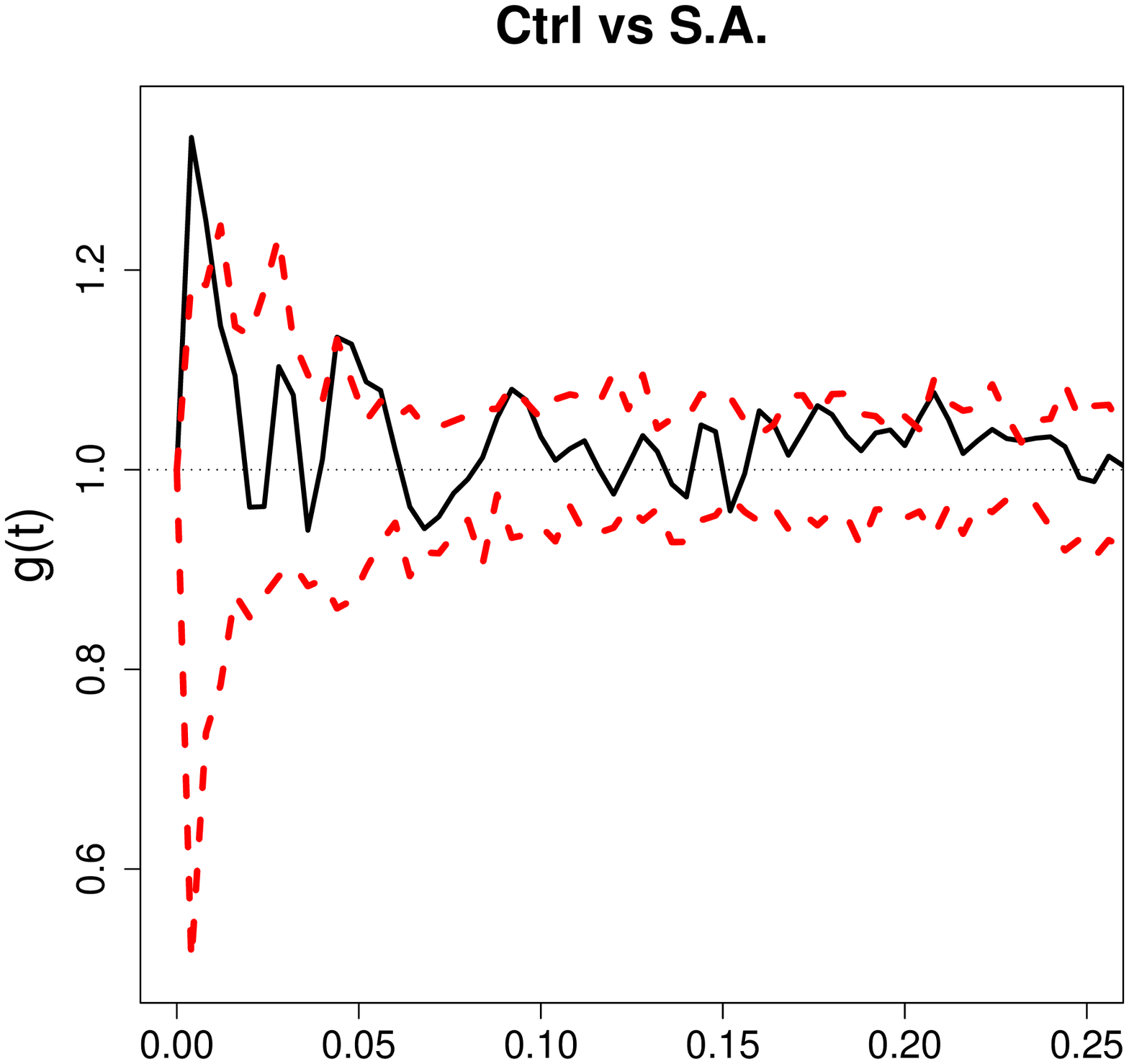} }}
\rotatebox{-90}{ \resizebox{2 in}{!}{\includegraphics{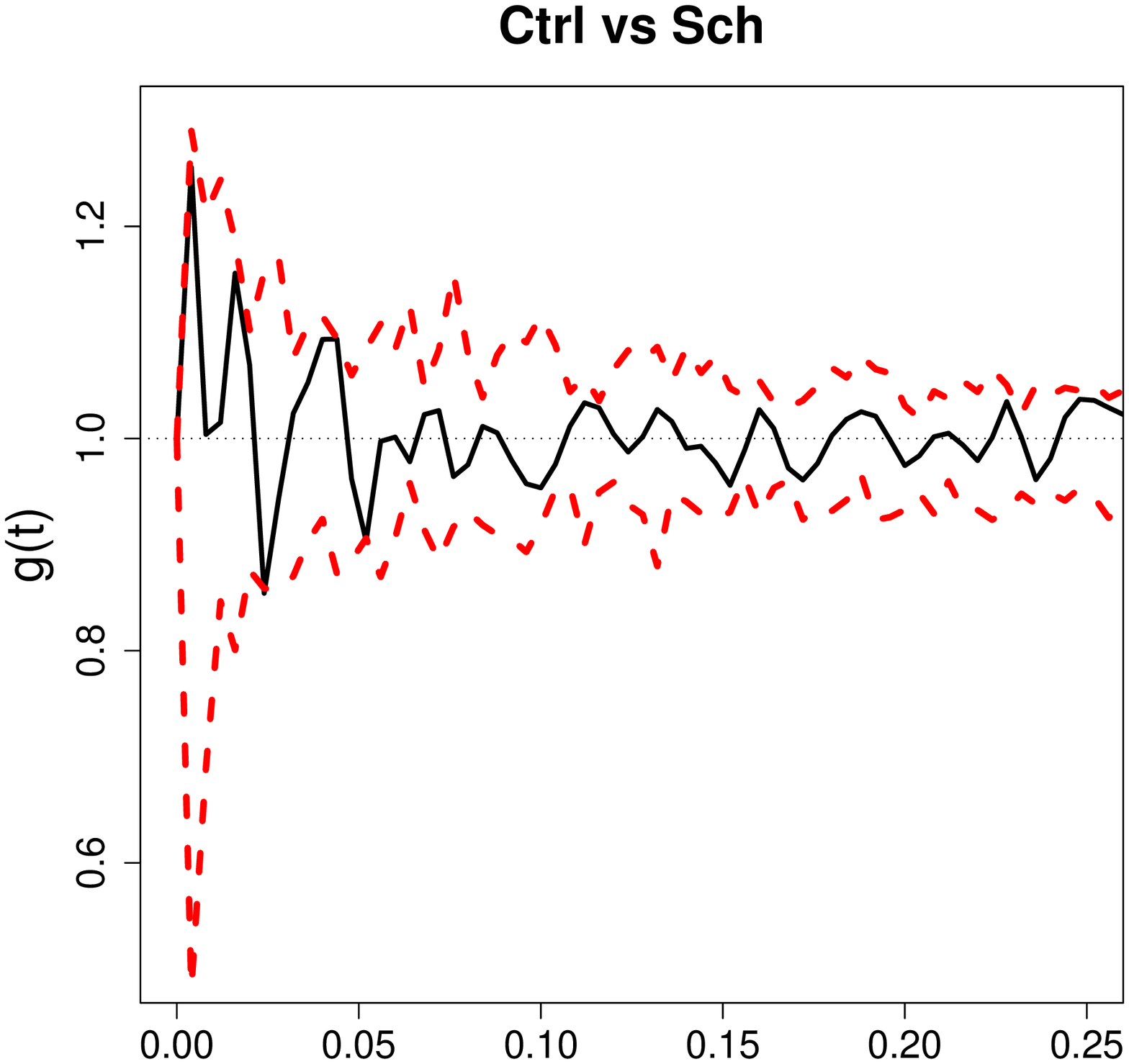} }}
\rotatebox{-90}{ \resizebox{2 in}{!}{\includegraphics{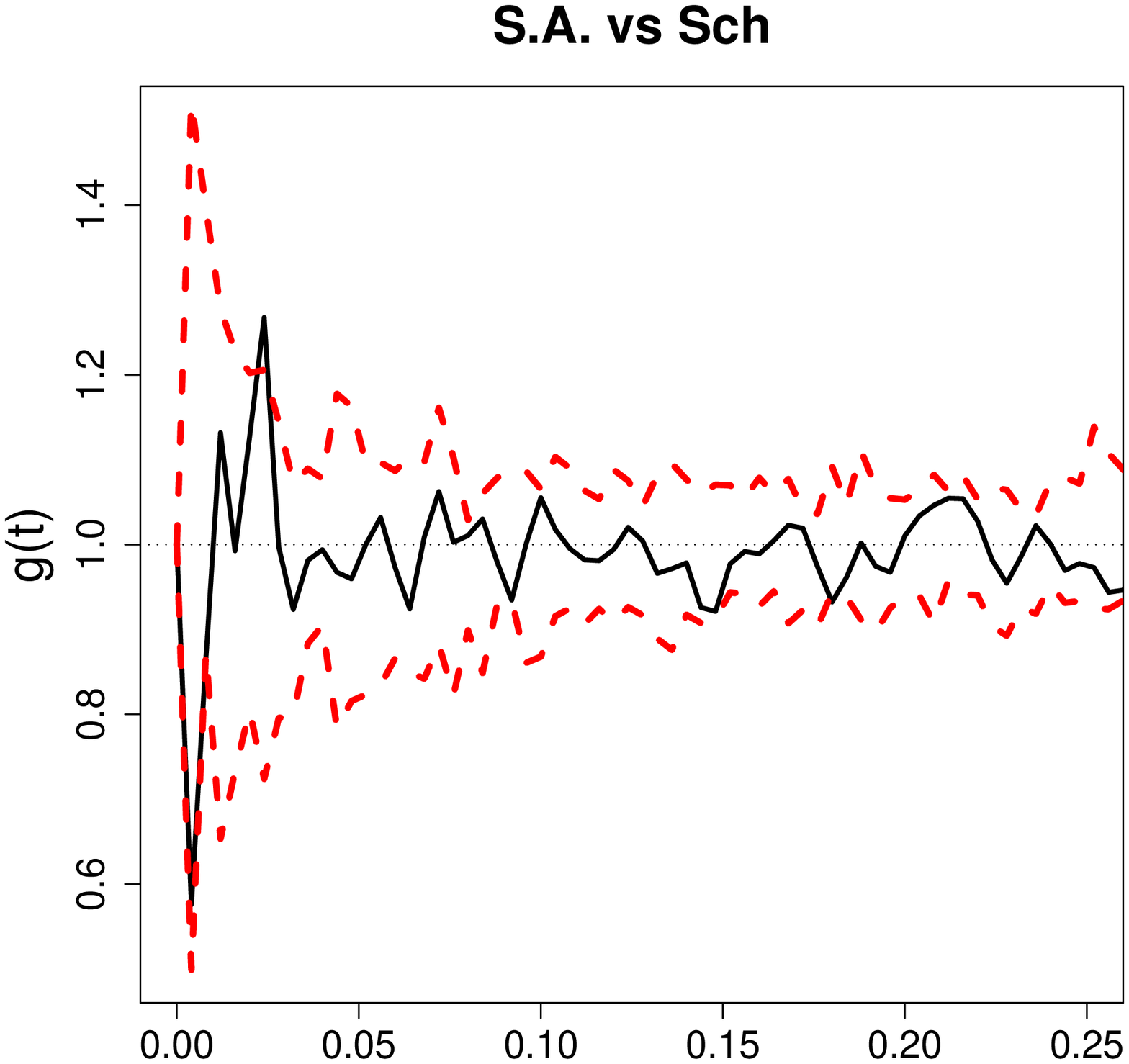} }}
\caption{
\label{fig:benes-PCFij}
Pair correlation functions for each pair of groups in the pyramidal neuron data.
Wide dashed lines around 1 (which is the theoretical value)
are the upper and lower (pointwise) 95 \% confidence bounds for the
$L$-functions based on Monte Carlo simulations under the CSR independence pattern.
Ctrl = Controls, S.A. = Schizoaffectives, Sch = Schizophrenics.}
\end{figure}

\section{Discussion and Conclusions}
\label{sec:disc-conc}
In this article we consider (Dixon's) overall and base-class-specific tests and
introduce NN-class-specific tests of segregation based on NNCTs.
We discuss the differences in these NNCT-tests,
present the asymptotic properties of them,
compare them using extensive Monte Carlo simulations
under CSR independence and RL and under various segregation and association alternatives
for two- and three-class cases.
We also illustrate the tests and compare them with Ripley's
$L$-function (\cite{ripley:2004}),
Diggle's $D$-function (\cite{diggle:2003} p. 131),
and pair correlation function (\cite{stoyan:1994})
on three data sets.

NNCT-tests (i.e., overall and class-specific tests of segregation)
are used in testing various forms of randomness in the
NN structure between two or more classes.
The overall test is used for testing any deviation
from randomness in the cells of the NNCT;
base-class-specific tests are used for testing any deviation from randomness
in the NN distribution of the base class in question
(i.e., any deviation in the corresponding row);
NN-class-specific tests are used for testing any deviation from randomness
in the base distribution of the classes to which the class in question serves as NN
(i.e., any deviation in the corresponding column).
The randomness in the NN structure is implied by the CSR independence or RL patterns.
However, we demonstrate that under the CSR independence pattern,
NNCT-tests are conditional on $Q$ and $R$,
while under the RL pattern, they are unconditional tests.
In the two-class case, both NN-class-specific tests and the overall test yield identical results.
Furthermore the three types of the NNCT-tests we consider
are all consistent under their respective null hypotheses.

The CSR independence pattern assumes that the study region
is unbounded for the analyzed pattern,
which is not the case in practice.
So the edge (or boundary) effects might confound the test results
in the analysis of empirical (i.e., bounded) data sets
if the null pattern is CSR independence and much effort has gone into the
development of edge correction methods (\cite{yamada:2003}).
Two correction methods for the edge effects on NNCT-tests,
namely \emph{buffer zone correction} and \emph{toroidal correction},
are investigated in (\cite{ceyhan:overall,ceyhan:cell-class-edge-correct})
where it is shown that the empirical sizes of the
NNCT-tests are not affected by the toroidal edge correction under CSR independence.
However, toroidal correction is biased for non-CSR patterns.
In particular if the pattern outside the plot (which is often unknown) is not the same as that inside it
it yields questionable results (\cite{haase:1995} and \cite{yamada:2003}).
The bias is more severe especially when there are clusters around the edges.
Under CSR independence, the (outer) buffer zone edge correction method
seems to have slightly stronger influence on the tests compared to toroidal correction.
But for these tests, buffer zone correction
does not change the sizes significantly for most sample size combinations.
This is in agreement with the findings of \cite{barot:1999}
who say that NN methods only require a small buffer area around the study region.
A large buffer area does not help much since one only
needs to be able to see far enough away from an event to find its NN.
Once the buffer area extends past the likely NN distances
(i.e., about the average NN distances),
it is not adding much helpful information for NNCTs.
Hence we recommend inner or outer buffer zone correction for NNCT-tests
with the width of the buffer area being about the average NN distance.
We do not recommend larger buffer areas,
since they are wasteful with little additional gain.
On the other hand, we recommend the use of toroidal edge correction
with points within the average NN distance in the additional copies
around the study region when there are no clusters around the edges
and it is reasonable to assume the replicated structure for the pattern.
For larger distances, the gain might not be worth the effort.
Empirical edge corrections are available for other nearest neighbor tests (\cite{sinclair:1985}).
They are potentially applicable to the NNCT-tests,
however the buffer zone correction analysis in (\cite{ceyhan:cell-class-edge-correct})
suggests that such a correction will not make much difference.

Based on our Monte Carlo simulations,
we conclude that the asymptotic approximation for the NNCT-tests
is appropriate only when each cell count in the NNCT is larger than 5;
when at least one cell count is $\leq 5$,
we recommend the Monte Carlo randomization version of the tests.
For large samples so that each cell count is larger than 5,
the NNCT-tests have similar empirical significance levels
which are about the nominal level.
When at least one sample size is small so that
some cell count(s) are $\leq 5$,
we compare the power of the tests using Monte Carlo critical values.
For large samples, the power comparisons can be made using both the asymptotic
or Monte Carlo critical values.
Under the segregation alternatives, for large samples,
we conclude that NN-class-specific tests have higher power;
when at least one sample is small, base-class-specific test
for the smaller (larger) class has the highest (lowest) power estimates.
Under the association alternatives,
we observe that when at least one sample size is small,
base-class-specific test for the smaller class has extremely poor performance,
while for the larger class it has higher power than the
NN-class-specific tests.
On the other hand, for larger samples, base-class-specific test for the
largest (smallest) class has the highest (lowest) power.
However base- and NN-class-specific tests answer different questions.
So it is not appropriate to compare the NNCT-tests,
as they provide related but different information about the NN structure in the data.

NNCT-tests summarize the pattern in the data set for small scales;
more specifically, they provide information on the pattern
around the average NN distance between the points.
On the other hand, pair correlation function $g(t)$
and Ripley's classical $K$ or $L$-functions and other variants provide
information on the pattern at various scales.
However, the classical $L$-function is not appropriate for the null pattern of RL
when locations of the points have spatial inhomogeneity.
For such cases, Diggle's $D$-function (\cite{diggle:2003} p. 131)
is more appropriate in testing the bivariate spatial clustering at various scales.

Ripley's classical $K$ or $L$-functions can be used when the null pattern
can be assumed to be CSR independence, that is, when the null pattern
assumes first-order homogeneity for each class.
When the null pattern is the RL of points from an inhomogeneous
Poisson process they are not appropriate (\cite{kulldorff:2006});
Cuzick-Edward's $k$-NN tests are designed for testing bivariate spatial interaction
and mostly used for spatial clustering of cases in epidemiology;
Diggle's $D$-function is a modified version of Ripley's $K$-function (\cite{diggle:2003})
and adjusts for any inhomogeneity in the locations of, e.g., cases and controls.
Furthermore, there are variants of $K(t)$ that explicitly correct for inhomogeneity
(see \cite{baddeley:2000b}).
Ripley's $K$-, Diggle's $D$-functions and pair correlation functions are designed to analyze
univariate or bivariate spatial interaction at various scales (i.e., inter-point distances).
Our examples illustrate that for distances around the average NN distance,
NNCT-tests and Ripley's $L$- and Diggle's $D$-functions yield similar results.
The NNCT-tests and the second-order analysis provide
very similar information for the two-class case at small scales.
However, for multi-class case with three or more classes,
NNCT-tests provide information on the multivariate spatial interaction
in one compound summary measure;
while the second-order analysis requires performing all bivariate
spatial interaction analysis.

The overall NNCT-test and Ripley's $L$-function provide
similar information in the two-class case at small scales.
For the $q$-class case with $q>2$ classes,
overall tests provide information on the (small-scale)
while the Ripley's $L$-function requires performing all bivariate
spatial interaction analysis.
On the other hand, the class-specific tests can serve as a mean of post hoc analysis
only when the overall test is significant.
Furthermore, the class-specific tests are testing
the spatial interaction of one class with all classes (including itself)
as part of the multivariate interaction between all the classes.
On the other hand, Ripley's univariate $K$- or $L$-functions
are restricted to one class and bivariate $K$- or $L$-functions
are restricted to two classes they pertain to,
ignoring the potentially important multivariate interaction
between all classes in the study area.
However, there are forms of the $J$-function
which is derived from the well-known $G$ and $F$ functions (\cite{lieshout:1999})
and deal with this multi-type setting (i.e., consider the pattern of type $i$ in
the context of the pattern of all other types).
\cite{lieshout:1999} define two basic types of $J$-functions.
First is a type-$i$-to-type-$j$ function
which considers the points of type $i$ in the context of the points of type $j$.
The second one is the type-$i$-to-any-type function
which considers the points of type $i$ in the context of points of
all types including type $i$.
Other forms can be derived from them by re-defining the types.
For example, if we want to consider the points of type $i$ in the context of
points of all other types,
then we collapse all the other types $j$
(i.e., all $j$ which are not equal to $i$) into a single
type $i'$ and then use the type-$i$-to-type-$i'$ function.
Several authors have written about the bivariate $K$-function,
which is of the type-$i$-to-type-$j$ form (\cite{diggle:1991}, \cite{haase:1995},
and \cite{diggle:2003}).
Type-$i$-to-type-$j$ $K$-function can easily be modified to
type-$i$-to-any-type $K$-function.
Thus, essentially there is only one family of multi-type $K$-functions in literature.
But type-$i$-to-type-$j$ $K$-function is comparable with a NNCT analysis
based on a $2\times2$ NNCT restricted to the classes $i$ and $j$.
Similarly, type-$i$-to-type-$i'$ $K$-function
is comparable with the NNCT analysis based on a
$2 \times 2$ NNCT with classes $i$ and the rest of the classes labeled as $i'$.
Furthermore, type-$i$-to-type-$i'$ $K$-function
may yield similar information with the base-class-specific test for class $i$
at small scales,
but the $K$-function ignores the interaction of class $i$ with itself.
On the other hand, by definition none of the variants of the $K$-function is comparable
with the NN-class-specific tests,
as they provide information on very different aspects of the spatial interaction between classes.

Since pairwise analysis of $q$ classes with $2 \times 2$ NNCTs
might yield conflicting results compared to $q \times q$ NNCT analysis (\cite{dixon:NNCTEco2002}),
Ripley's $L$-function and NNCT-tests might also yield conflicting results at small distances.
Hence Ripley's $L$-function and NNCT-tests may provide similar but not identical information
about the spatial pattern and the latter might provide
small-scale interaction that is not detected by the former.
Since the pair correlation functions are derivatives of Ripley's $K$-function,
most of the above discussion holds for them also,
except $g(t)$ is reliable only for large scale interaction analysis.
Hence NNCT-tests and pair correlation function are not comparable but
provide complimentary information about the pattern in question.

For a data set for which CSR independence is the reasonable null pattern,
we recommend the overall segregation test
if the question of interest is the spatial interaction at small scales
(i.e., about the mean NN distance).
If it yields a significant result, then to determine which
classes have significant spatial interaction with themselves and all other classes,
the class-specific tests can be performed.
One can also perform Ripley's $K$ or $L$-function
and only consider distances up to around the average NN distance
and compare the results with those of NNCT analysis.
If the spatial interaction at higher scales is of interest,
pair correlation function is recommended (\cite{loosmore:2006}),
due to the cumulative nature of Ripley's $K$- or $L$-functions for larger distances.
On the other hand, if the RL pattern is the reasonable null pattern for the data,
we recommend the NNCT-tests if the small-scale interaction is of interest
and Diggle's $D$-function if the spatial interaction at higher scales is also of interest.

\section*{Acknowledgments}
I would like to thank the associate editor and the referees,
whose constructive comments and suggestions greatly improved the
presentation and flow of the paper.
Most of the Monte Carlo simulations presented in this article
were executed on the Hattusas cluster of
Ko\c{c} University High Performance Computing Laboratory.


\begin{thebibliography}{}

\bibitem[Armstrong and Irvine, 1989]{armstrong:1989}
Armstrong, J.~E. and Irvine, A.~K. (1989).
\newblock Flowering, sex ratios, pollen-ovule ratios, fruit set, and
  reproductive effort of a dioecious tree, \emph{{M}yristica {I}nsipida}
  (\emph{Myristicacea}), in two different rain forest communities.
\newblock {\em American Journal of Botany}, 76:75--85.

\bibitem[Baddeley et~al., 2000]{baddeley:2000b}
Baddeley, A., M{\o}ller, J., and Waagepetersen, R. (2000).
\newblock Non- and semi-parametric estimation of interaction in inhomogeneous
  point patterns.
\newblock {\em Statistica Neerlandica}, 54(3):329–--350.

\bibitem[Barot et~al., 1999]{barot:1999}
Barot, S., Gignoux, J., and Menaut, J.~C. (1999).
\newblock Demography of a savanna palm tree: predictions from comprehensive
  spatial pattern analyses.
\newblock {\em Ecology}, 80:1987--2005.

\bibitem[Ceyhan, 2007]{ceyhan:cell-class-edge-correct}
Ceyhan, E. (2007).
\newblock Edge correction for cell- and class-specific tests of segregation
  based on nearest neighbor contingency tables.
\newblock In {\em Proceedings of the International Conference on Environment:
  Survival and Sustainability, Near East University.}

\bibitem[Ceyhan, 2008a]{ceyhan:overall}
Ceyhan, E. (2008a).
\newblock On the use of nearest neighbor contingency tables for testing spatial
  segregation.
\newblock {\em Environmental and Ecological Statistics.
  doi:10.1007/s10651-008-0104-x}.

\bibitem[Ceyhan, 2008b]{ceyhan:2008cell}
Ceyhan, E. (2008b).
\newblock Overall and pairwise segregation tests based on nearest neighbor
  contingency tables.
\newblock {\em Computational Statistics and Data Analysis.
  doi:10.1016/j.csda.2008.08.002}.

\bibitem[Coomes et~al., 1999]{coomes:1999}
Coomes, D.~A., Rees, M., and Turnbull, L. (1999).
\newblock Identifying aggregation and association in fully mapped spatial data.
\newblock {\em Ecology}, 80(2):554--565.

\bibitem[Cox, 1981]{cox:1981}
Cox, T.~F. (1981).
\newblock Reflexive nearest neighbours.
\newblock {\em Biometrics}, 37(2):367--369.

\bibitem[Cressie, 1993]{cressie:1993}
Cressie, N. A.~C. (1993).
\newblock {\em Statistics for Spatial Data}.
\newblock Wiley, New York.

\bibitem[Cuzick and Edwards, 1990]{cuzick:1990}
Cuzick, J. and Edwards, R. (1990).
\newblock Spatial clustering for inhomogeneous populations (with discussion).
\newblock {\em Journal of the Royal Statistical Society, Series B}, 52:73--104.

\bibitem[Diggle, 2003]{diggle:2003}
Diggle, P.~J. (2003).
\newblock {\em Statistical Analysis of Spatial Point Patterns}.
\newblock Hodder Arnold Publishers, London.

\bibitem[Diggle and Chetwynd, 1991]{diggle:1991}
Diggle, P.~J. and Chetwynd, A.~G. (1991).
\newblock Second-order analysis of spatial clustering for inhomogeneous
  populations.
\newblock {\em Biometrics}, 47:1155--1163.

\bibitem[Diggle et~al., 1991]{diggle-benes:1991}
Diggle, P.~J., Lange, N., and Benes, F. (1991).
\newblock Analysis of variance for replicated spatial point patterns in
  clinical neuroanatomy.
\newblock {\em Journal of American Statistical Association}, 86:618--625.

\bibitem[Dixon, 1994]{dixon:1994}
Dixon, P.~M. (1994).
\newblock Testing spatial segregation using a nearest-neighbor contingency
  table.
\newblock {\em Ecology}, 75(7):1940--1948.

\bibitem[Dixon, 2002a]{dixon:NNCTEco2002}
Dixon, P.~M. (2002a).
\newblock Nearest-neighbor contingency table analysis of spatial segregation
  for several species.
\newblock {\em Ecoscience}, 9(2):142--151.

\bibitem[Dixon, 2002b]{dixon:EncycEnv2002}
Dixon, P.~M. (2002b).
\newblock Nearest neighbor methods.
\newblock {\em Encyclopedia of Environmetrics, edited by Abdel H. El-Shaarawi
  and Walter W. Piegorsch, John Wiley \& Sons Ltd., NY}, 3:1370--1383.

\bibitem[Gavrikov and Stoyan, 1995]{gavrikov:1995}
Gavrikov, V. and Stoyan, D. (1995).
\newblock The use of marked point processes in ecological and environmental
  forest studies.
\newblock {\em Environmental and Ecological Statistics}, 2(4):331--344.

\bibitem[Good and Whipple, 1982]{good:1982}
Good, B.~J. and Whipple, S.~A. (1982).
\newblock Tree spatial patterns: {S}outh {C}arolina bottomland and swamp
  forests.
\newblock {\em Bulletin of the Torrey Botanical Club}, 109:529--536.

\bibitem[Goreaud and P\'{e}lissier, 2003]{goreaud:2003}
Goreaud, F. and P\'{e}lissier, R. (2003).
\newblock Avoiding misinterpretation of biotic interactions with the intertype
  ${K}_{12}$-function: population independence vs. random labelling hypotheses.
\newblock {\em Journal of Vegetation Science}, 14(5):681–--692.

\bibitem[Haase, 1995]{haase:1995}
Haase, P. (1995).
\newblock Spatial pattern analysis in ecology based on {R}ipley's
  {$K$}-function: {I}ntroduction and methods of edge correction.
\newblock {\em The Journal of Vegetation Science}, 6:575--582.

\bibitem[Hamill and Wright, 1986]{hamill:1986}
Hamill, D.~M. and Wright, S.~J. (1986).
\newblock Testing the dispersion of juveniles relative to adults: A new
  analytical method.
\newblock {\em Ecology}, 67(2):952--957.

\bibitem[Herler and Patzner, 2005]{herler:2005}
Herler, J. and Patzner, R.~A. (2005).
\newblock Spatial segregation of two common {G}obius species ({T}eleostei:
  {G}obiidae) in the {N}orthern {A}driatic {S}ea.
\newblock {\em Marine Ecology}, 26(2):121--129.

\bibitem[Herrera, 1988]{herrera:1988}
Herrera, C.~M. (1988).
\newblock Plant size, spacing patterns, and host-plant selection in
  \emph{{O}syris quadripartita}, a hemiparasitic dioecious shrub.
\newblock {\em Journal of Ecology}, 76:995--1006.

\bibitem[Kulldorff, 2006]{kulldorff:2006}
Kulldorff, M. (2006).
\newblock Tests for spatial randomness adjusted for an inhomogeneity: A general
  framework.
\newblock {\em Journal of the American Statistical Association},
  101(475):1289--1305.

\bibitem[Lahiri, 1996]{lahiri:1996}
Lahiri, S.~N. (1996).
\newblock On consistency of estimators based on spatial data under infill
  asymptotics.
\newblock {\em Sankhya: The Indian Journal of Statistics, Series A},
  58(3):403--417.

\bibitem[Loosmore and Ford, 2006]{loosmore:2006}
Loosmore, N. and Ford, E. (2006).
\newblock Statistical inference using the $g$ or $k$ point pattern spatial
  statistics.
\newblock {\em Ecology}, 87:1925--1931.

\bibitem[Meagher and Burdick, 1980]{meagher:1980}
Meagher, T.~R. and Burdick, D.~S. (1980).
\newblock The use of nearest neighbor frequency analysis in studies of
  association.
\newblock {\em Ecology}, 61(5):1253--1255.

\bibitem[Moran, 1948]{moran:1948}
Moran, P. A.~P. (1948).
\newblock The interpretation of statistical maps.
\newblock {\em Journal of the Royal Statistical Society, Series B},
  10:243--251.

\bibitem[Nanami et~al., 1999]{nanami:1999}
Nanami, S.~H., Kawaguchi, H., and Yamakura, T. (1999).
\newblock Dioecy-induced spatial patterns of two codominant tree species,
  \emph{{P}odocarpus nagi} and \emph{{N}eolitsea aciculata}.
\newblock {\em Journal of Ecology}, 87(4):678--687.

\bibitem[Penttinen et~al., 1992]{penttinen:1992}
Penttinen, A., Stoyan, D., and Henttonen, H. (1992).
\newblock Marked point processes in forest statistics.
\newblock {\em Forest Science}, 38(4):806--824.

\bibitem[Pielou, 1961]{pielou:1961}
Pielou, E.~C. (1961).
\newblock Segregation and symmetry in two-species populations as studied by
  nearest-neighbor relationships.
\newblock {\em Journal of Ecology}, 49(2):255--269.

\bibitem[Ripley, 2004]{ripley:2004}
Ripley, B.~D. (2004).
\newblock {\em Spatial Statistics}.
\newblock Wiley-{I}nterscience, New York.

\bibitem[Schlather et~al., 2004]{schlather:2004}
Schlather, M., Ribeiro~Jr, P., and Diggle, P. (2004).
\newblock Detecting dependence between marks and locations of marked point
  processes.
\newblock {\em Journal of the Royal Statistical Society: Series B (Statistical
  Methodology)}, 66(1):79–--93.

\bibitem[Searle, 2006]{searle:2006}
Searle, S.~R. (2006).
\newblock {\em Matrix Algebra Useful for Statistics}.
\newblock Wiley-{I}ntersciences.

\bibitem[Sinclair, 1985]{sinclair:1985}
Sinclair, D.~F. (1985).
\newblock On tests of spatial randomness using mean nearest neighbor distance.
\newblock {\em Ecology}, 66(3):1084--1085.

\bibitem[Stoyan and Stoyan, 1994]{stoyan:1994}
Stoyan, D. and Stoyan, H. (1994).
\newblock {\em Fractals, random shapes and point fields: methods of geometrical
  statistics.}
\newblock John Wiley and Sons, New York.

\bibitem[Stoyan and Stoyan, 1996]{stoyan:1996}
Stoyan, D. and Stoyan, H. (1996).
\newblock Estimating pair correlation functions of planar cluster processes.
\newblock {\em Biometrical Journal}, 38(3):259--271.

\bibitem[van Lieshout and Baddeley, 1999]{lieshout:1999}
van Lieshout, M.~N.~M. and Baddeley, A.~J. (1999).
\newblock Indices of dependence between types in multivariate point patterns.
\newblock {\em Scandinavian Journal of Statistics}, 26:511--532.

\bibitem[Waller and Gotway, 2004]{waller:2004}
Waller, L.~A. and Gotway, C.~A. (2004).
\newblock {\em Applied Spatial Statistics for Public Health Data}.
\newblock Wiley-Interscience, NJ.

\bibitem[Whipple, 1980]{whipple:1980}
Whipple, S.~A. (1980).
\newblock Population dispersion patterns of trees in a {S}outhern {L}ouisiana
  hardwood forest.
\newblock {\em Bulletin of the Torrey Botanical Club}, 107:71--76.

\bibitem[Wiegand et~al., 2007]{wiegand:2007}
Wiegand, T., Gunatilleke, S., and Gunatilleke, N. (2007).
\newblock Species associations in a heterogeneous {S}ri {L}ankan dipterocarp
  forest.
\newblock {\em The {A}merican {N}aturalist}, 170(4):77--95.

\bibitem[Yamada and Rogersen, 2003]{yamada:2003}
Yamada, I. and Rogersen, P.~A. (2003).
\newblock An empirical comparison of edge effect correction methods applied to
  {$K$}-function analysis.
\newblock {\em Geographical Analysis}, 35(2):97--109.

\end{thebibliography}

\end{document}